\documentclass{emulateapj}

\slugcomment{}

\shorttitle{AGN Outflows \& Star Formation}
\shortauthors{Farrah et al.}

\begin{document}

\title{Direct Evidence for Termination of Obscured Star Formation by Radiatively Driven Outflows in Reddened QSOs}

\author{Duncan Farrah\altaffilmark{1}, Tanya Urrutia\altaffilmark{2}, Mark Lacy\altaffilmark{3}, Andreas Efstathiou\altaffilmark{4}, Jose Afonso\altaffilmark{5,6}, Kristen Coppin\altaffilmark{7}, Patrick B. Hall\altaffilmark{8}, Carol Lonsdale\altaffilmark{3}, Tom Jarrett\altaffilmark{9}, Carrie Bridge\altaffilmark{10}, Colin Borys\altaffilmark{9}, Sara Petty\altaffilmark{11}}
\altaffiltext{1}{University of Sussex, Falmer, Brighton BN1 9QH, UK}
\altaffiltext{2}{Leibniz Institute for Astrophysics, An der Sternwarte 16, 14482 Potsdam, Germany}
\altaffiltext{3}{NRAO, 520 Edgemont Road, Charlottesville, VA 22903, USA}
\altaffiltext{4}{School of Sciences, European University Cyprus, Diogenes Street, Engomi, Nicosia, Cyprus}
\altaffiltext{5}{Observat\'{o}rio Astron\'{o}mico de Lisboa, Faculdade de Ci\^{e}ncias, Universidade de Lisboa, Tapada da Ajuda, 1349-018 Lisbon, Portugal}
\altaffiltext{6}{Centro de Astronomia e Astrof\'{\i}sica da Universidade de Lisboa, Lisbon, Portugal}
\altaffiltext{7}{Department of Physics, McGill University, Ernest Rutherford Building, 3600 Rue University, Montr\'{e}al, Qu\'{e}bec, H3A 2T8, Canada}
\altaffiltext{8}{Department of Physics \& Astronomy, York University, 4700 Keele Street, Toronto, ON M3J 1P3, Canada}
\altaffiltext{9}{Infrared Processing and Analysis Center, MS220-6, California Institute of Technology, Pasadena, CA 91125, USA}
\altaffiltext{10}{Division of Physics, Mathematics, and Astronomy, California Institute of Technology, Pasadena, CA 91125, USA}
\altaffiltext{11}{UCLA, Physics and Astronomy Building, 430 Portola Plaza, Box 951547, Los Angeles, CA 90095, USA}

\begin{abstract}
We present optical to far-infrared photometry of 31 reddened QSOs that show evidence for radiatively driven outflows originating from AGN in their rest-frame UV spectra. We use these data to study the relationships between the AGN-driven outflows, and the AGN and starburst infrared luminosities. We find that FeLoBAL QSOs are invariably IR-luminous, with IR luminosities exceeding $10^{12}$L$_{\odot}$ in all cases. The AGN supplies 76\% of the total IR emission, on average, but with a range from 20\% to 100$\%$. We find no evidence that the absolute luminosity of obscured star formation is affected by the AGN-driven outflows. Conversely, we find an anticorrelation between the strength of AGN-driven outflows, as measured from the range of outflow velocities over which absorption exceeds a minimal threshold, and the contribution from star formation to the total IR luminosity, with a much higher chance of seeing a starburst contribution in excess of 25\% in systems with weak outflows than in systems with strong outflows. Moreover, we find no convincing evidence that this effect is driven by the IR luminosity of the AGN. We conclude that radiatively driven outflows from AGN can have a dramatic, negative impact on luminous star formation in their host galaxies. We find that such outflows act to curtail star formation such that star formation contributes less than $\sim 25\%$ of the total IR luminosity. We also propose that the degree to which termination of star formation takes place is not deducible from the IR luminosity of the AGN. 
\end{abstract}

\keywords{accretion, accretion discs. quasars: absorption lines. galaxies: starburst}

\section{Introduction}
The last decade has seen substantial progress in understanding how galaxies assemble their stars and central supermassive black holes (SMBHs). Recent results almost all point to the same conclusion; that the bulk of galaxy assembly occurred at $z\gtrsim0.7$, and that a significant fraction of it took place in obscured `bursts' of intense star formation and SMBH accretion. Indirect evidence for this comes from, for example, the stellar ages of low-redshift galaxies \citep{heavens04}, the existence of evolved elliptical galaxies at high redshifts \citep{dun96,ell97,rak07}, and studies of stellar mass assembly over wide redshift ranges \citep{glaz04,fon06,per08,march09,ilb10}. Direct evidence mostly comes from extragalactic infrared and millimetre imaging surveys, which find that the number density of IR-luminous galaxies increases dramatically with increasing redshift (e.g. \citealt{leflo05,chapman05,babbedge06,coppin06,shupe08,aus09,eales10}), and that much of the growth period of SMBHs was shrouded in dust \citep{mart05,alex08}. 

The importance of obscured starburst and AGN activity in assembling galaxies suggests that the two phenomena may affect each other. A link between them is implied by, for example, the tight correlation between the mass of the SMBH and both stellar bulge mass \citep{mag98,ferrarese00,tre02,marc03,shields03,har04} and dark matter halo mass \citep{ferrarese02}, and the coeval presence of both starburst and AGN activity in many IR-luminous systems at all redshifts (see e.g. \citealt{blain02,lag05,lonsdale06,hernan09}). 

\begin{deluxetable*}{clcccccccccccc}
\tabletypesize{\scriptsize}
\tablecaption{The Sample. \label{tablesample}}
\tablewidth{0pt}
\tablehead{
\colhead{ID} & \colhead{Name} & \colhead{z} & \colhead{SDSS} & \multicolumn{3}{c}{2MASS (or UKIDSS) } & \multicolumn{4}{c}{WISE} \\
   &                          &      & $m_z$ & $m_J$ &    $m_H$ &    $m_K$ & 3.4 & 4.6 & 12 & 22 
}
\startdata
1  & SDSS J011117.34+142653.6 & 1.15 & 17.37 & 		        16.05 &                  15.53 &                     14.78 & 13.40 & 12.15 & 9.22  & 6.80   \\
2  & SDSS J024254.66-072205.6 & 1.22 & 19.03 & 		        --    &                  --    &                     --    & 14.36 & 13.04 & 10.42 & 8.97 (2.2$\sigma$)    \\  
3  & SDSS J030000.57+004828.0 & 0.89 & 16.13 &		        15.09 &                  14.59 &                     14.11 & 12.68 & 11.40 & 8.31  & 5.96    \\ 
4  & SDSS J033810.84+005617.7 & 1.63 & 18.38 & 17.77\tablenotemark{a} & 17.32\tablenotemark{a} &    16.93\tablenotemark{a} & 16.06 & 14.42 & 11.07 & 8.34    \\   
5  & SDSS J081312.60+432640.0 & 1.09 & 18.82 &                  --    &   	         --    &                     --    & 15.15 & 13.81 & 10.65 & 8.51    \\  
6  & SDSS J083522.76+424258.3 & 0.81 & 17.24 & 		        15.95 &   	         15.70 &                     15.05 & 13.41 & 12.17 & 9.43  & 6.88    \\  
7  & SDSS J084044.41+363327.8 & 1.23 & 16.11 & 		        15.02 &   	         14.39 &                     13.89 & 12.74 & 11.46 & 8.74  & 6.11    \\  
8  & SDSS J091103.49+444630.4 & 1.30 & 19.19 & 		        --    &   	         --    &                     --    & 14.43 & 12.65 & 8.97  & 6.57    \\  
9  & SDSS J091854.48+583339.7 & 1.32 & 18.89 &	                --    &   	         --    &                     --    & 15.44 & 14.35 & 12.38 & 8.76 (1.3$\sigma$)  \\ 
10 & SDSS J100605.66+051349.0 & 0.97 & 18.28 & 16.62\tablenotemark{a} & 16.39\tablenotemark{a} &    15.26\tablenotemark{a} & 13.37 & 12.06 & 9.14  & 6.72    \\ 
11 & SDSS J101927.36+022521.4 & 1.36 & 18.00 & 		        16.37 &                  15.22 &    15.04\tablenotemark{a} & 14.03 & 12.78 & 9.49  & 6.74    \\
12 & SDSS J102036.10+602338.9 & 0.99 & 17.91 & 		        16.56 &	              $<$15.60 &                     15.40 & 13.99 & 12.57 & 9.54  & 7.14    \\ 
13 & SDSS J102358.97+015255.8 & 1.08 & 18.91 &		        --    & 17.40\tablenotemark{a} &    16.81\tablenotemark{a} & --    & --    & --    & --      \\
14 & SDSS J105748.63+610910.8 & 1.28 & 19.55 & 		        --    &                  --    &                     --    & 15.02 & 13.94 & 11.06 & 8.93    \\  
15 & SDSS J112526.12+002901.3 & 0.86 & 17.68 & 	 	        16.51 &               $<$16.32 &                     15.36 & --    & --    & --    & --      \\ 
16 & SDSS J112828.31+011337.9 & 0.89 & 18.02 & 16.55\tablenotemark{a} & 16.55\tablenotemark{a} &    15.91\tablenotemark{a} & 14.69 & 13.50 & 10.60 & 8.22    \\ 
17 & SDSS J112901.71+050617.0 & 1.29 & 18.93 & 		        --    & 16.15\tablenotemark{a} &    15.54\tablenotemark{a} & 14.67 & 13.74 & 11.55 & 8.58 (1.4$\sigma$)    \\  
18 & SDSS J114556.26+110018.4 & 0.93 & 18.53 & 		        --    & 16.85\tablenotemark{a} &    16.32\tablenotemark{a} & --    & --    & --    & --      \\ 
19 & SDSS J115436.60+030006.3 & 1.39 & 17.46 &		        16.05 &                  15.42 &                     15.23 & 13.87 & 12.38 &  9.27 & 7.29    \\
20 & SDSS J115852.86-004302.0 & 0.98 & 18.81 &			17.06 &                  16.61 &                     15.55 & --    & --    & --    & --      \\ 
21 & SDSS J120049.55+632211.8 & 0.89 & 18.47 &                  --    &                  --    &                     --    & 15.12 & 14.25 & 11.24 & 8.37    \\  
22 & SDSS J120627.62+002335.3 & 1.11 & 18.55 & 17.59\tablenotemark{a} & 16.76\tablenotemark{a} &    15.98\tablenotemark{a} & 14.50 & 13.23 & 10.65 & 8.51    \\
23 & SDSS J121441.42-000137.8 & 1.05 & 18.52 & 17.58\tablenotemark{a} & 16.86\tablenotemark{a} &    15.98\tablenotemark{a} & 14.15 & 12.76 & 10.04 & 8.11    \\ 
24 & SDSS J123549.95+013252.6 & 1.29 & 18.82 &                  --    & 16.41\tablenotemark{a} &    16.19\tablenotemark{a} & 15.00 & 13.89 & 10.93 & 8.61    \\ 
25 & SDSS J132401.53+032020.6 & 0.93 & 18.48 & 16.68\tablenotemark{a} & 16.58\tablenotemark{a} &    15.83\tablenotemark{a} & 14.41 & 13.34 & 10.84 & 7.28 (1.8$\sigma$)     \\ 
26 & SDSS J142703.62+270940.3 & 1.17 & 17.96 &                  --    &                  --    &                     --    & 14.32 & 12.93 & 10.04 & 7.73    \\
27 & SDSS J155633.77+351757.3 & 1.50 & 17.60 &                  15.91 &                  14.91 &                     14.79 & 13.19 & 11.72 &  8.75 & 6.67   \\
28 & SDSS J173753.97+553604.8 & 1.10 & 19.82 &                  --    &                  --    &                     --    & 15.73 & 14.36 & 11.43 & 8.77    \\  
29 & SDSS J210712.77+005439.4 & 0.92 & 19.69 &                  --    &                  --    &                     --    & 13.24 & 11.60 &  8.55 & 6.34     \\ 
30 & SDSS J221511.93-004549.9 & 1.48 & 16.59 &                  15.65 &                  14.89 &                     14.69 & --    & --    & --    & --     \\       
31 & SDSS J233646.20-010732.6 & 1.29 & 18.68 &                  17.24 &                  --    &                     16.23 & --    & --    & --    & --      
\enddata
\tablecomments{Positions, redshifts and SDSS magnitudes are taken from the SDSS DR6. For the SDSS magnitudes we used the PSF magnitudes in the AB system. For 2MASS, we used the `default' magnitudes in the Vega system, taken from the public 2MASS All-Sky Point Source Catalogue. WISE magnitudes from the operational database as of June 2011. An `--' indicates that the source is not in the catalogue in that band. }
\tablenotetext{a}{UKIDSS LAS magnitude}
\end{deluxetable*}

Recently, interest into the relationship between starburst and AGN activity has been stimulated by what first appeared to be irreconcilable differences between observational and theoretical results on the assembly history of galaxies. Foremost among these were the difficulties that models faced in explaining the observed galaxy luminosity function (LF) at low and high redshifts simultaneously; if the models were tuned to match the local galaxy LF then they underpredicted the number of massive galaxies observed at high redshifts, whereas the models that matched the number of high redshift galaxies gave poor fits to the local galaxy LF \citep{kauff99,cole00,somer01,benson03}. Other problems included: (1) the prediction that a higher fraction of rich galaxy clusters should harbour cooling flows than is observed \citep{peterson03,xu02,peterson06}, (2) the difficulties that early semi-analytic models faced in reproducing the large number of IR-luminous galaxies observed at high redshift (e.g. \citealt{baugh05}), and (3) the expectation that the mass return rate from stars would cause central SMBHs to be larger than is observed \citep{ciotti91,ciotti07}.

\begin{deluxetable*}{lccccccccccc}
\tabletypesize{\scriptsize}
\tablecaption{MIPS fluxes, absorption strength measures (see Equation \ref{eqn:bal} \& Table \ref{absmeasures}), and IR luminosities. \label{tableasandlirs}}
\tablewidth{0pt}
\tablehead{
 ID&\multicolumn{3}{c}{MIPS Fluxes (mJy)}           &\multicolumn{4}{c}{Absorption Strengths (km s$^{-1}$)} &\multicolumn{3}{c}{Infrared Luminosities (log (L$_{\odot}$))} & $\chi^{2}_{red}$ \\
   & 24$\mu$m     & 70$\mu$m      & 160$\mu$m      & AS$_{0}$ & AS$_{0}^{Gib}$ &AS$_{2}$ & AS$_{4}$ & L$_{Tot}$\tablenotemark{a} & L$_{AGN}$\tablenotemark{b} & L$_{SB}$\tablenotemark{c} & 
}
\startdata
1  & $16.9\pm0.8$ & $26.4\pm7.9$  & $15.4\pm17.1$  &        0 & --    &        0 &      348 & 13.10$^{+0.03}_{-0.04}$ & 13.06$^{+0.03}_{-0.02}$ & 11.97$^{+0.42}_{-0.40}$ &  0.7 \\ 
2  & $6.4 \pm0.3$ & $10.8\pm4.5$  & $16.2\pm14.6$  &        0 & 0     &      241 &     2030 & 12.70$^{+0.06}_{-0.07}$ & 12.62$^{+0.03}_{-0.09}$ & 12.00$^{+0.26}_{-0.91}$ &  2.6 \\ 
3  & $29.4\pm2.9$ & $56.3\pm10.6$ & $-8.8\pm16.0$  &     7001 & 15310 &     7098 &    12722 & 13.12$^{+0.03}_{-0.04}$ & 13.11$^{+0.02}_{-0.12}$ & $<11.95$                &  1.8 \\
4  & $2.4 \pm0.3$ & $3.6\pm6.5$   & $18.2\pm18.0$  &        0 & 0     &        0 &      749 & 12.76$^{+0.22}_{-0.15}$ & 12.62$^{+0.07}_{-0.11}$ & $<$12.23$^{+0.49}$      &  1.1 \\
5  & $5.4\pm0.5$  & $10.5\pm3.8$  & $12.7\pm14.7$  &        0 & --    &        0 &      143 & 12.51$^{+0.08}_{-0.05}$ & 12.41$^{+0.05}_{-0.10}$ & 11.86$^{+0.37}_{-0.54}$ &  0.7 \\ 
6  & $12.3\pm0.6$ & $33.1\pm7.2$  & $16.2\pm14.9$  &        0 & --    &        0 &     1335 & 12.69$^{+0.03}_{-0.03}$ & 12.60$^{+0.06}_{-0.13}$ & 11.90$^{+0.39}_{-0.21}$ &  1.9 \\ 
7  & $26.8\pm1.3$ & $43.7\pm7.8$  & $21.4\pm9.7$   &     5958 & 4863  &     8707 &    12164 & 13.33$^{+0.06}_{-0.02}$ & 13.30$^{+0.09}_{-0.01}$ & 12.04$^{+0.35}_{-0.34}$ &  1.4 \\ 
8  & $16.3\pm0.6$ & $51.1\pm7.8$  & $22.0\pm7.9$   &     3623 & 0     &     4095 &     8156 & 13.25$^{+0.08}_{-0.16}$ & 13.23$^{+0.04}_{-0.20}$ & 12.29$^{+0.09}_{-1.01}$ &  1.7 \\ 
9  & $1.0\pm0.3$  & $1.7\pm6.6$   & $24.4\pm14.3$  &      770 & --    &     1437 &     3479 & 12.29$^{+0.49}_{-0.22}$ & 12.09$^{+0.05}_{-0.26}$ & $<12.72$                &  1.0 \\ 
10 & $15.3\pm0.8$ & $49.2\pm16.4$ & $73.6\pm21.0$  &        0 & 0     &      267 &     2540 & 12.97$^{+0.01}_{-0.10}$ & 12.85$^{+0.02}_{-0.11}$ & 12.37$^{+0.17}_{-0.80}$ &  2.3 \\ 
11 & $16.0\pm0.8$ & $61.4\pm9.9$  & $28.7\pm11.5$  &     3360 & 1803  &     4255 &     8361 & 13.27$^{+0.06}_{-0.13}$ & 13.20$^{+0.09}_{-0.16}$ & 12.40$^{+0.19}_{-0.22}$ &  1.9 \\
12 & $8.8\pm0.9$  & $23.6\pm8.1$  & $10.9\pm12.4$  &     5206 & 5253  &     7372 &    10775 & 12.75$^{+0.05}_{-0.06}$ & 12.70$^{+0.07}_{-0.07}$ & 11.69$^{+0.32}_{-1.12}$ &  0.6 \\
13 & $0.9\pm0.1$  & $3.7\pm7.6$   & $14.1\pm11.7$  &        0 & --    &        0 &     2202 & 12.11$^{+0.20}_{-0.24}$ & 11.85$^{+0.03}_{-0.20}$ & 11.75$^{+0.50}_{-1.28}$ &  0.2 \\ 
14 & $3.4\pm0.3$  & $10.6\pm6.6$  & $36.7\pm15.4$  &        0 & --    &        0 &      509 & 12.65$^{+0.12}_{-0.21}$ & 12.39$^{+0.07}_{-0.10}$ & 12.30$^{+0.26}_{-1.25}$ &  1.1 \\ 
15 & $6.8\pm0.4$  & $15.5\pm5.7$  & $10.2\pm15.1$  &     4630 & 3716  &     5357 &     9705 & 12.52$^{+0.04}_{-0.05}$ & 12.48$^{+0.02}_{-0.07}$ & 11.41$^{+0.66}_{-1.21}$ &  0.3 \\
16 & $3.6\pm0.4$  & $4.4\pm6.5$   & $8.9\pm12.7$   &     4348 & 3769  &     5185 &     9020 & 12.28$^{+0.06}_{-0.05}$ & 12.16$^{+0.08}_{-0.04}$ & $<$11.66$^{+0.27}$      &  1.5 \\ 
17 & $4.0\pm0.3$  & $14.0\pm5.4$  & $21.8\pm11.5$  &     1572 & 673   &     1770 &     5249 & 12.60$^{+0.14}_{-0.11}$ & 12.40$^{+0.01}_{-0.04}$ & 12.15$^{+0.27}_{-0.25}$ &  3.0 \\
18 & $2.2\pm0.3$  & $1.5\pm6.4$   & $-13.4\pm12.2$ &     1498 & --    &     1500 &     4674 & 12.08$^{+0.11}_{-0.03}$ & 12.07$^{+0.10}_{-0.05}$ & $<11.57$                &  0.4 \\
19 & $9.3\pm0.5$  & $13.5\pm5.6$  & $15.0\pm16.5$  &    10369 & 10422 &    12585 &    17989 & 13.12$^{+0.06}_{-0.10}$ & 13.01$^{+0.12}_{-0.03}$ & $<$12.17$^{+12.52}$     &  1.3 \\ 
20 & $3.7\pm0.3$  & $-5.5\pm7.5$  & $15.2\pm14.2$  &        0 & --    &      377 &      972 & 12.38$^{+0.15}_{-0.06}$ & 12.31$^{+0.05}_{-0.09}$ & $<$11.31$^{+0.91}$      &  0.7 \\ 
21 & $4.6\pm0.4$  & $19.0\pm7.3$  & $9.9\pm14.0$   &        0 & 37    &      637 &     4627 & 12.12$^{+0.10}_{-0.08}$ & 12.02$^{+0.04}_{-0.13}$ & $<$11.46$^{+0.55}$      &  2.9 \\
22 & $3.7\pm0.3$  & $4.0\pm6.5$   & $10.5\pm14.4$  &     6322 & 5476  &     7889 &    12930 & 12.48$^{+0.10}_{-0.12}$ & 12.34$^{+0.13}_{-0.02}$ & $<$11.78$^{+0.42}$      &  0.6 \\ 
23 & $4.9\pm0.4$  & $36.8\pm9.4$  & $78.9\pm23.1$  &     1400 & 1531  &     1400 &     3914 & 12.83$^{+0.05}_{-0.17}$ & 12.47$^{+0.06}_{-0.18}$ & 12.58$^{+0.17}_{-0.43}$ &  1.4 \\ 
24 & $4.7\pm0.3$  & $30.0\pm7.5$  & $22.7\pm10.8$  &     1427 & 734   &     2365 &     5068 & 12.76$^{+0.13}_{-0.09}$ & 12.56$^{+0.06}_{-0.10}$ & 12.31$^{+0.28}_{-0.21}$ &  1.5 \\ 
25 & $7.2\pm0.3$  & $0.6\pm7.9$   & $34.4\pm13.2$  &      309 & 0     &      901 &     4472 & 12.45$^{+0.08}_{-0.07}$ & 12.38$^{+0.09}_{-0.21}$ & $<$11.66$^{+0.40}$      &  1.8 \\
26 & $5.6\pm0.3$  & $36.8\pm6.3$  & $69.0\pm14.4$  &      732 & --    &     2141 &     4917 & 12.96$^{+0.03}_{-0.04}$ & 12.59$^{+0.10}_{-0.02}$ & 12.69$^{+0.11}_{-0.10}$ &  0.7 \\ 
27 & $16.5\pm0.5$ & $31.5\pm5.1$  & $35.4\pm18.6$  &    11891 & 13053 &    13932 &    19000 & 13.43$^{+0.02}_{-0.10}$ & 13.39$^{+0.03}_{-0.11}$ & 12.39$^{+0.35}_{-0.35}$ &  0.8 \\
28 & $3.5\pm0.5$  & $1.8\pm7.6$   & $27.3\pm11.4$  &        0 & --    &        0 &      573 & 12.32$^{+0.13}_{-0.15}$ & 12.14$^{+0.06}_{-0.13}$ & $<$11.83$^{+0.37}$      &  1.1 \\ 
29 & $20.1\pm0.4$ & $44.1\pm8.2$  & $60.3\pm20.8$  &     2422 & 0     &     3742 &     7250 & 13.06$^{+0.02}_{-0.03}$ & 13.02$^{+0.02}_{-0.03}$ & 12.09$^{+0.15}_{-0.86}$ &  0.4 \\
30 & $12.2\pm0.6$ & $33.6\pm5.8$  & $42.1\pm21.0$  &        0 & 551   &        0 &     1926 & 13.35$^{+0.03}_{-0.04}$ & 13.18$^{+0.01}_{-0.02}$ & 12.84$^{+0.08}_{-0.12}$ &  1.4 \\ 
31 & $0.9\pm0.2$  & $2.2\pm6.7$   & $20.2\pm16.3$  &        0 & --    &        0 &     1318 & 12.74$^{+0.11}_{-0.39}$ & 12.00$^{+0.17}_{-0.39}$ & 12.67$^{+0.11}_{-0.83}$ &  0.8 
\enddata
\tablecomments{Errors on the absorption strengths from uncertainties in the fits are 100-200 km s$^{-1}$, but see \S\ref{discusssboutflow} \& the appendix. The AS$_{0}^{Gib}$ values are an alternative measure of AS$_{0}$ (Table \ref{absmeasures}), taken from \citet{gibson09} (a `--' indicates that the object is not in the \citet{gibson09} sample). Errors are the 90\% confidence intervals, see \S\ref{sectfitting}.}
\tablenotetext{a}{Total infrared luminosity, integrated over 1-1000$\mu$m in the rest-frame.}
\tablenotetext{b}{Infrared luminosity of the AGN component, integrated over 1-1000$\mu$m in the rest-frame.}
\tablenotetext{c}{Infrared luminosity of the starburst component, integrated over 1-1000$\mu$m in the rest-frame.}
\end{deluxetable*}

Several solutions have been proposed to resolve these issues. Among the most promising is an idea termed `AGN feedback'. AGN feedback is the exertion of influence of an SMBH on kpc to Mpc scales to curtail star formation in the host galaxy, and/or further accretion onto the SMBH itself\footnote{The idea that SMBHs may exert `negative' feedback on inflowing gas to regulate their own growth and that of the host galaxy predated its use to reconcile galaxy evolution models with observations, see e.g. \citealt{haiman96,silk98}}. This can occur in four ways. Radiation from regions immediately local to the SMBH can (1) heat gas in the ISM so it cannot collapse to form stars, and/or (2) radiatively drive gas out of the galaxy (or, equivalently, stop it falling in from the IGM), thus emptying the galaxy of fuel for further star formation. Additionally, matter expelled from regions immediately local to the SMBH can (3) heat gas in the ISM, and/or (4) drive gas out of the galaxy via kinetic pressure.

Most discussions of AGN feedback condense the four feedback mechanisms described above into two simplified paradigms; `quasar' mode feedback and `radio' mode feedback. Quasar mode feedback assumes that radiation from an accretion disk terminates star formation in the host galaxy, usually by coupling some fraction of the QSO luminosity to kinetic energy injected into the ISM. Quasar mode feedback is thought to be short-term, acting for only the $\sim10^{8}$ years over which the quasar is accreting rapidly. Radio mode feedback on the other hand is assumed to occur via a relativistic jet that transfers momentum to the ISM. Radio mode feedback can occur over longer timescales ($\sim10^{9}$ years) than quasar mode as it requires only a low accretion rate to produce jets sufficiently powerful to affect the ISM. Nevertheless, radio mode and quasar mode feedback are not mutually exclusive. 

\begin{figure*}
\includegraphics[width=78mm,angle=0]{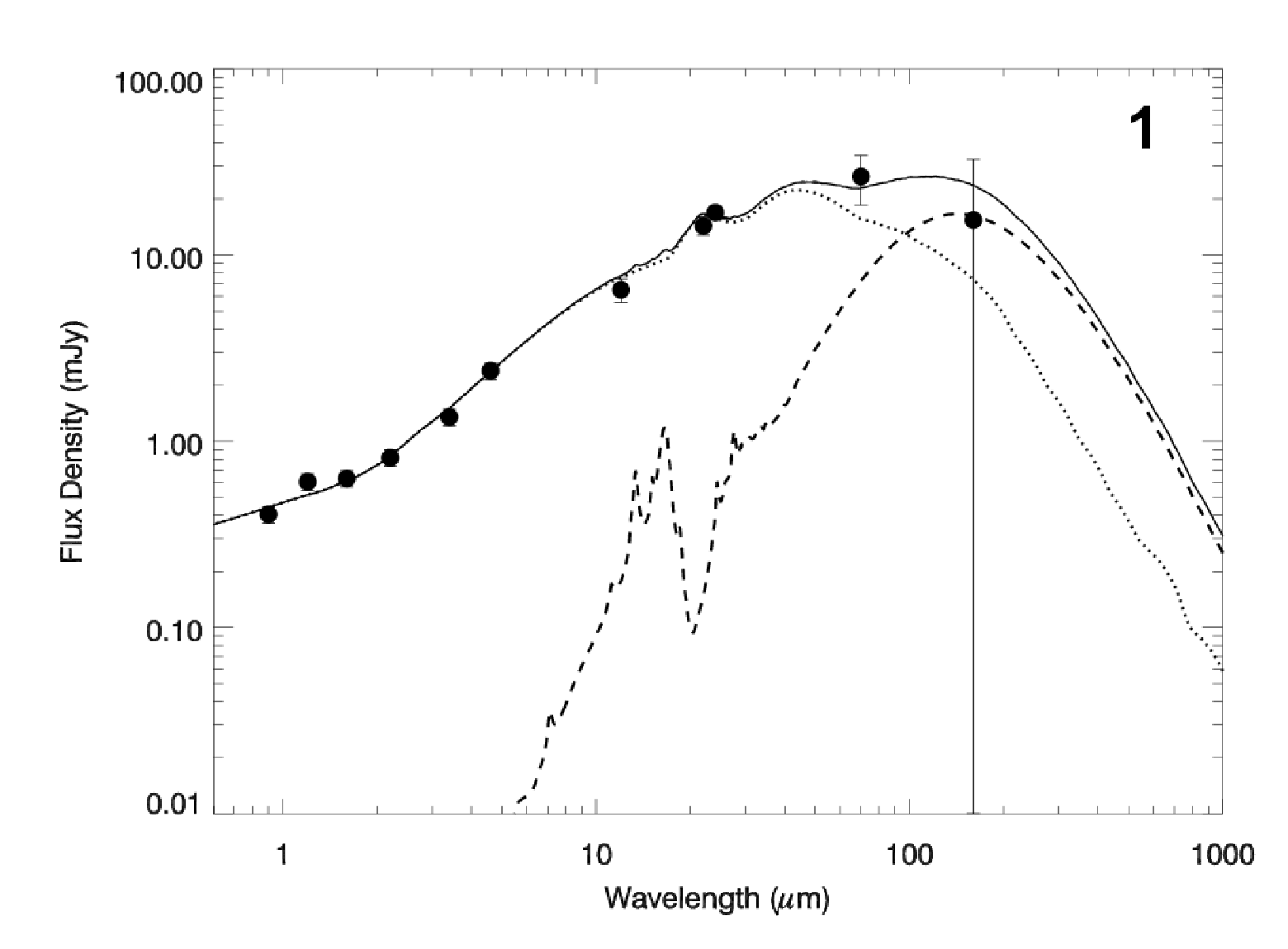}
\includegraphics[width=78mm,angle=0]{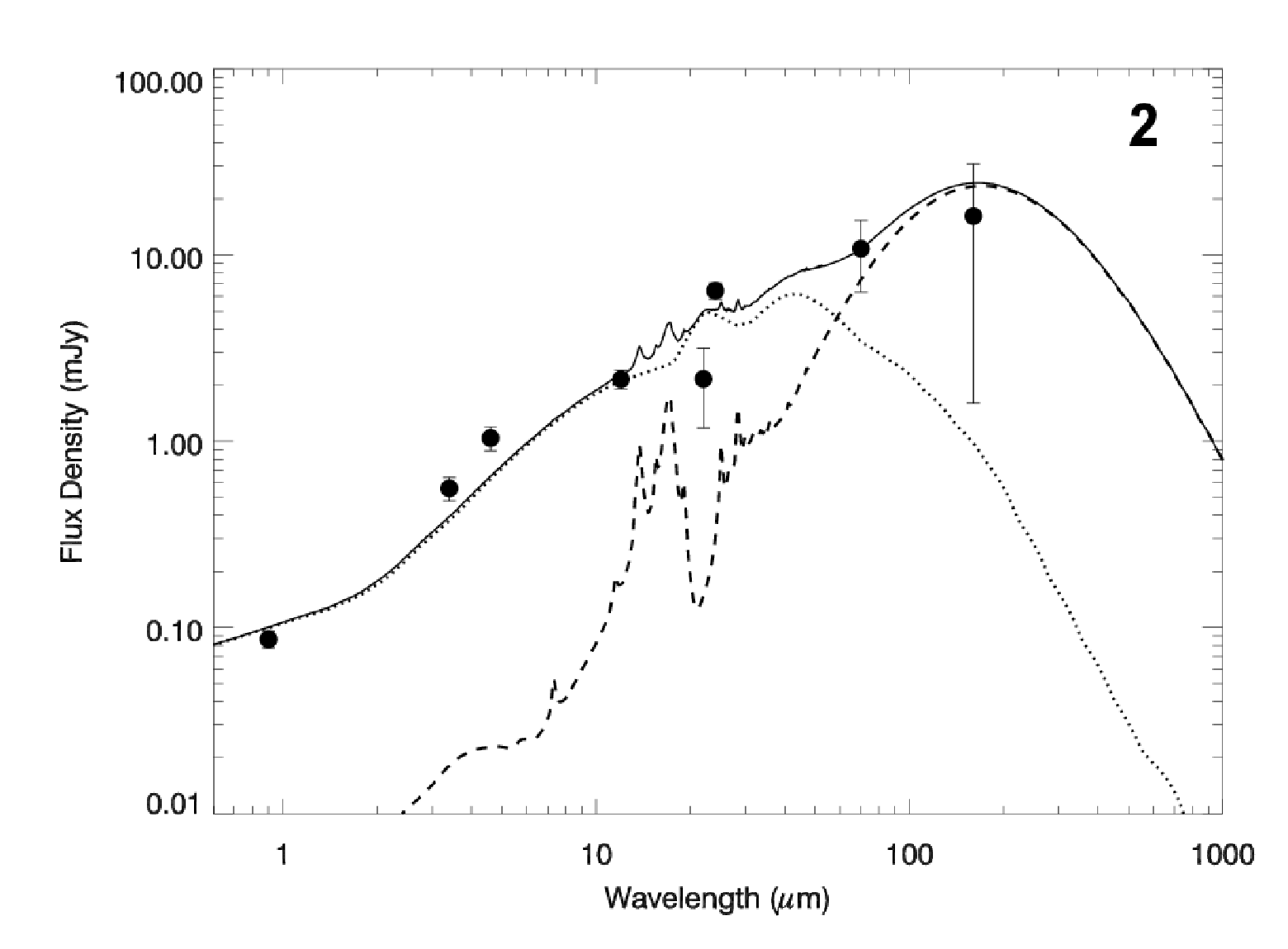}\\
\includegraphics[width=78mm,angle=0]{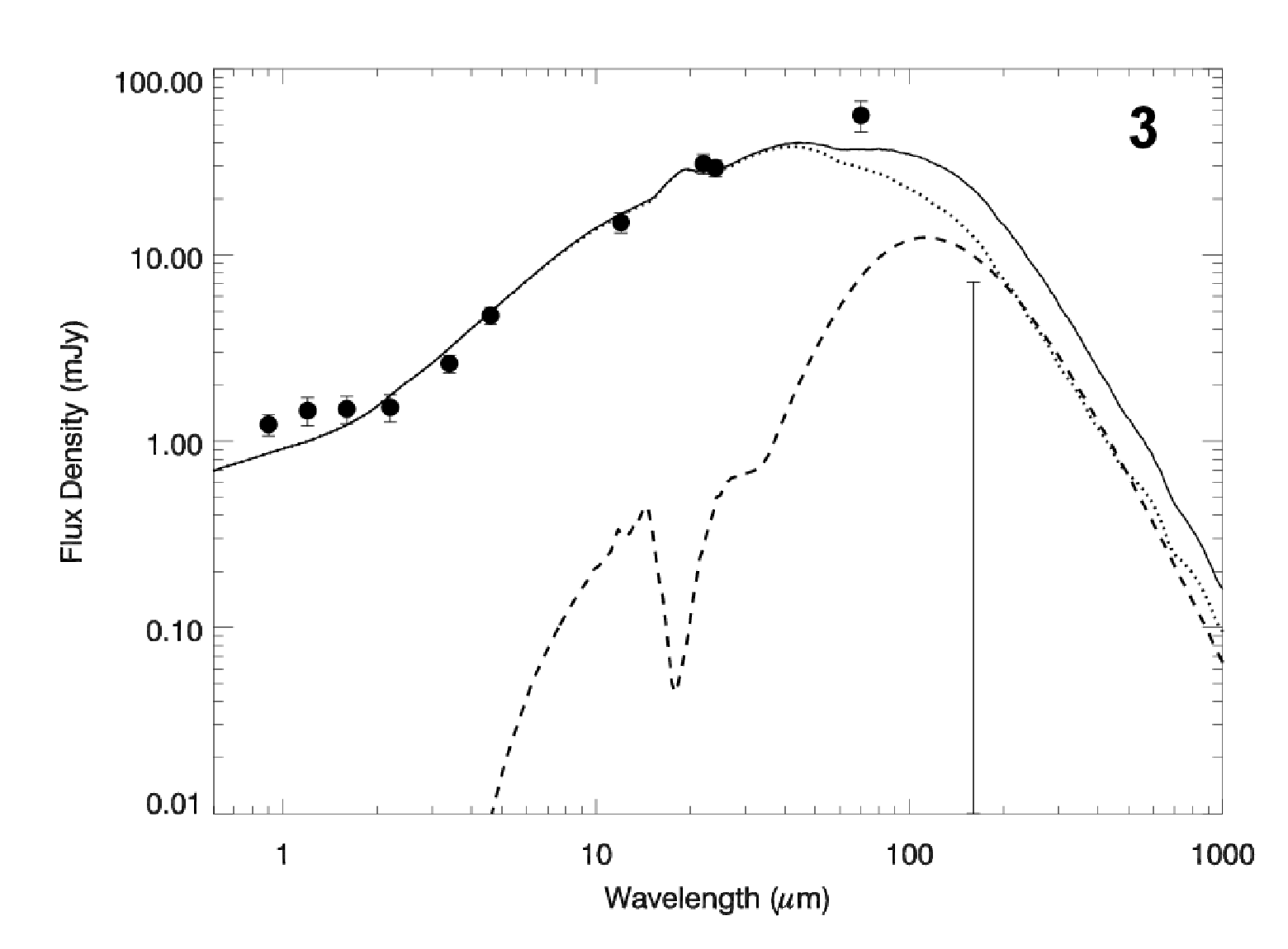}
\includegraphics[width=78mm,angle=0]{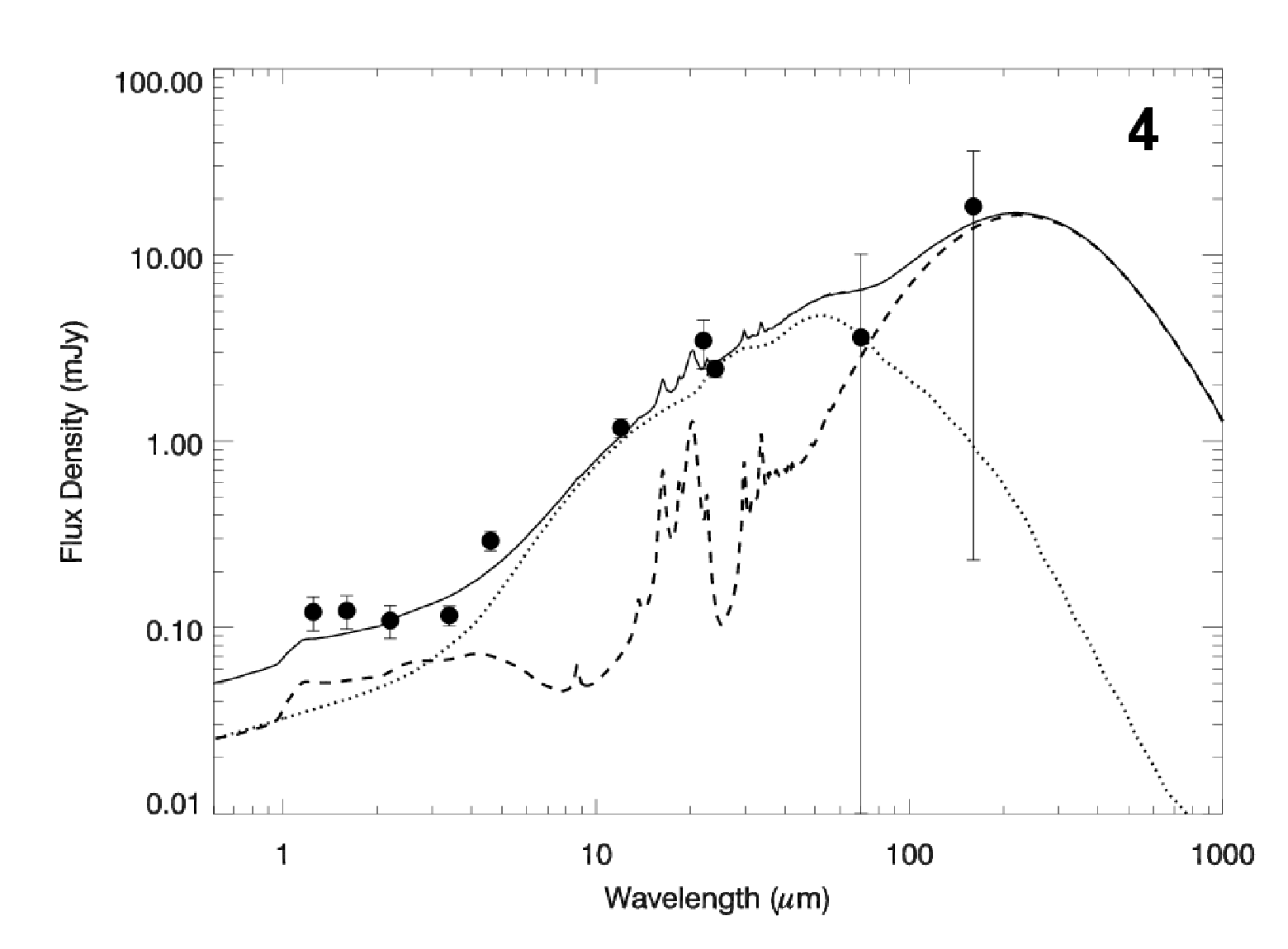}\\
\includegraphics[width=78mm,angle=0]{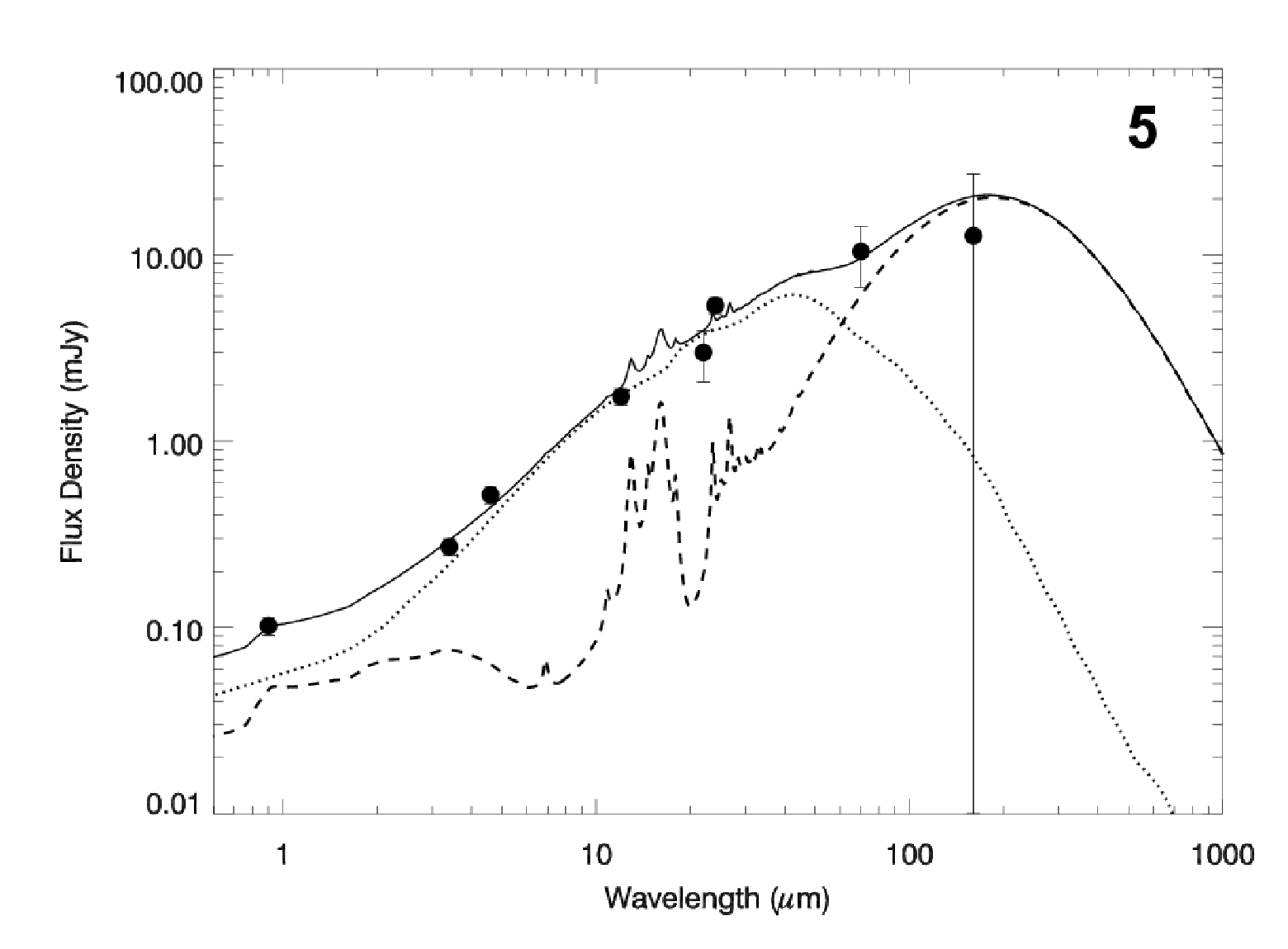}
\includegraphics[width=78mm,angle=0]{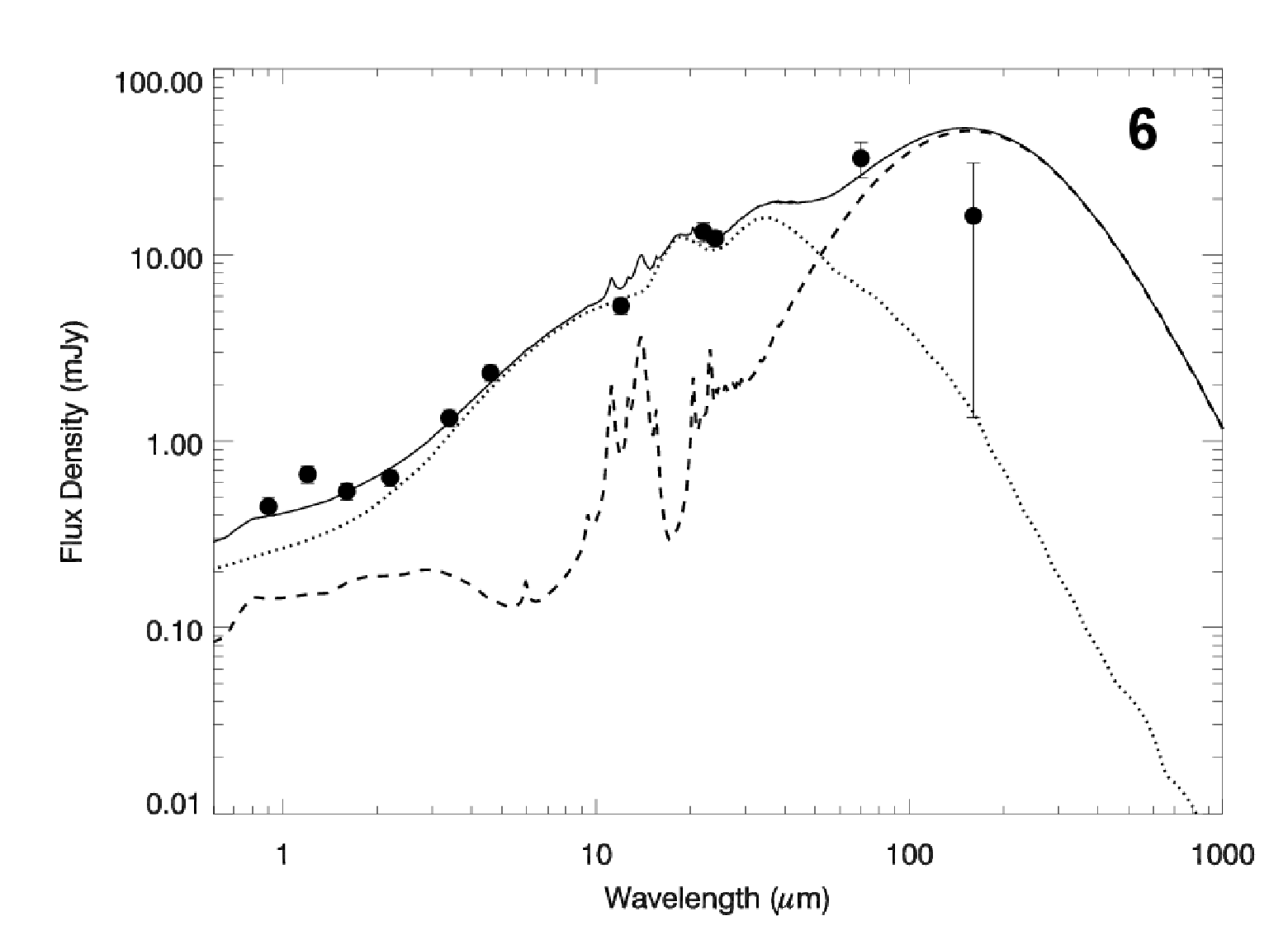}
\caption{Observed-frame fits to the optical through MIPS SEDs for objects 1-6 in Table \ref{tablesample}. The solid line is the combined best-fit model, while the dashed and dotted lines are the starburst and AGN components, respectively. A description of the models is in \S\ref{subsmodels}. The number in the top right hand corner of each plot is the ID number in Table \ref{tablesample}.}\label{figexamplefits1}
\end{figure*}

\begin{figure*}
\includegraphics[width=78mm,angle=0]{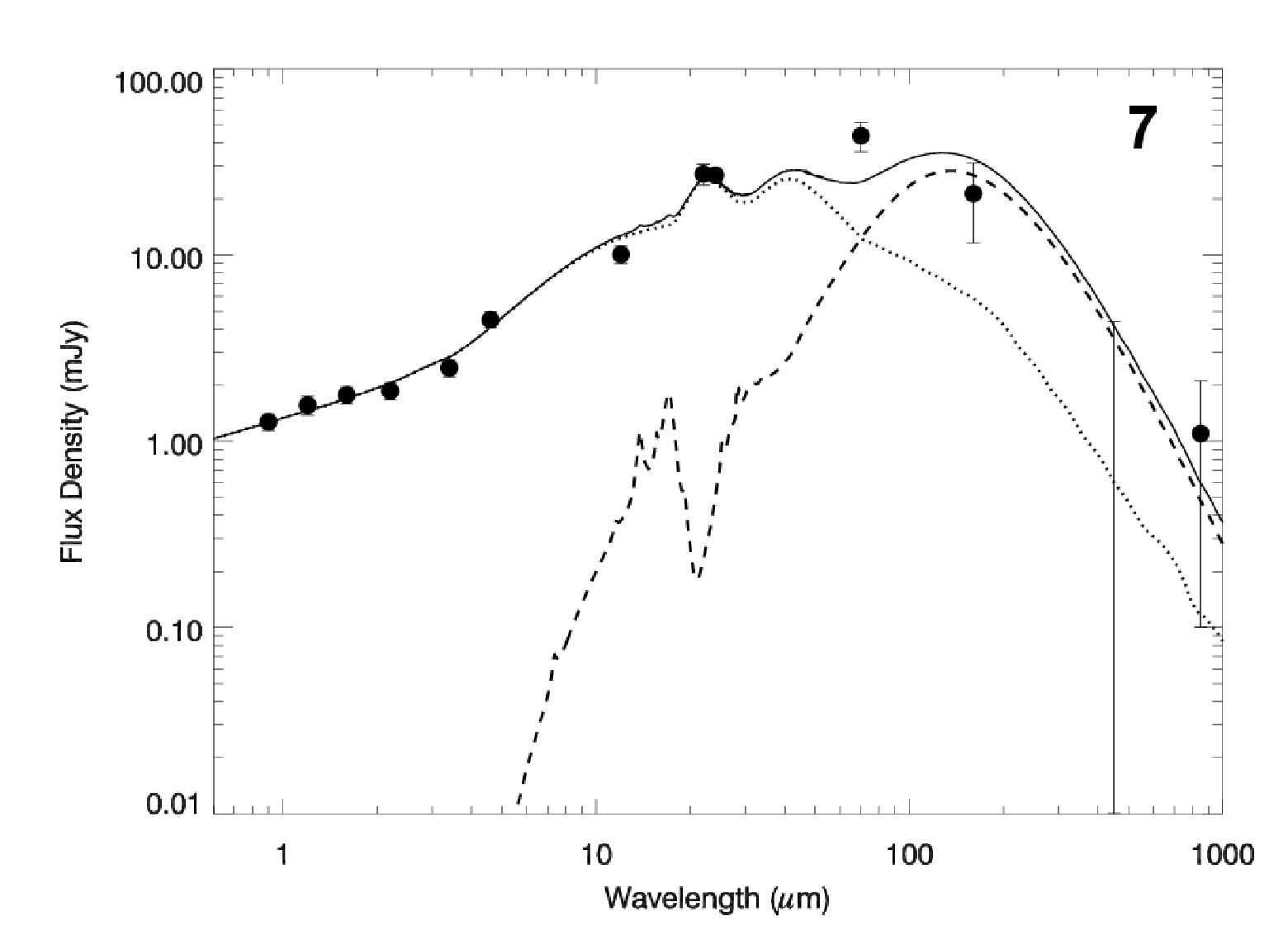}
\includegraphics[width=78mm,angle=0]{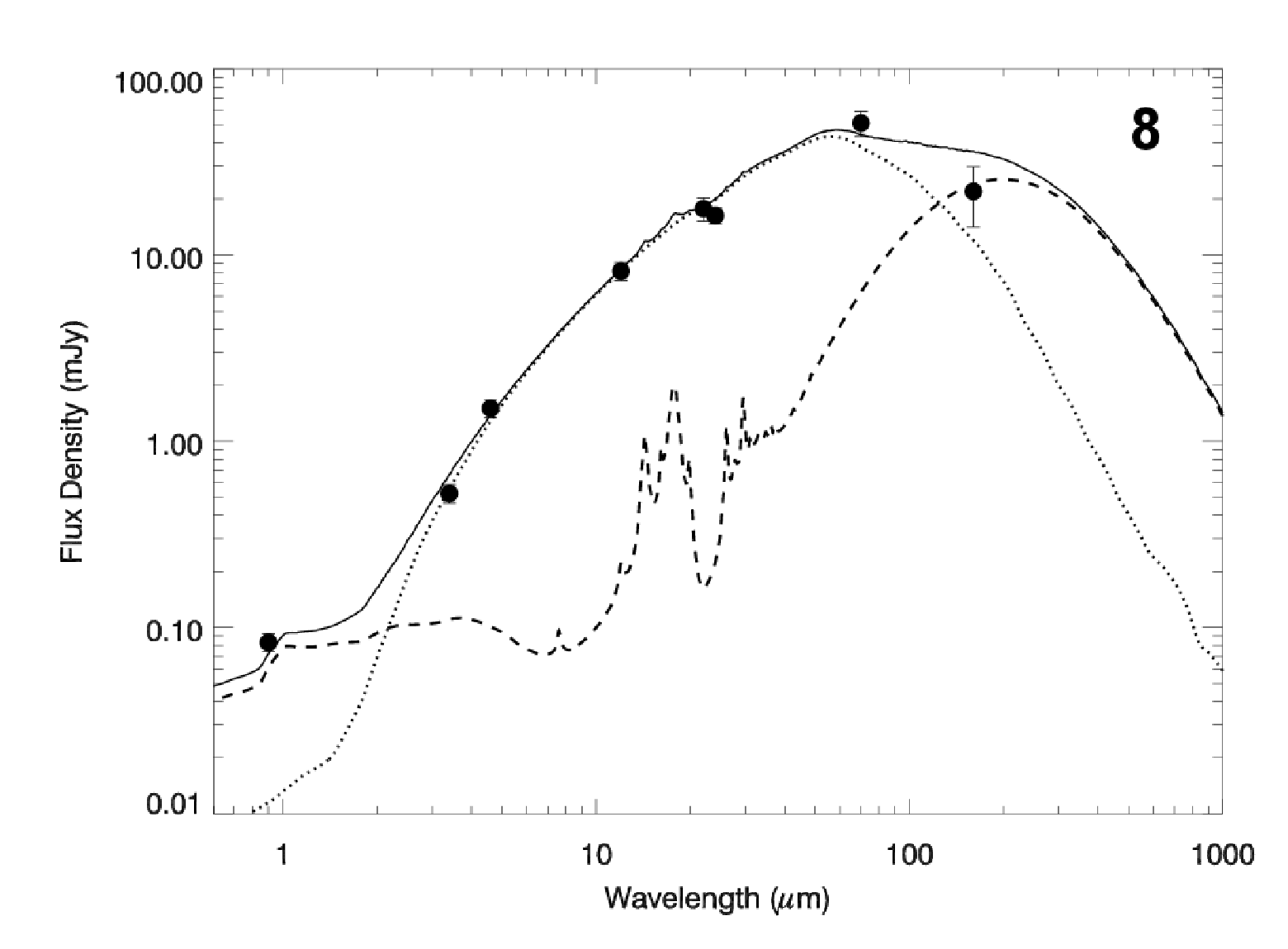}\\
\includegraphics[width=78mm,angle=0]{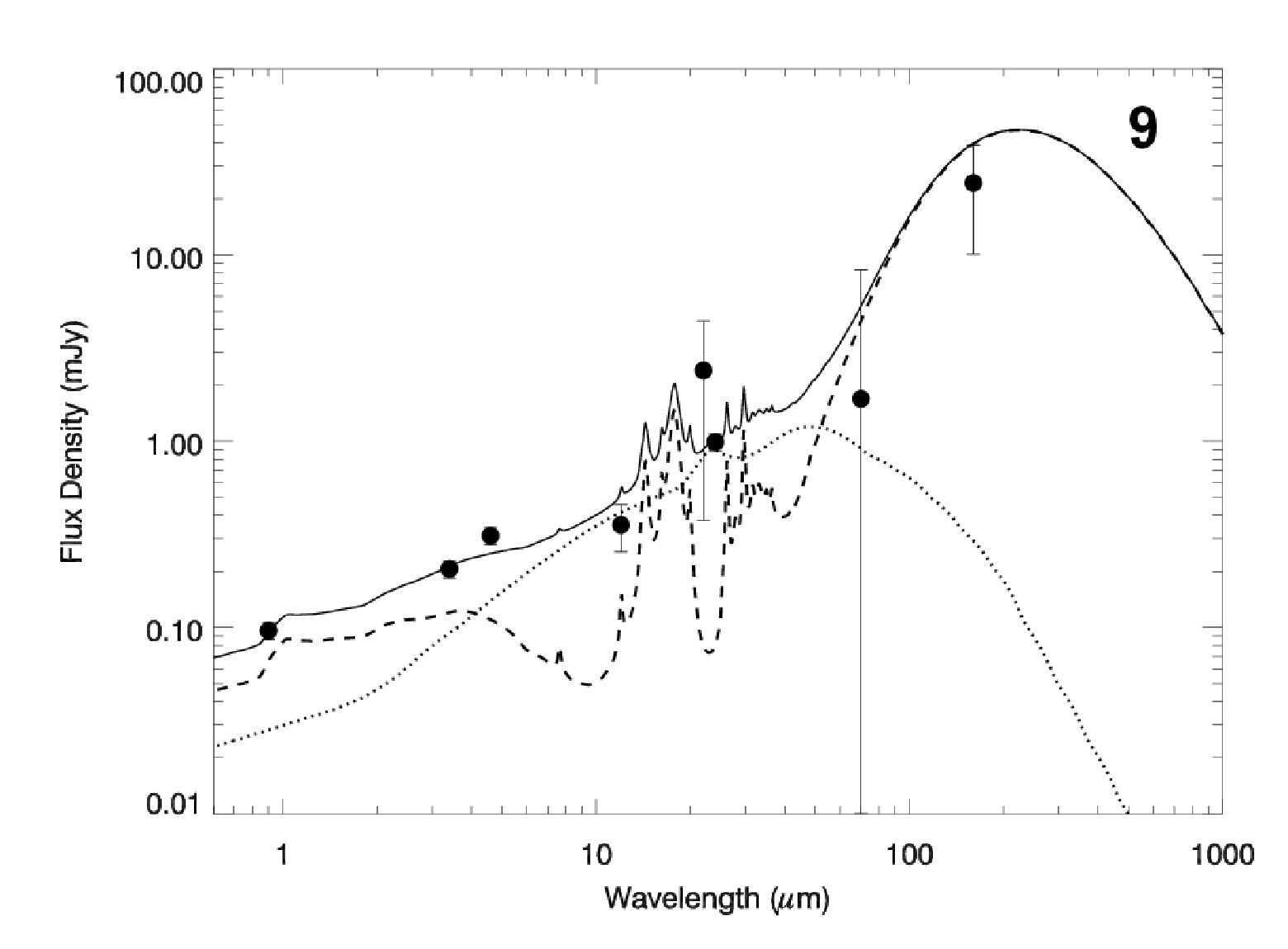}
\includegraphics[width=78mm,angle=0]{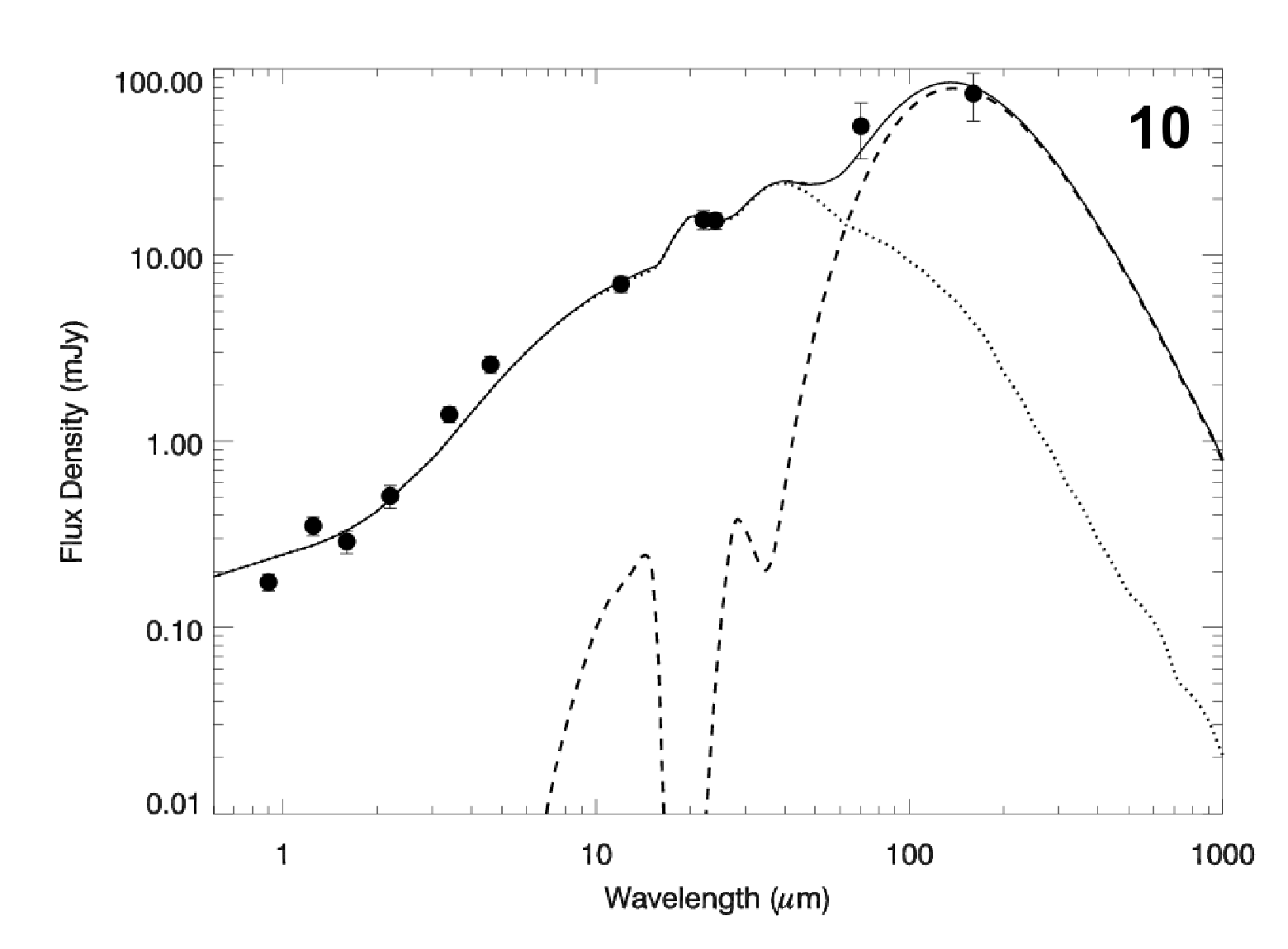}\\
\includegraphics[width=78mm,angle=0]{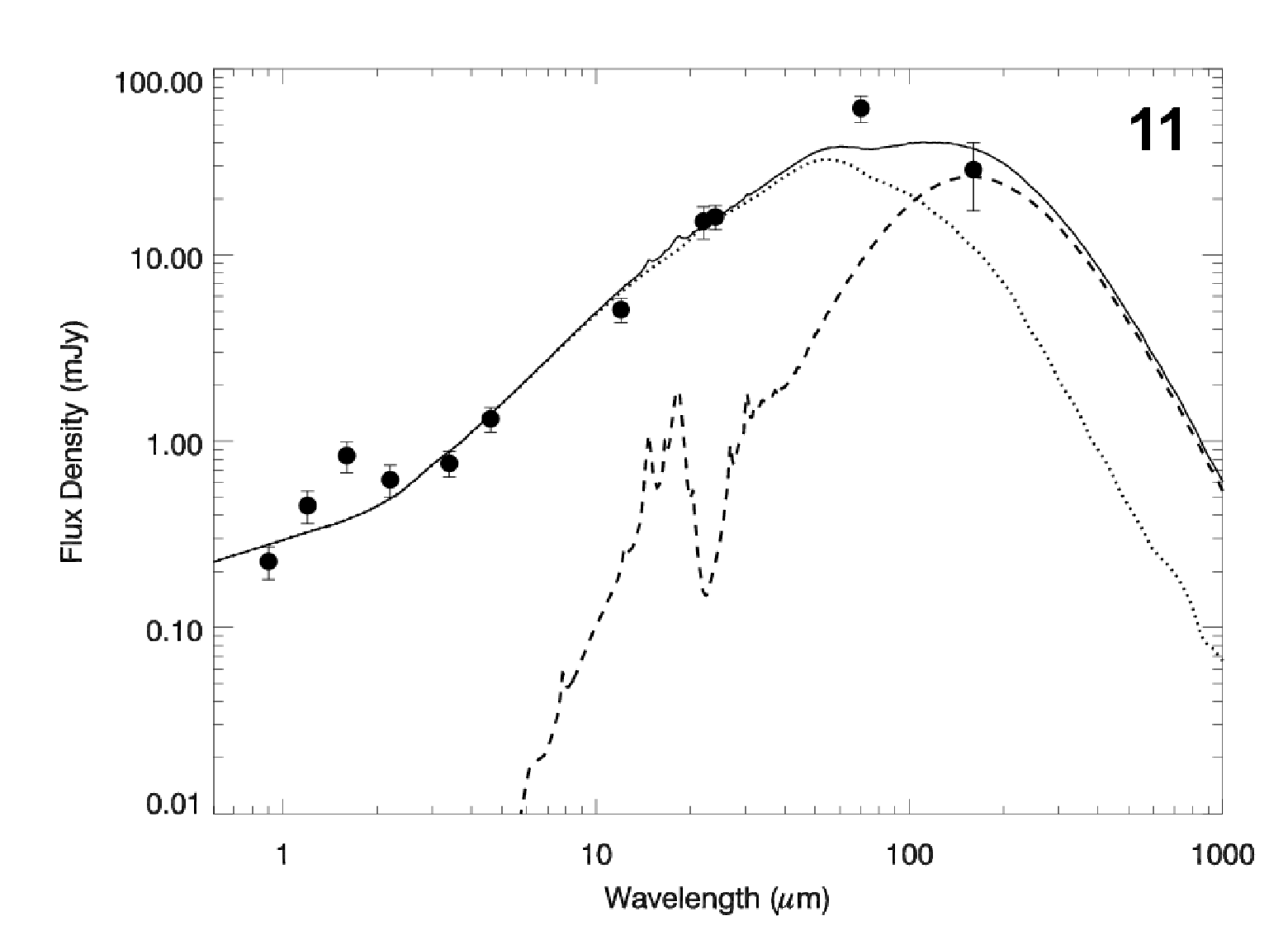}
\includegraphics[width=78mm,angle=0]{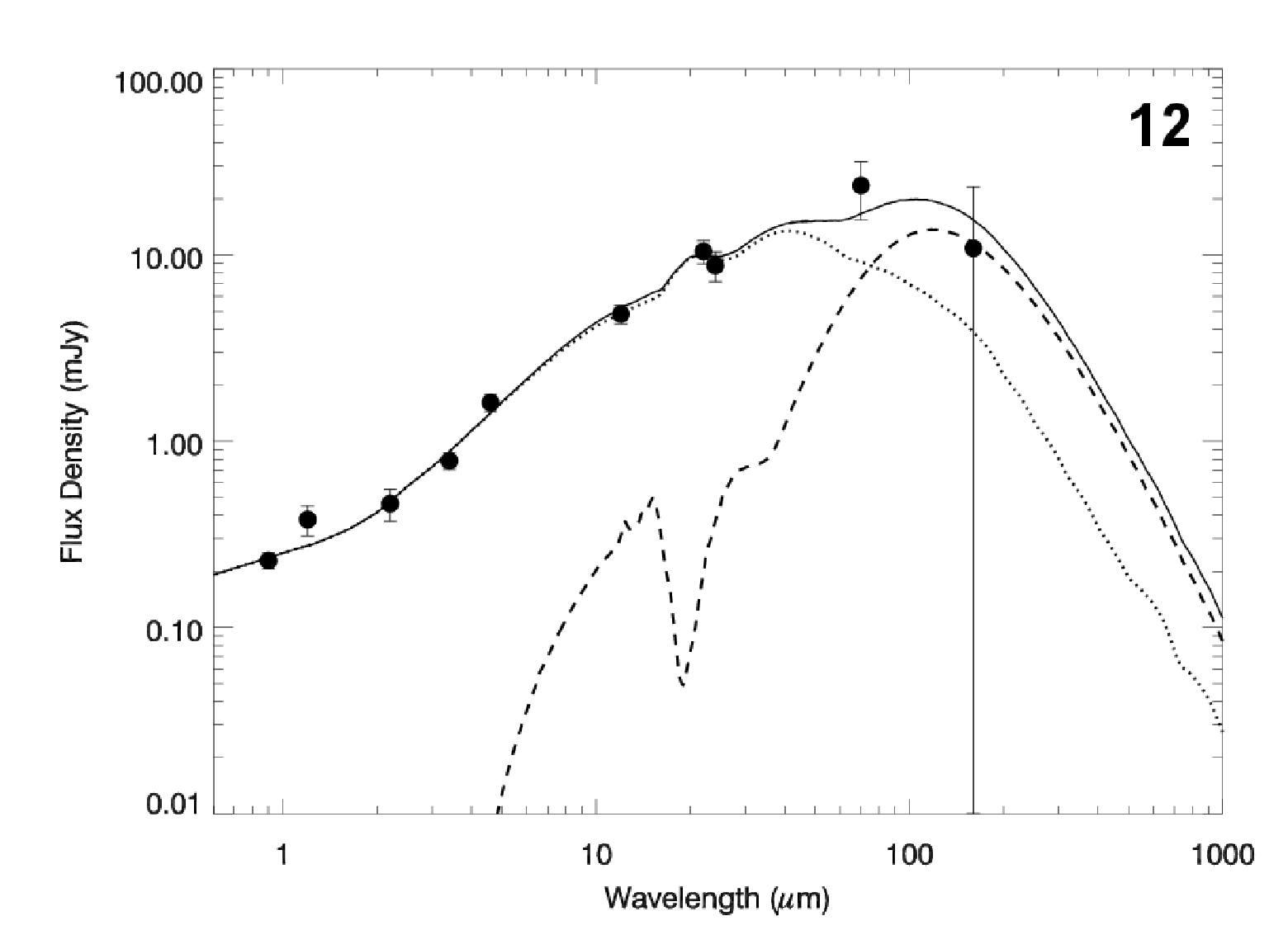}
\caption{Observed-frame fits to the optical through MIPS SEDs for objects 7-12 in Table \ref{tablesample}. Details are the same as for figure \ref{figexamplefits1}.}\label{figexamplefits2}
\end{figure*}

\begin{figure*}
\includegraphics[width=78mm,angle=0]{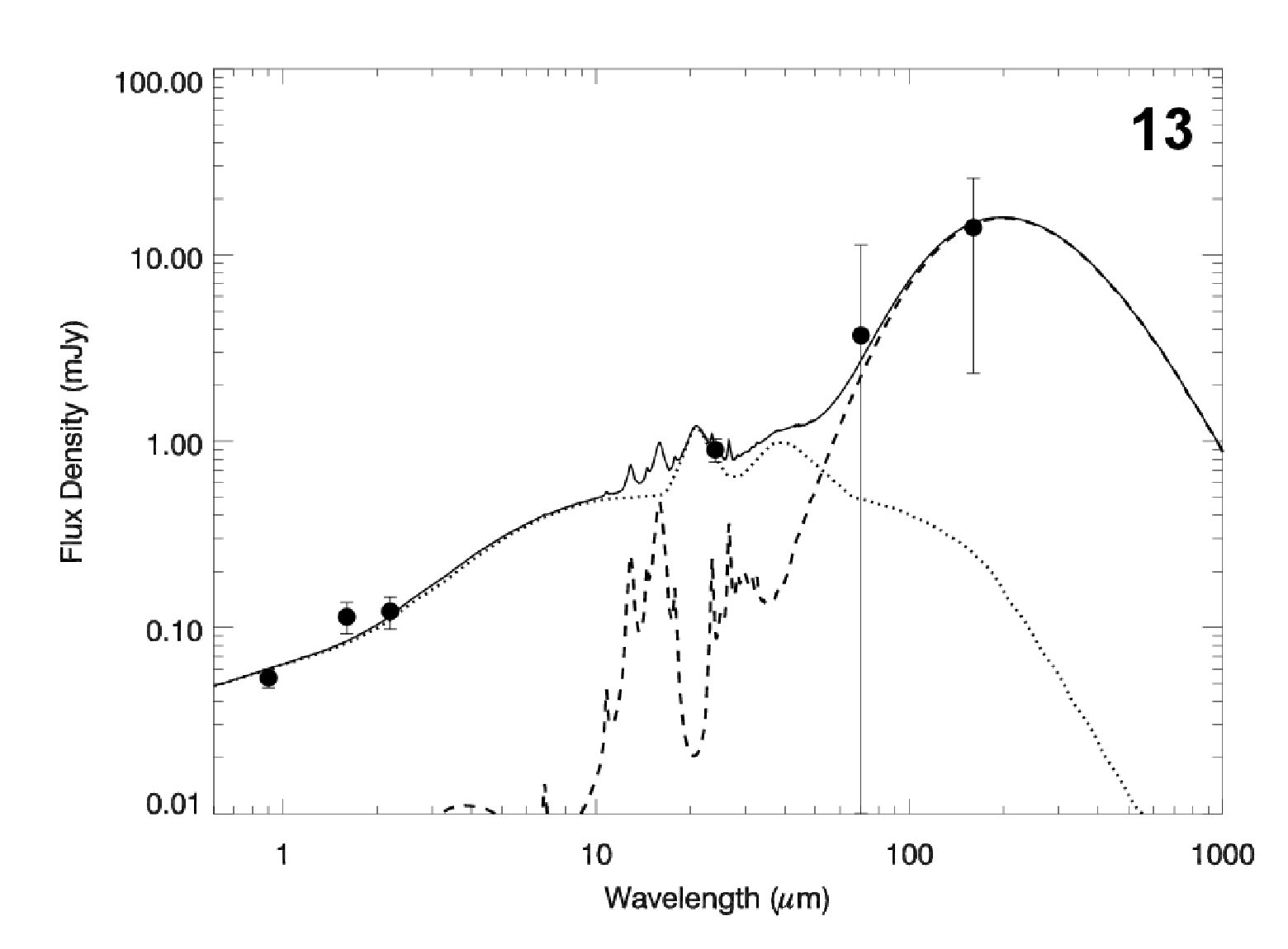}
\includegraphics[width=78mm,angle=0]{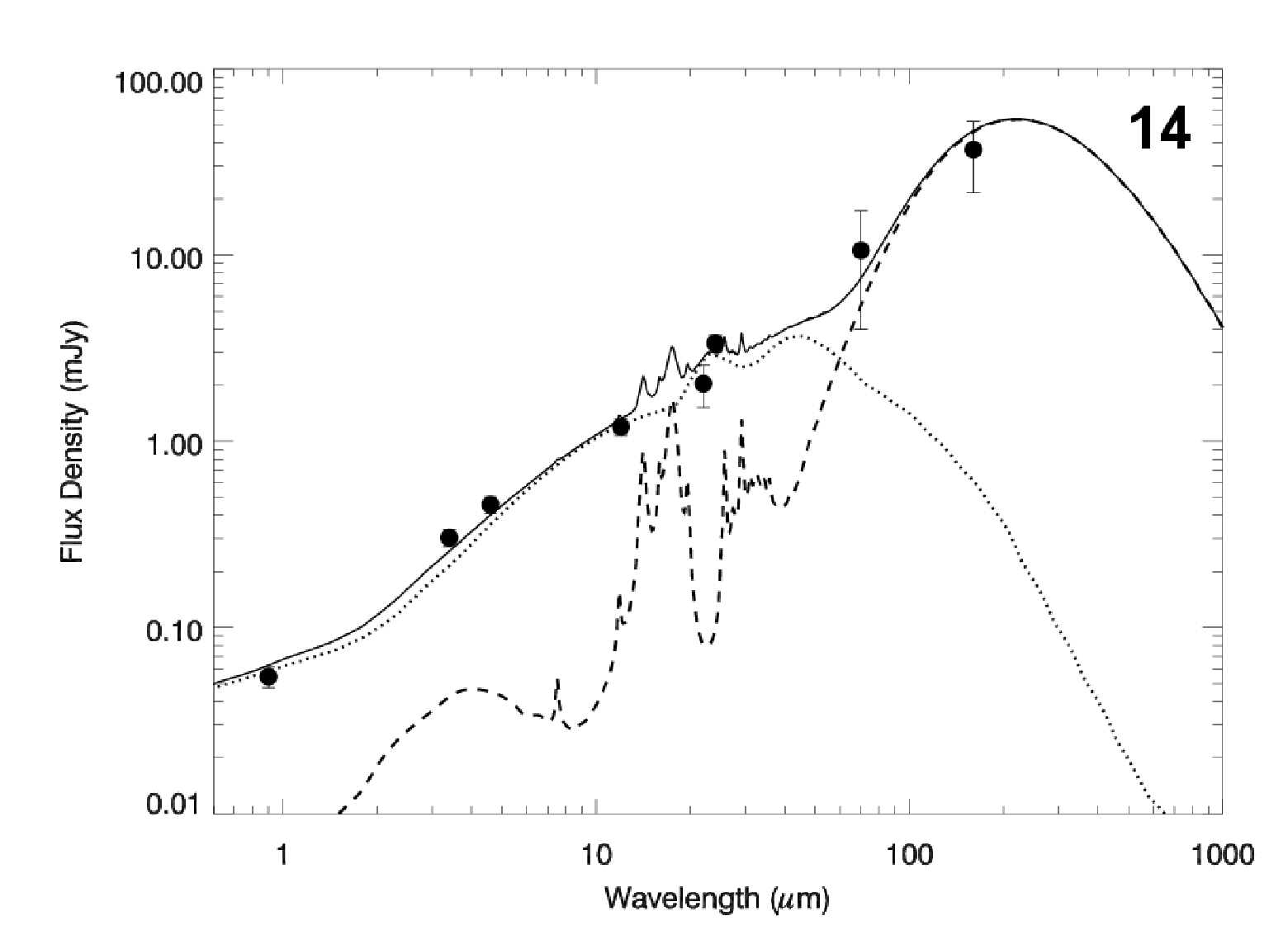}\\
\includegraphics[width=78mm,angle=0]{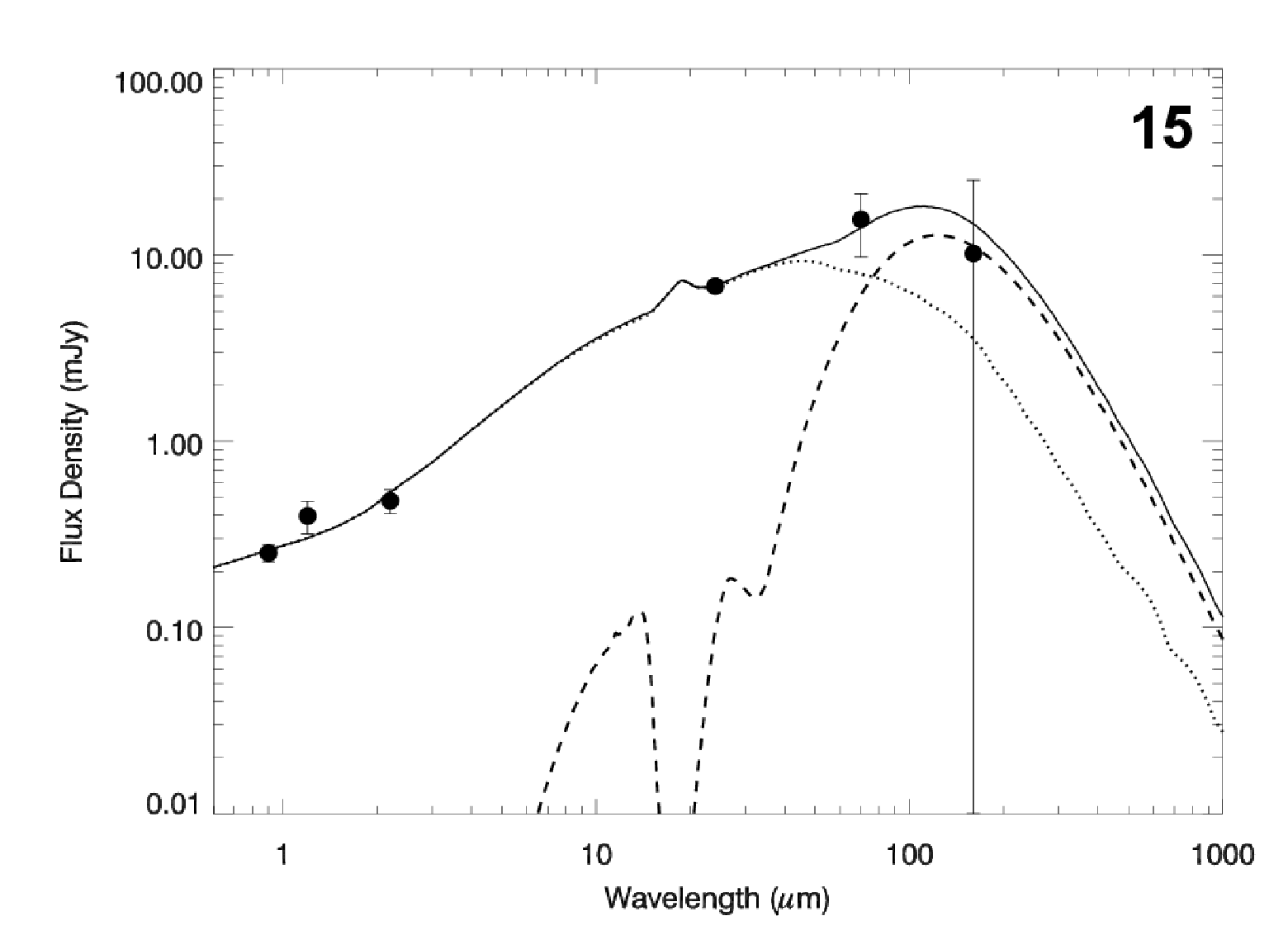}
\includegraphics[width=78mm,angle=0]{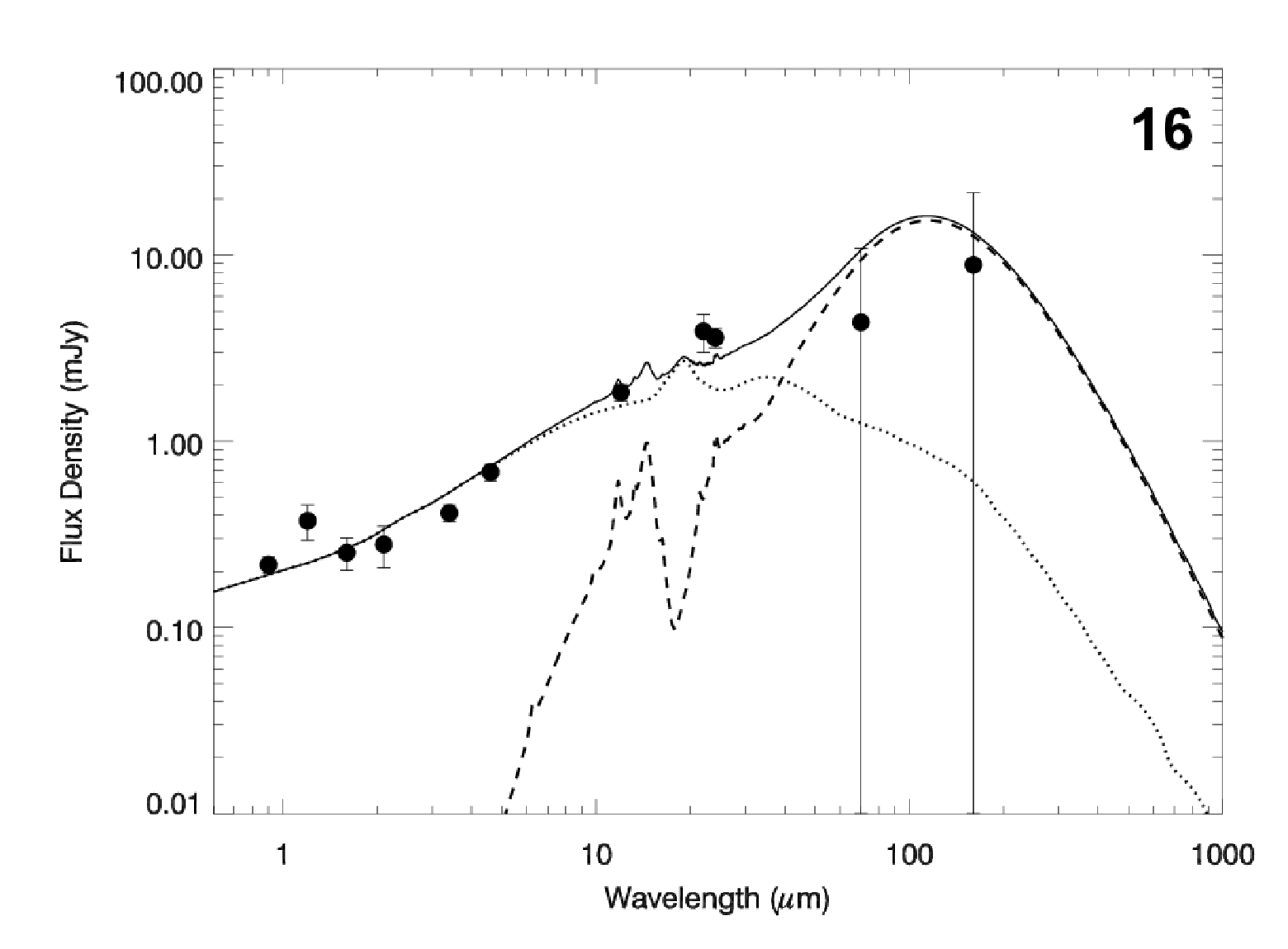}\\
\includegraphics[width=78mm,angle=0]{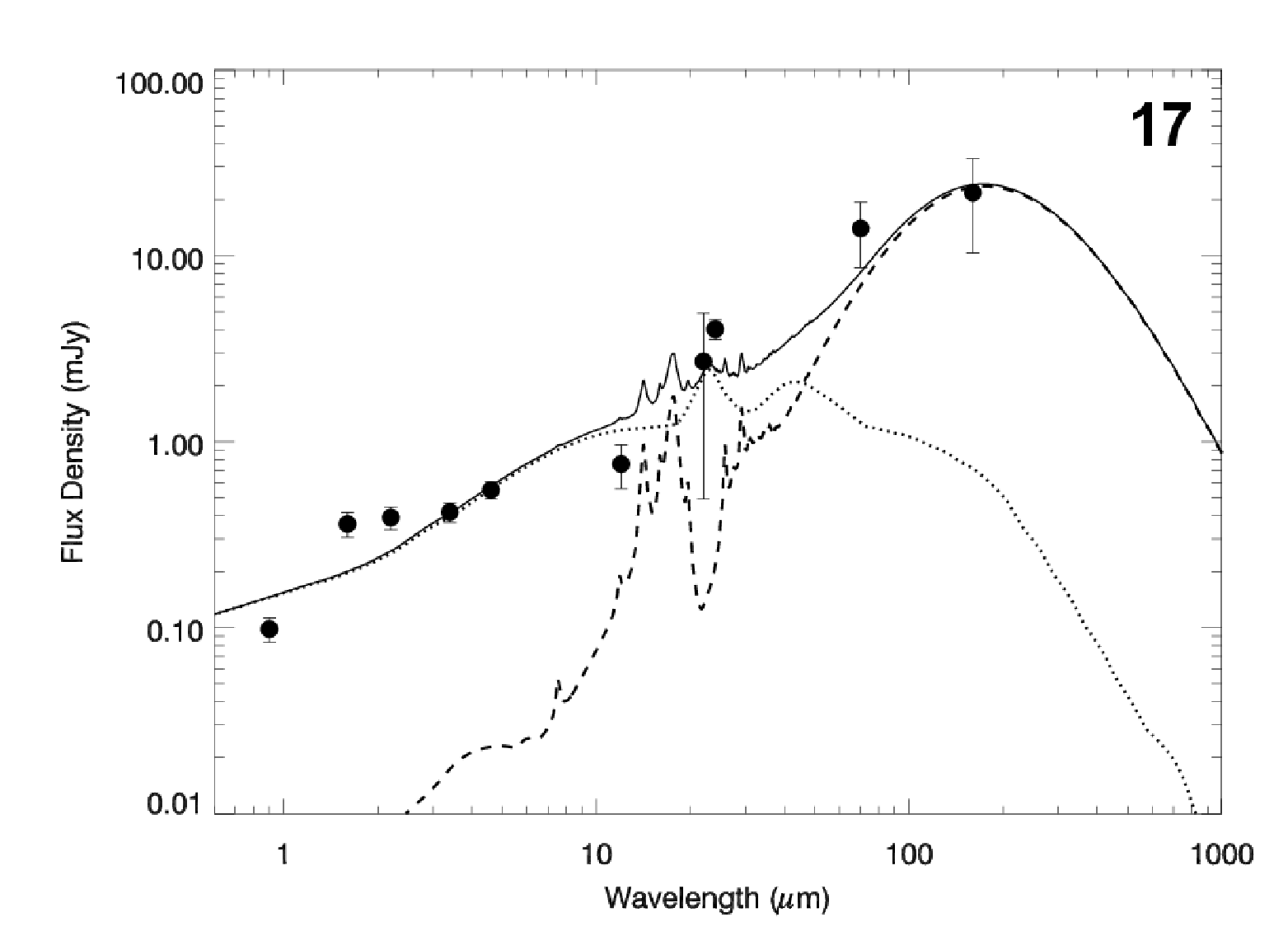}
\includegraphics[width=78mm,angle=0]{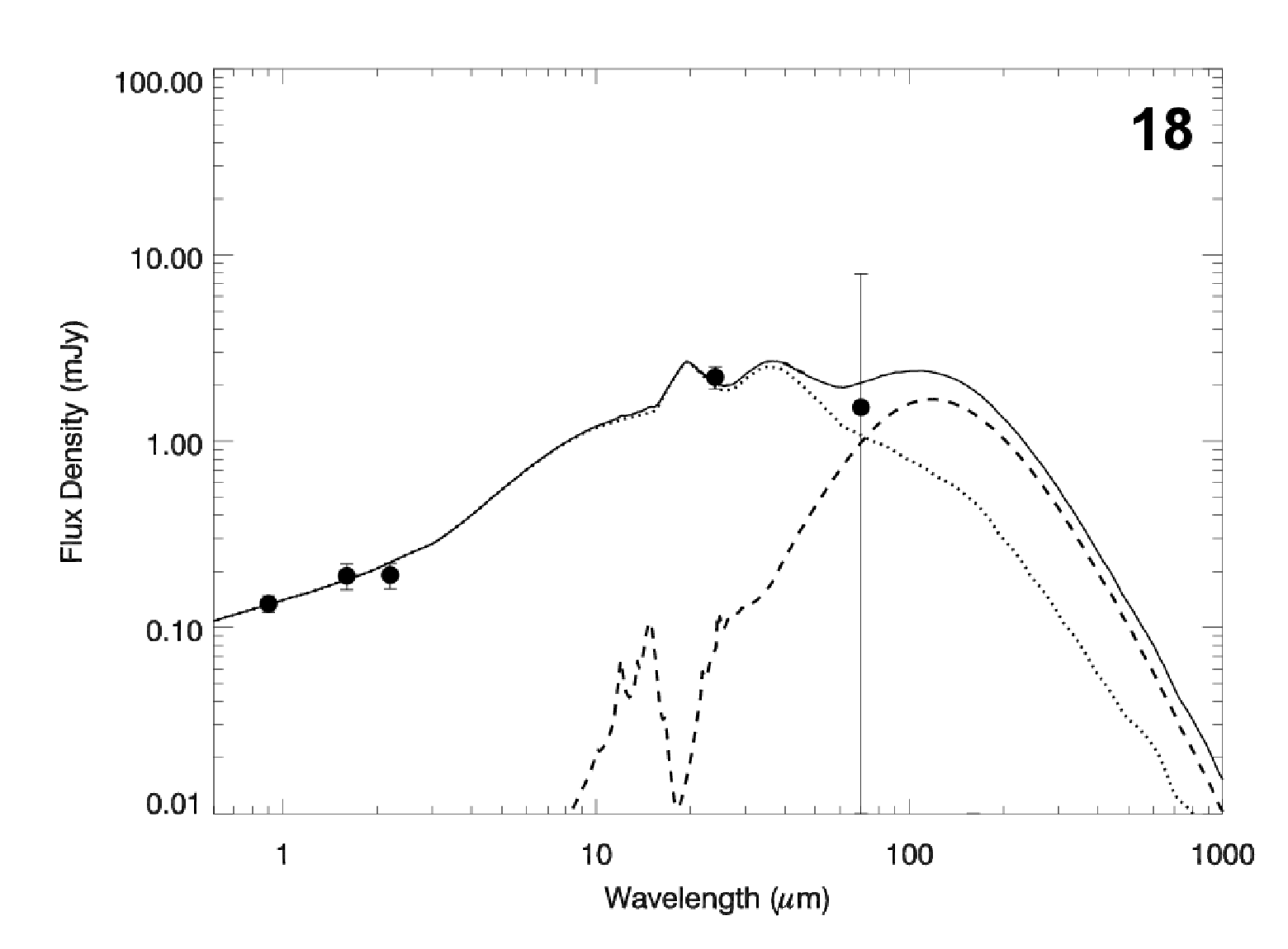}
\caption{Observed-frame fits to the optical through MIPS SEDs for objects 13-18 in Table \ref{tablesample}. Details are the same as for figure \ref{figexamplefits1}.}\label{figexamplefits3}
\end{figure*}

\begin{figure*}
\includegraphics[width=78mm,angle=0]{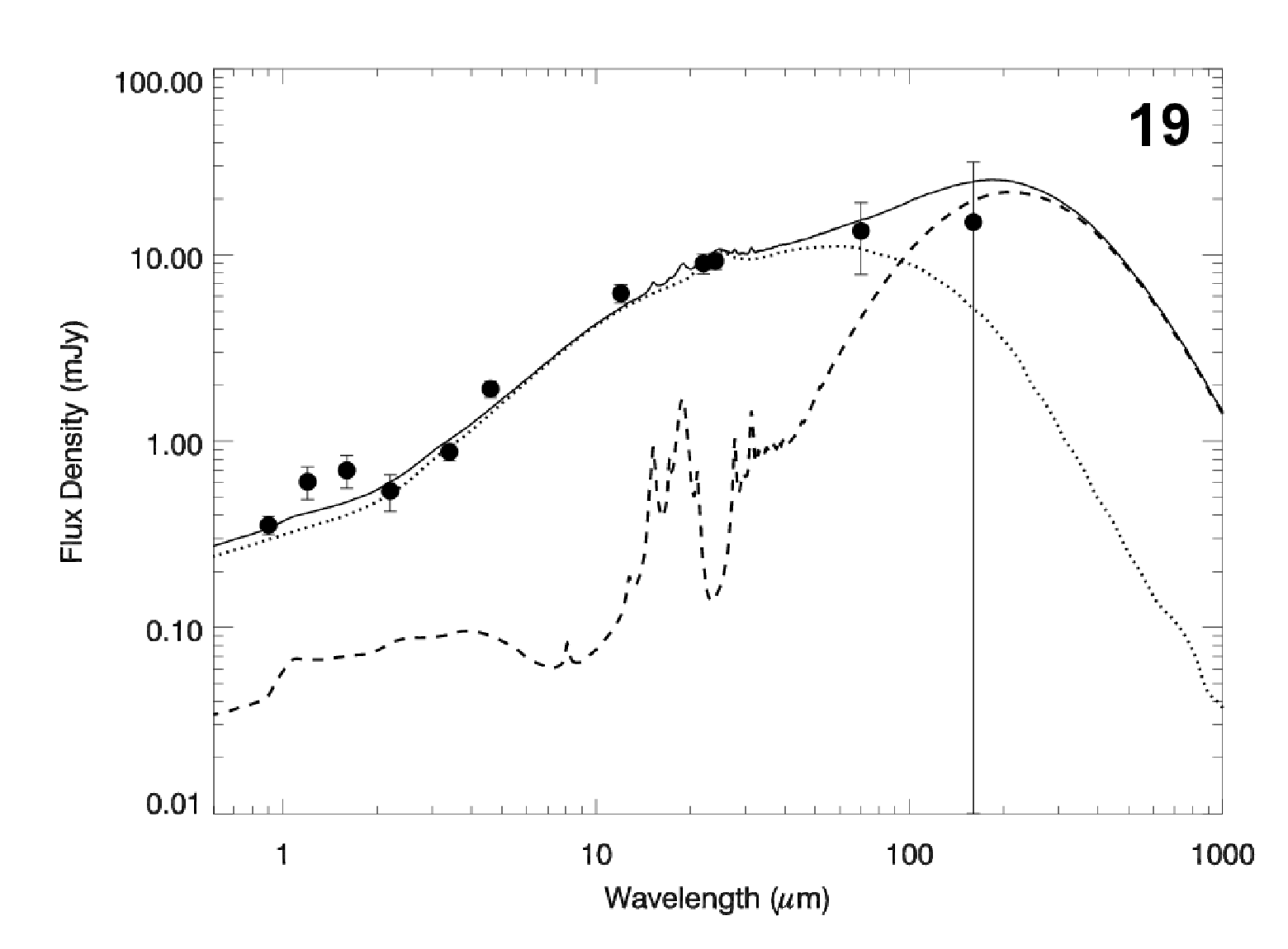}
\includegraphics[width=78mm,angle=0]{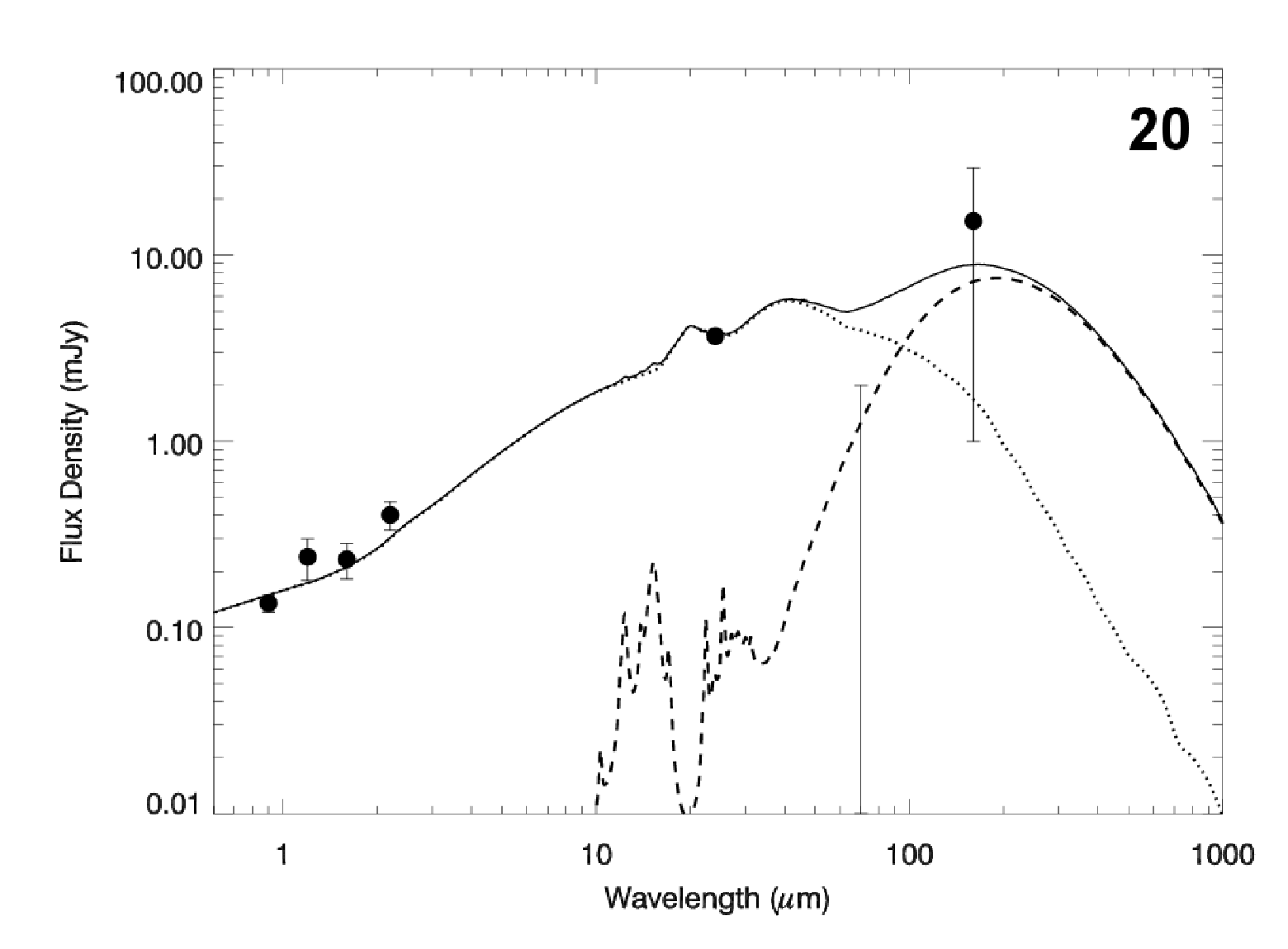}\\
\includegraphics[width=78mm,angle=0]{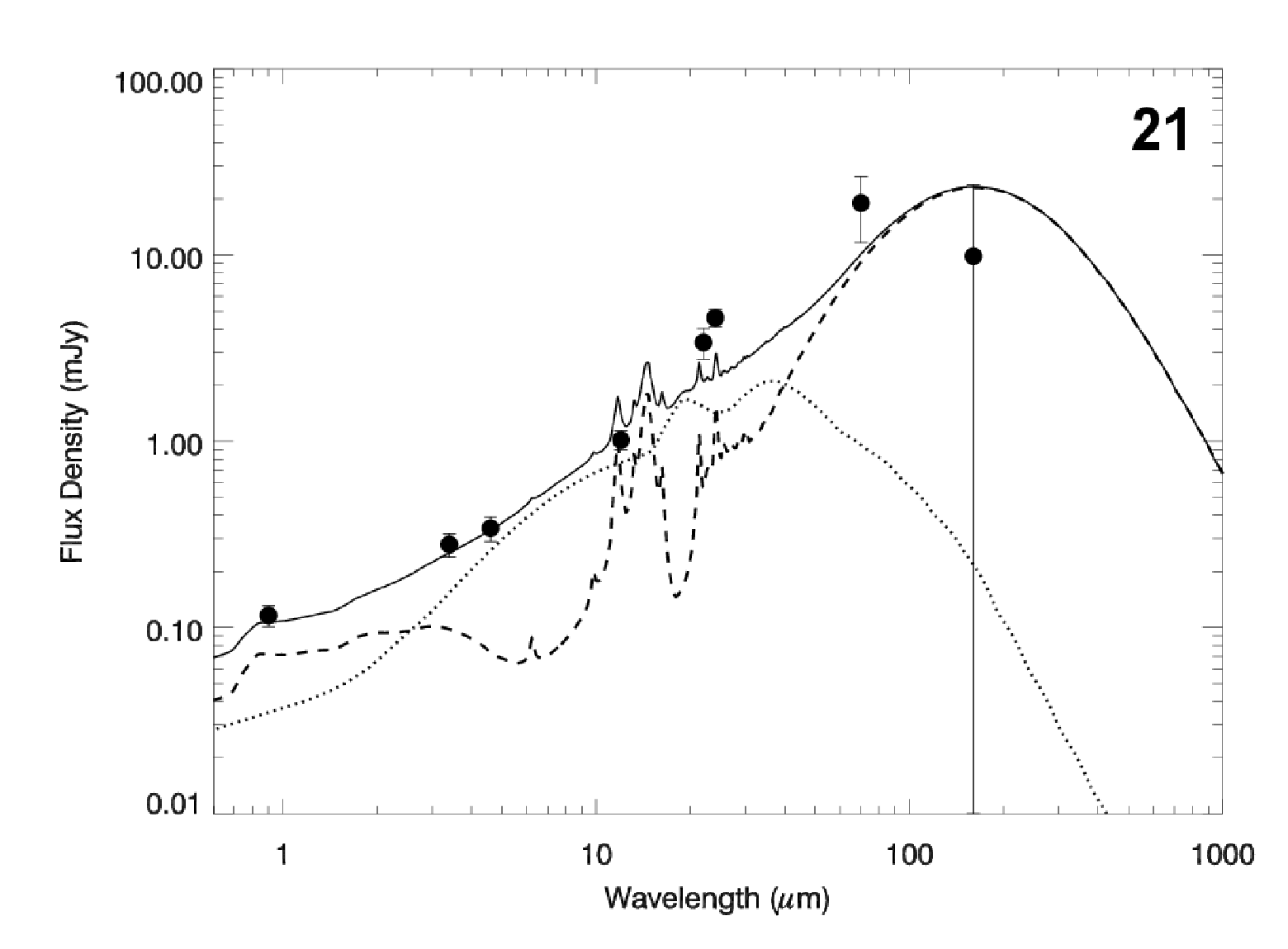}
\includegraphics[width=78mm,angle=0]{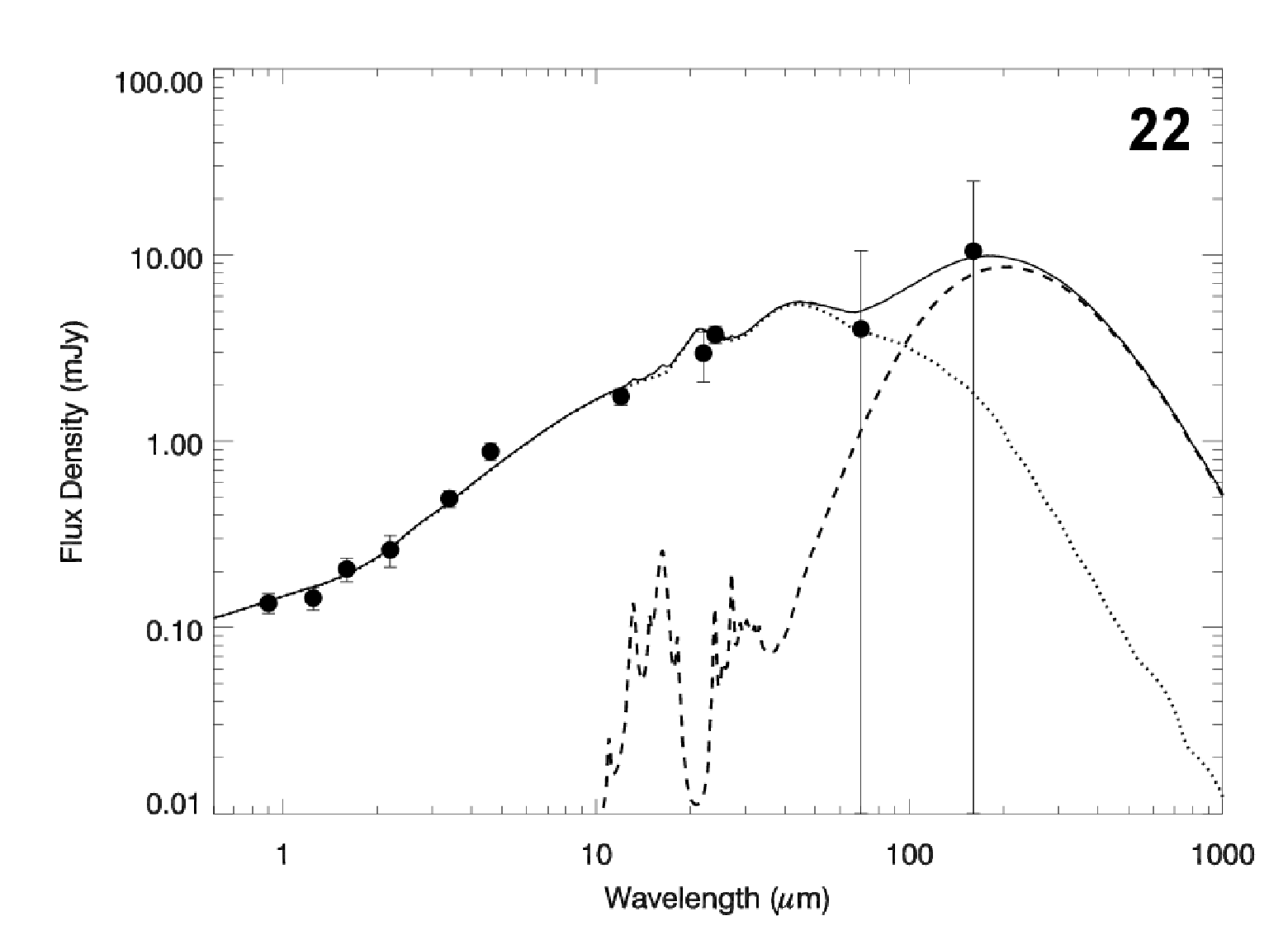}\\
\includegraphics[width=78mm,angle=0]{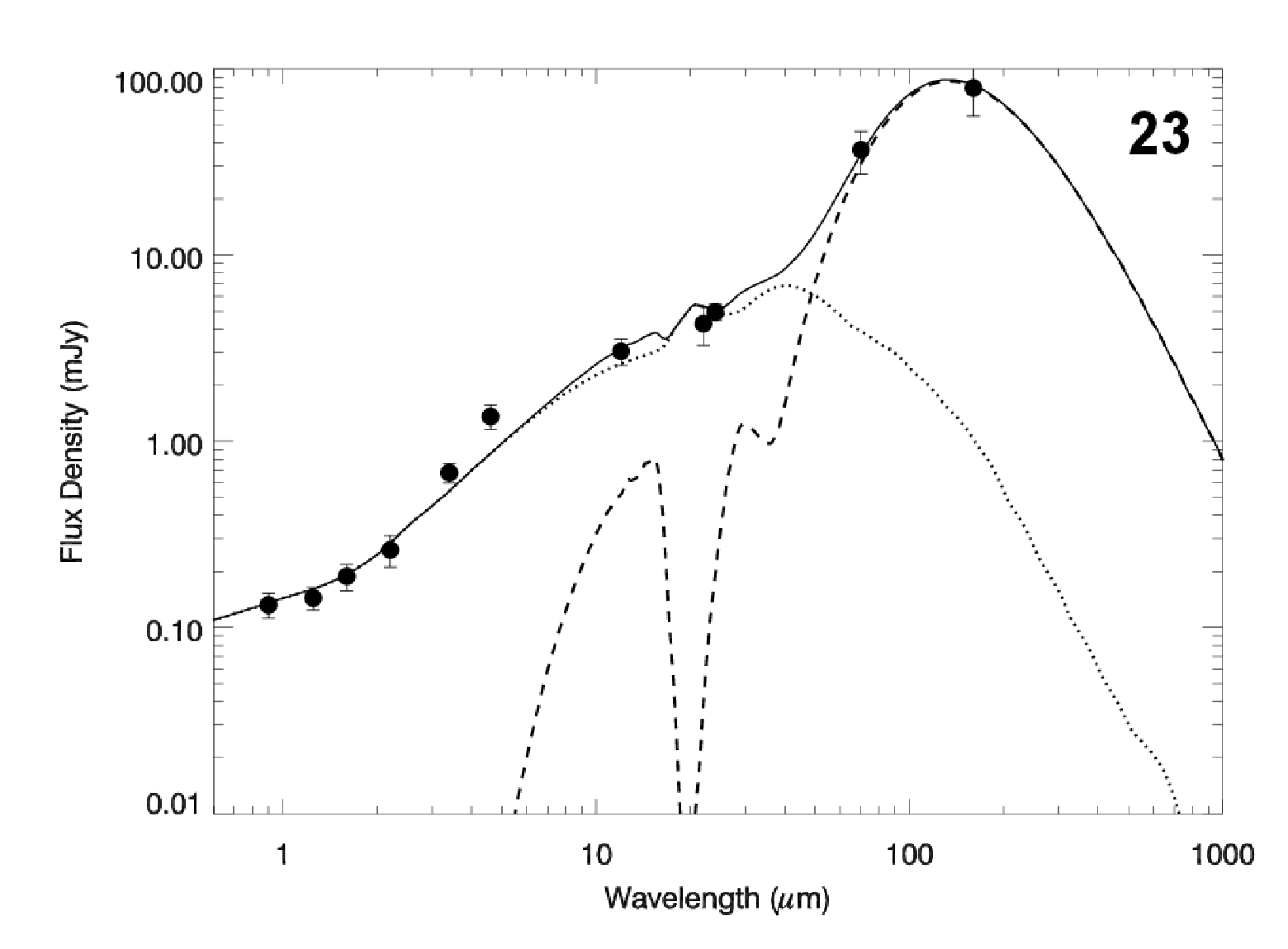}
\includegraphics[width=78mm,angle=0]{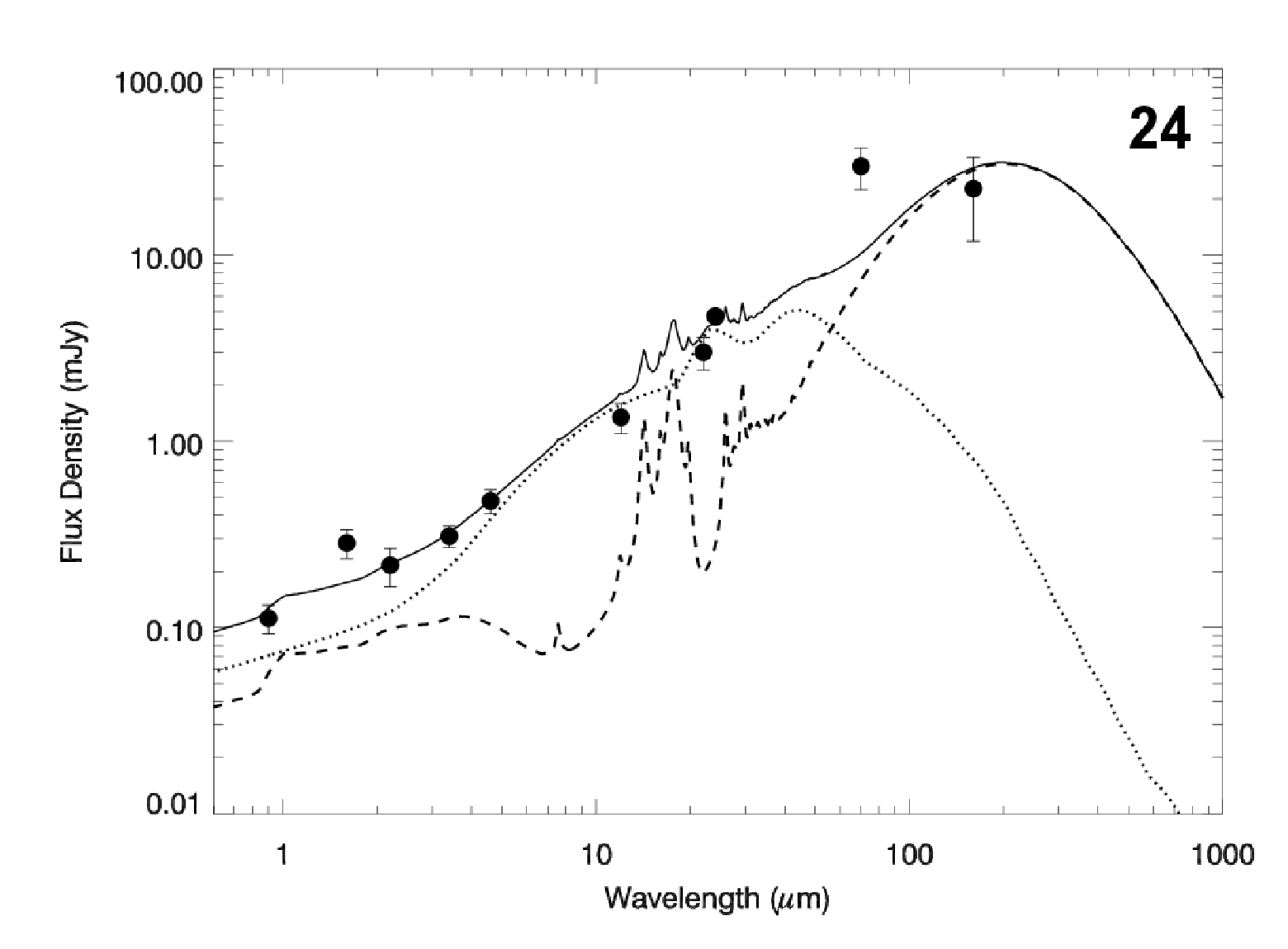}
\caption{Observed-frame fits to the optical through MIPS SEDs for objects 19-24 in Table \ref{tablesample}. Details are the same as for figure \ref{figexamplefits1}.}\label{figexamplefits4}
\end{figure*}

\begin{figure*}
\includegraphics[width=78mm,angle=0]{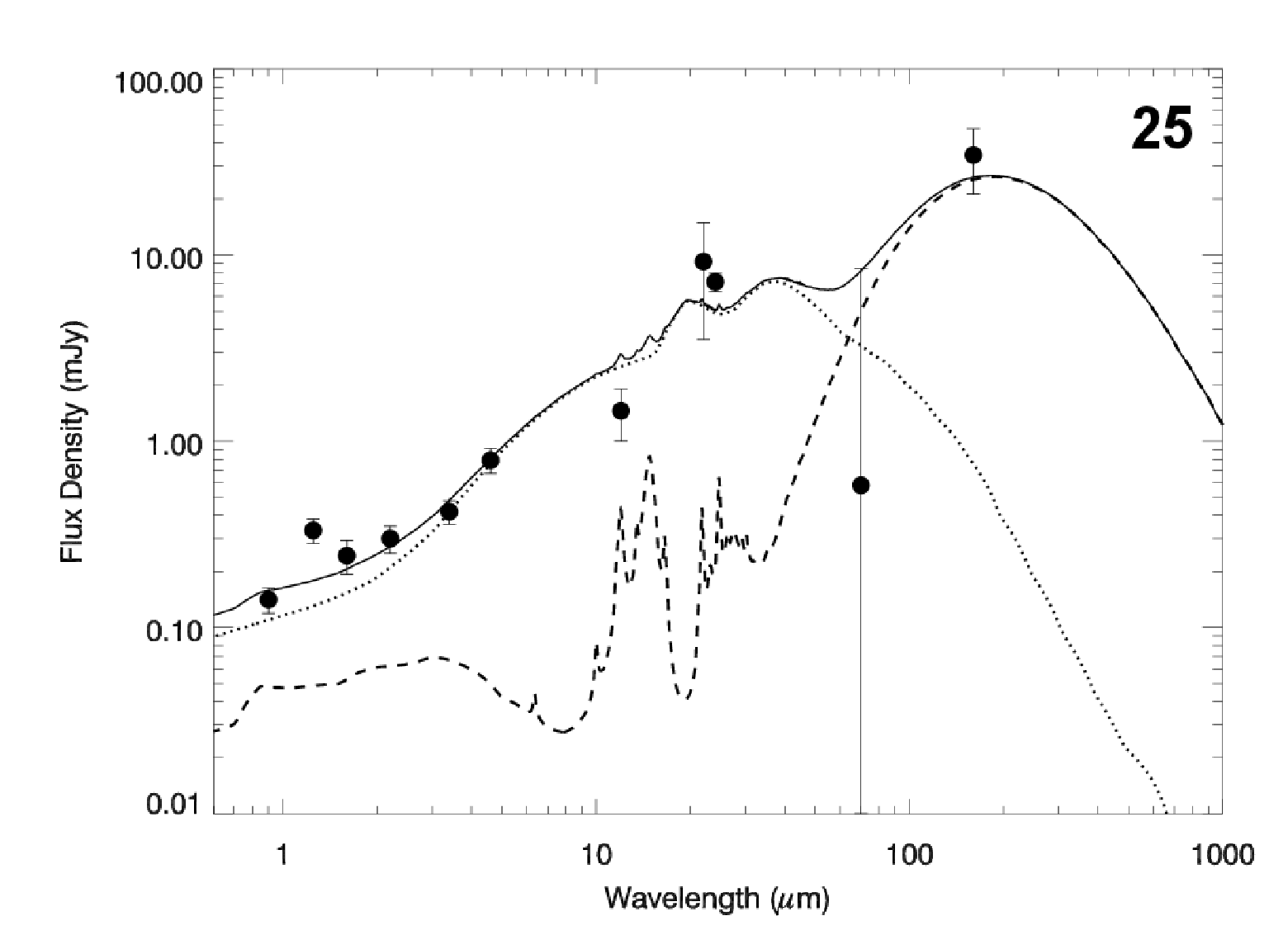}
\includegraphics[width=78mm,angle=0]{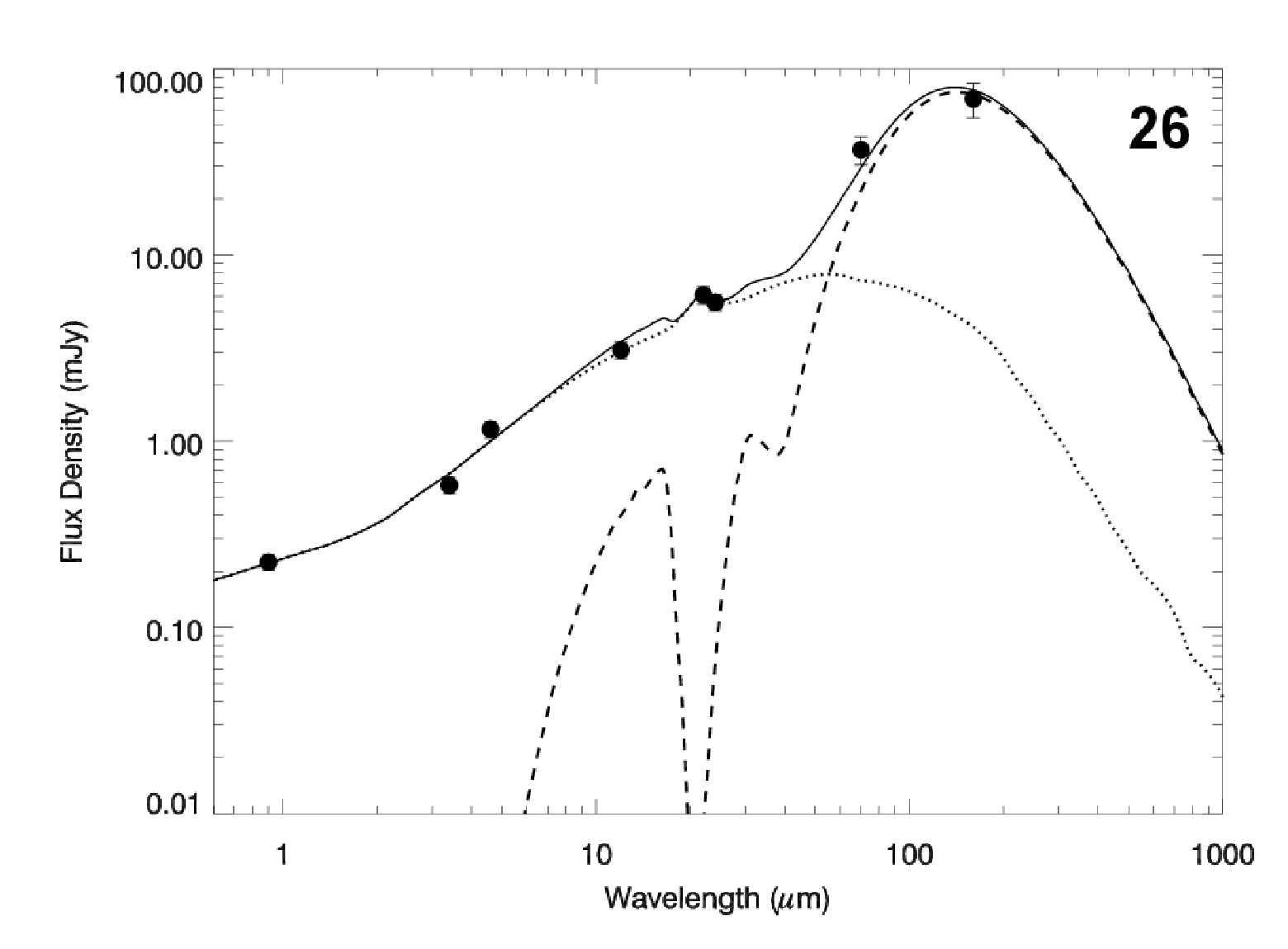}\\
\includegraphics[width=78mm,angle=0]{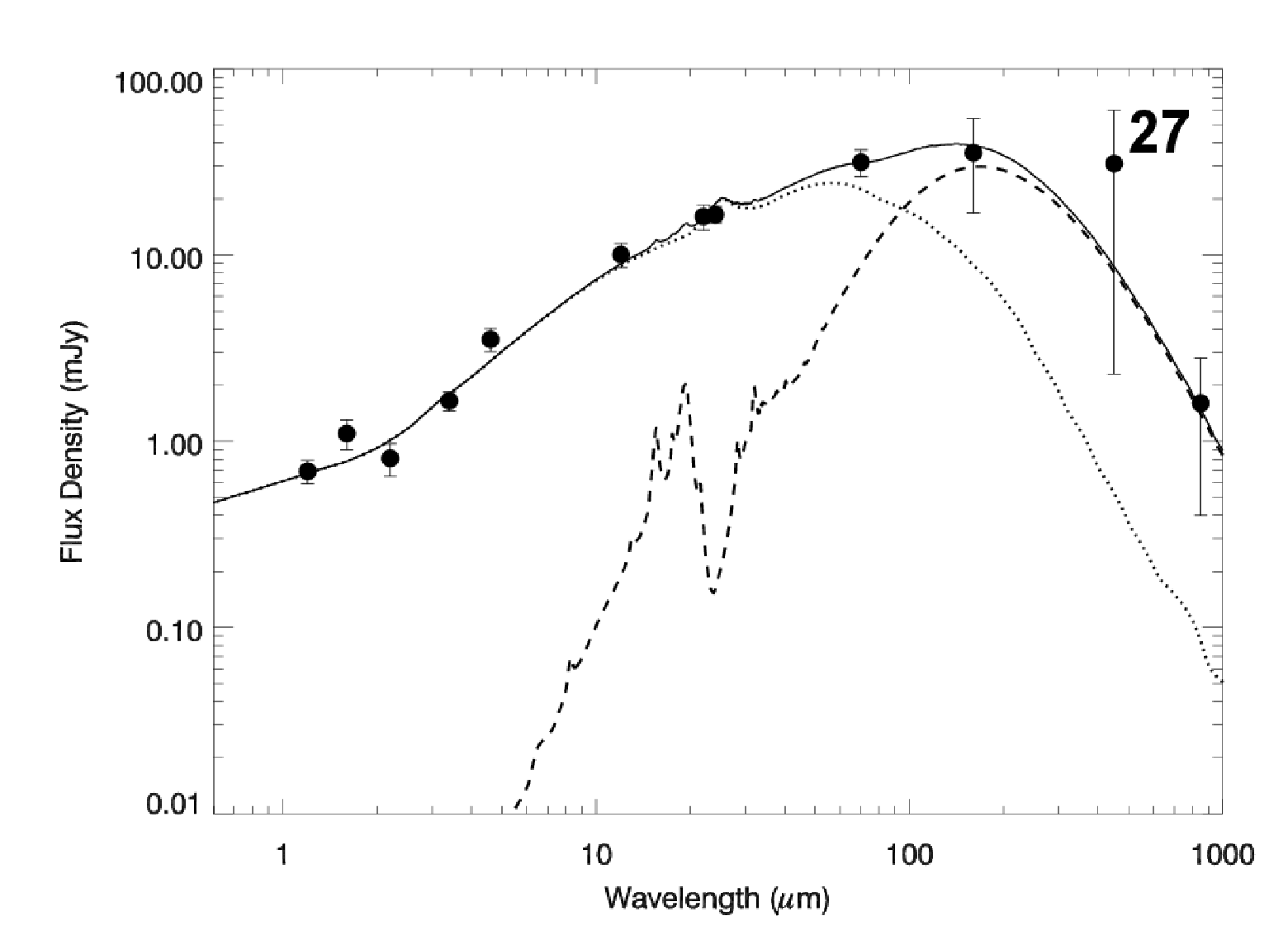}
\includegraphics[width=78mm,angle=0]{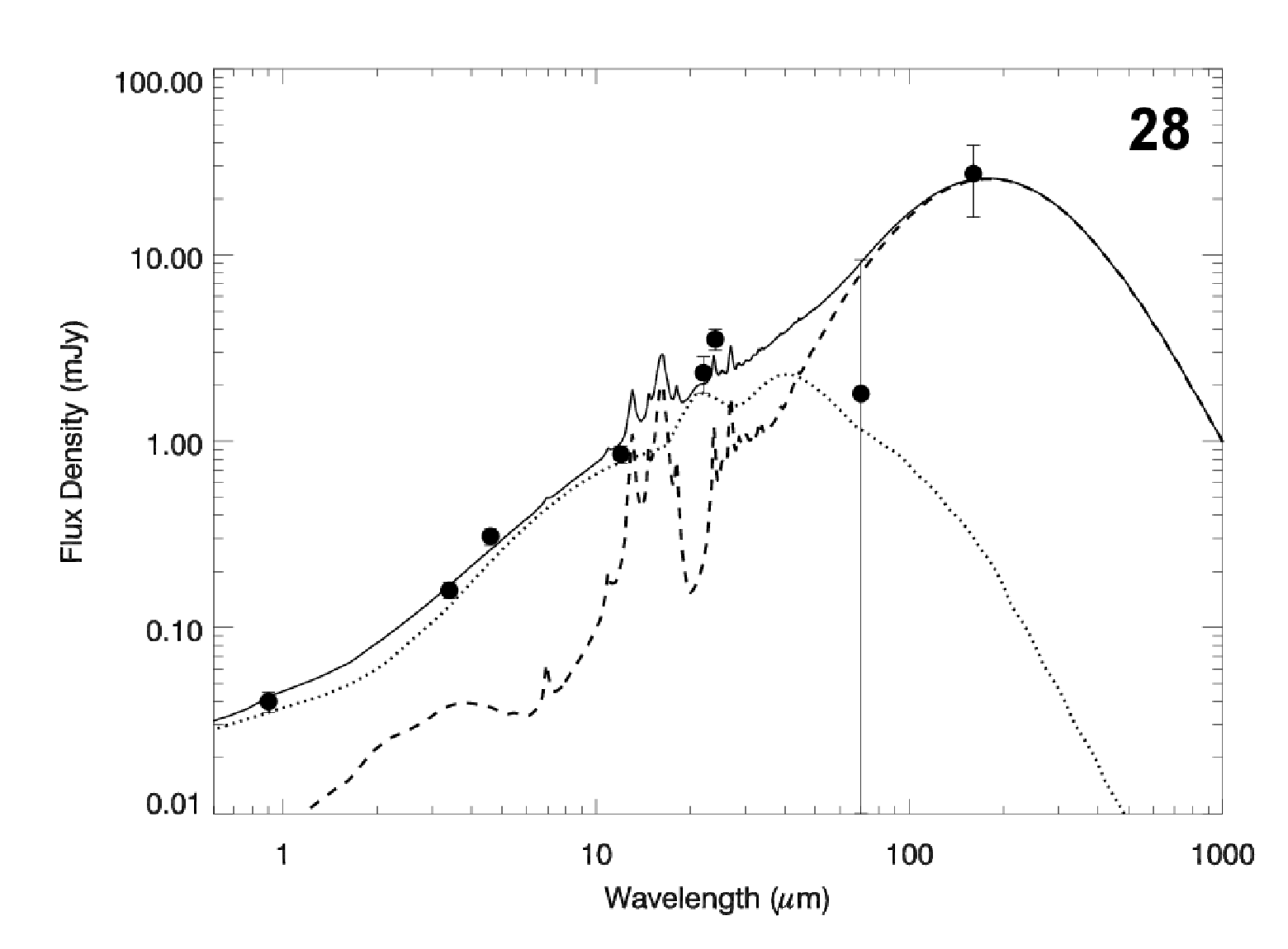}\\
\includegraphics[width=78mm,angle=0]{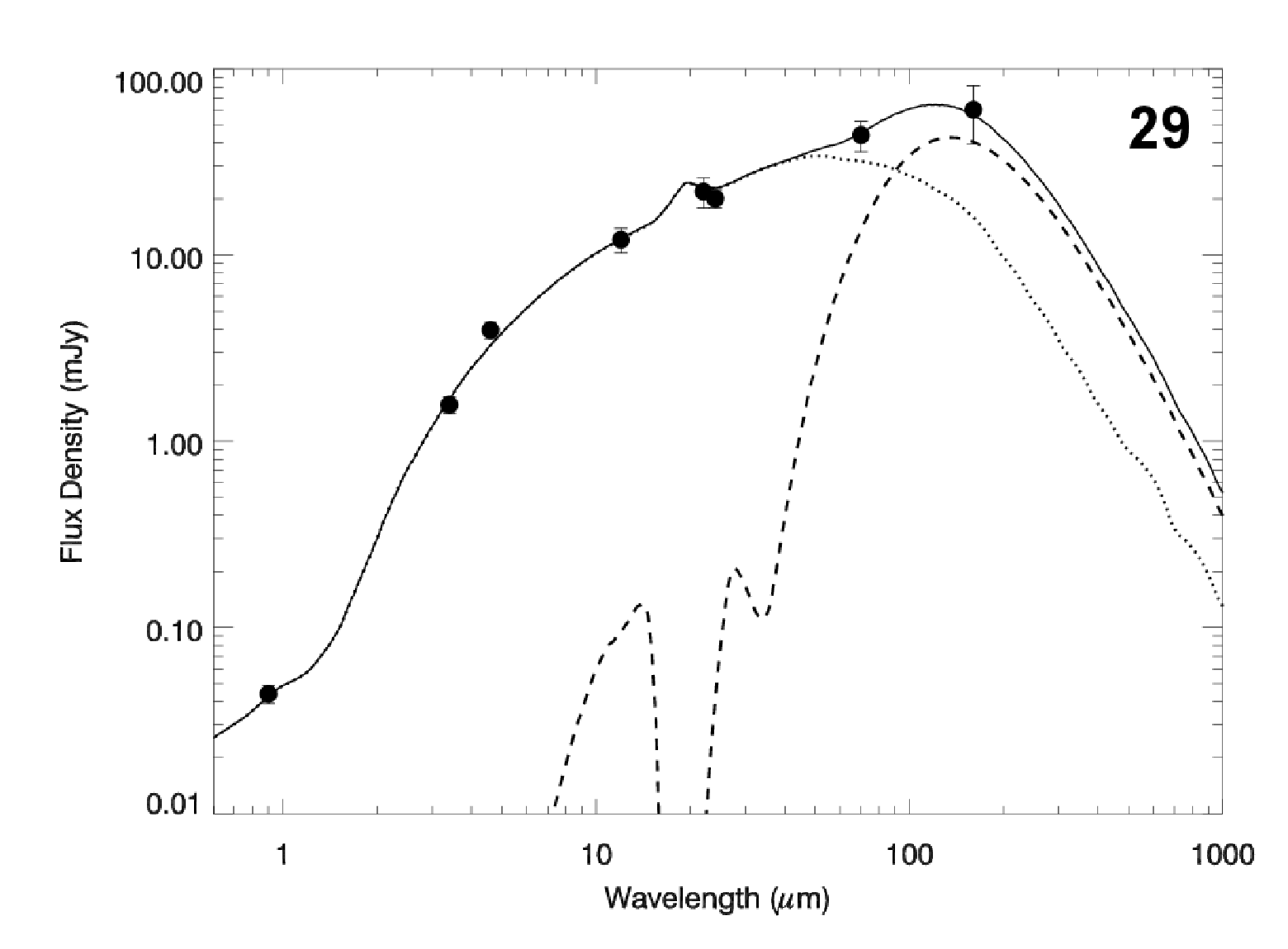}
\includegraphics[width=78mm,angle=0]{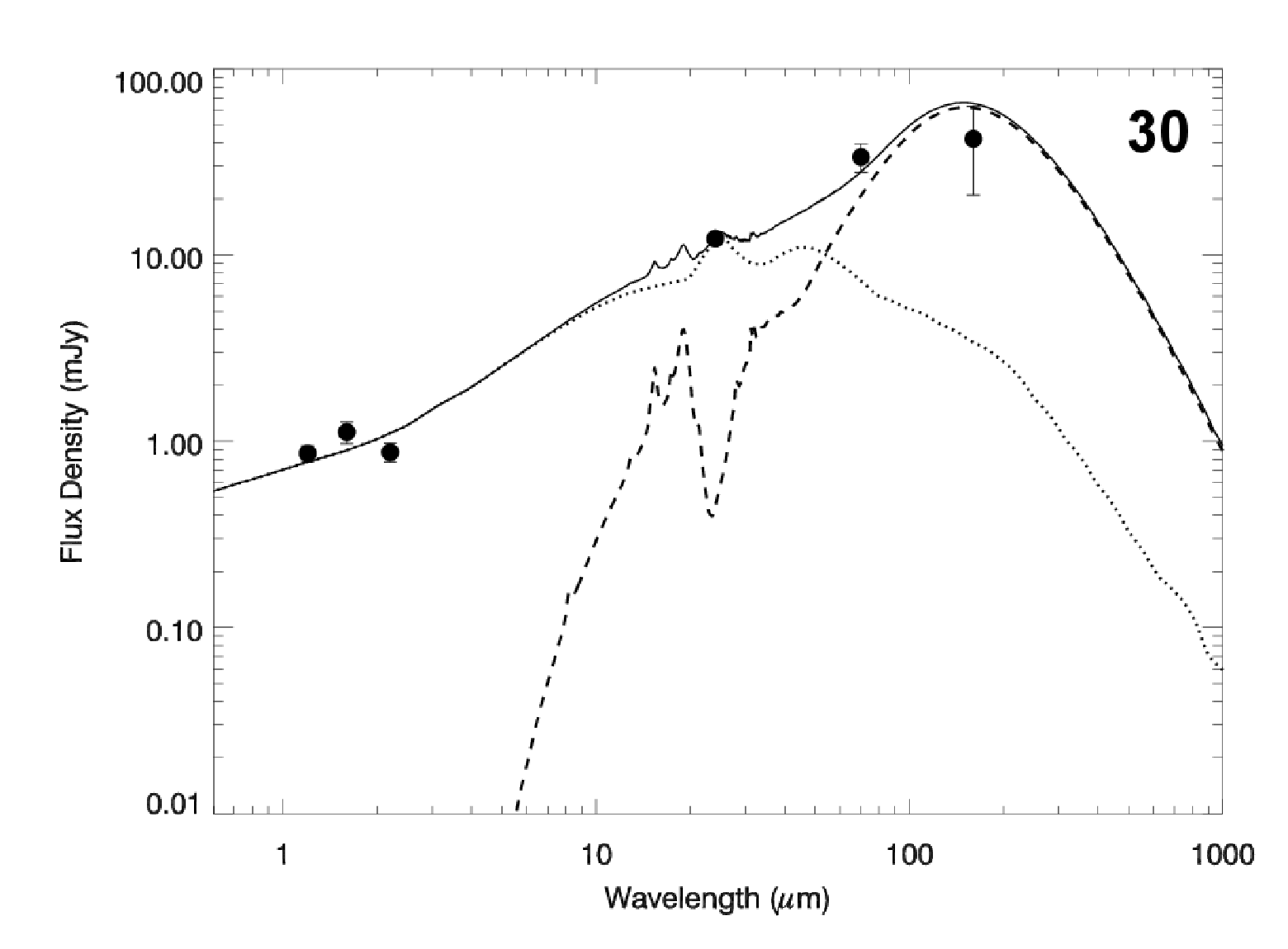}
\caption{Observed-frame fits to the optical through MIPS SEDs for objects 25-30 in Table \ref{tablesample}. Details are the same as for figure \ref{figexamplefits1}.}\label{figexamplefits5}
\end{figure*}

\begin{figure}
\includegraphics[width=78mm,angle=0]{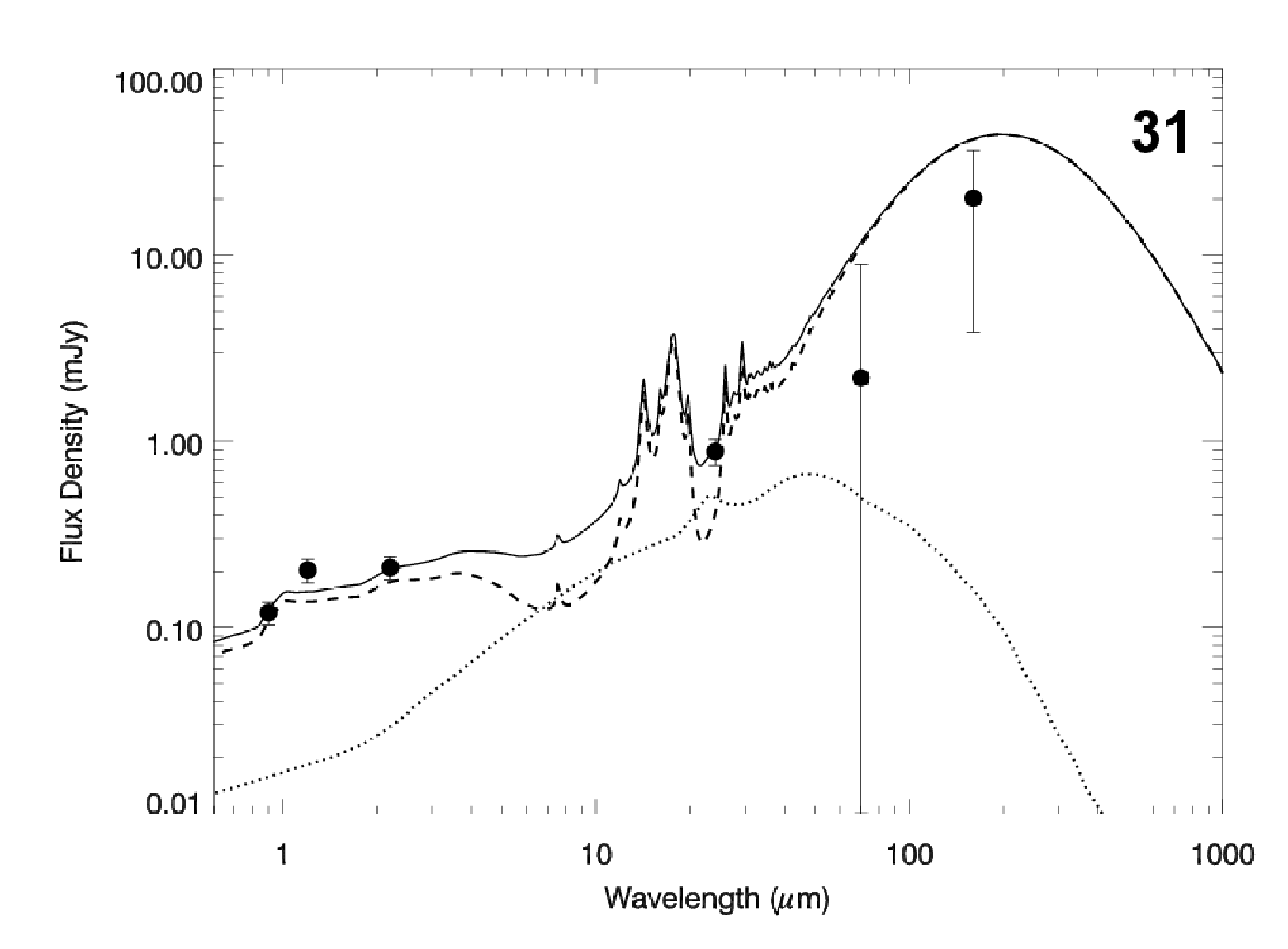}
\caption{Observed-frame fit to the optical through MIPS SEDs for object 31 in Table \ref{tablesample}. Details are the same as for figure \ref{figexamplefits1}.}\label{figexamplefits6}
\end{figure}

For individual galaxies, AGN feedback is predicted to have a significant impact, both from analytic \citep{fabian99,wyithe03,saz05,fabian06,pope09,power11,kav11} and numerical studies \citep{omma04,ciotti07,ant08,tort09,ost10,wagner11,hambrick11}. For galaxy mergers, quasar mode feedback is predicted to have a profound influence \citep{dimatteo05,sprdimher05,debuhr11,snyd11}. Models for the cosmological evolution of galaxies and clusters have incorporated one or both of these feedback paradigms, often with noticeable improvements in reproducing observations. These include semi-analytic models using quasar mode \citep{granato04,menci08}, radio mode \citep{bower06,bower08,gonzalez09}, or both \citep{somer08}, and numerical models with quasar mode \citep{booth09,sales10,mccar11,teyss11,chat11}, or both forms of feedback \citep{croton06,deluc06,sija07,kitz07,monaco07,puch08,short09,gasp11}.

Observationally, evidence for AGN feedback is mounting. Studies of early type galaxies suggest that some form of feedback may have terminated their star formation (\citealt{schaw07,rose09}, but see also \citealt{shin11}). Far-IR observations have also found powerful, plausibly AGN driven outflows in OH$^{-}$ and CO 1-0 \citep{ferug10,chung11,sturm11} in local ULIRGs and QSOs that could exhaust the fuel supply for star formation within $\sim$10$^{7}$ years, while X-ray observations have found mildly relativistic outflows in radio-quiet and radio-loud AGN, with an origin close to the SMBH \citep{chartas03,braito07,tombesi10}. Further observations have provided indirect evidence for `quasar mode' feedback in some galaxies \citep{dunn10,alex10,muller11,rupke11}, and for radio-mode feedback in galaxies and clusters \citep{best06,macnam07,mit09,nesv10,werner10,ehlert11,shab11,ma11,wern11}. Controversies do, however, remain. For example, \citet{jahnke11} suggest that there is no need for AGN feedback to explain the Bulge-SMBH mass relation, \citet{ammons11} suggest that only a small fraction of intermediate luminosity AGN at $z\sim1$ can be undergoing galaxy-wide AGN feedback, and  \citet{lutz10} find that there is no simple inverse relation between star formation rate and AGN luminosity in X-ray selected AGN. It is also worth mentioning that the effectiveness of AGN feedback may depend on the mode of star formation, since spatially compact, `bulge' star formation is likely easier to turn off than spatially extended, `disk' star formation. 

Our group has been examining the role of AGN feedback by looking at systems in which such feedback may be ongoing. To do so, we have been studying the `FeLoBAL' class of QSOs \citep{hazard87}, which comprise part of the Broad Absorption Line (BAL) QSO population\footnote{BAL QSOs show broad, deep absorption troughs in their rest-frame UV spectra. They come in three subtypes. High Ionization BAL QSOs (HiBALs) show absorption in CIV $\lambda$1549\AA, NV $\lambda$1240\AA, SiIV $\lambda$1394\AA\ \& Ly$\alpha$. Low Ionization BAL QSOs (LoBALs) additionally show absorption in MgII $\lambda$2799\AA\ and other lower ionization species. Finally, the FeLoBAL QSOs, in addition to showing all the absorption lines seen in LoBALs, also show weak absorption from any {\it excited} term of FeII (e.g. FeII* $\lambda$2400,2600\AA), \& also FeII $\lambda$2750\AA, FeIII $\lambda$1895,1914,1926\AA. See e.g. \citealt{lyn67,weymann91,green96,becker97,sch99,hall02,lacy02,trump06,gallagher07,case08,gibson09,leigh11}.}. We selected this population for three reasons. First, the UV absorption troughs are unambiguous signatures of radiatively driven outflows powered by an AGN. They have velocities of up to 0.2$c$, widths of at least a few thousand km s$^{-1}$), and are usually very deep or black, implying high column densities moving at high velocities, which plausibly implies high mass-loss rates \citep{arav94,proga00,chartas03,crenshaw03}. Therefore, they cannot be driven by even the most extreme starbursts. Furthermore, in at least some FeLoBAL QSOs the outflows may extend up to several Kpc into the host galaxy \citep{arav08,moe09,bautista10,dunn10}. Second, they are invariably reddened objects with high IR luminosities \citep{farrah07}, and sometimes harbour intense starbursts \citep{farrah10}, suggesting they may be a transition phase in the lifetime of an AGN (FeLoBAL features are also occasionally seen in ULIRGs, see e.g. \citealt{farrah05}). Third, recent results have shown that FeLoBAL QSOs are much more common at $z\gtrsim0.5$ than was originally thought, by up to a factor of ten \citep{dai08,f2ms,allen11}. 

In this paper, we compare the strengths of the AGN-driven outflows to the luminosities of obscured star formation in a sample of 31 FeLoBAL QSOs, selected purely on the basis of their rest-frame UV spectral properties. We combine data from the Spitzer space telescope \citep{werner04} with data from the Sloan Digital Sky Survey (SDSS, \citealt{york00}), the Two Micron All-Sky Survey (2MASS, \citealt{skrutskie06}), the UKIRT Infrared Deep Sky Survey (UKIDSS, \citealt{law07}), and from the Wide-Field Infrared Survey Explorer (WISE, \citealt{wright10,jarr11}) in order to measure the luminosities of both obscured AGN activity and star formation. We measure the strengths of the AGN-driven outflows from the UV absorption troughs in the SDSS spectra. We assume a spatially flat cosmology, with H$_{0}$ = 70 km s$^{-1}$ Mpc$^{-1}$, $\Omega = 1$, and $\Omega_{\Lambda} = 0.7$. We use the term "IR luminosity" to refer to the luminosity integrated over 1-1000$\mu$m in the rest-frame. We quote luminosities in units of bolometric Solar luminosities, where L$_{\odot} = 3.826\times10^{26}$ Watts.

\section{Analysis}

\subsection{Sample Selection}\label{secselect}
We aimed to select a sample of FeLoBAL QSOs purely on the basis of their rest-frame UV spectral features. Furthermore, since measuring BAL properties is not straightforward (see \S\ref{discusssboutflow} and \citealt{hall02,trump06,gibson09,allen11}) we required the sample to have optical spectra from the same source, and BAL measurements already in the literature. 

Accordingly, we chose our sample from the SDSS. We started with the six SDSS objects in \citealt{farrah07} (hereafter F07), excluding ISO J005645.1-273816 and LBQS 0059-2735 as neither lie within the SDSS survey. These six objects were originally selected with the only constraint being that their redshifts satisfied $1.0<z<1.8$. We then selected a further 25 FeLoBAL QSOs from \citet{trump06}. We imposed the same upper redshift cut of $z=1.8$ to ensure that observed-frame 160$\mu$m remains close to the peak of the far-IR emission for the (monolithic) dust temperatures of $\gtrsim 50$K expected within $\sim100$pc of an AGN. We slightly reduced the lower redshift cut from the F07 sample to $z=0.8$ to ensure that the SDSS spectra always contain the Mg II BAL (see \S\ref{discstrengths}). Hence, the final sample comprises 31 objects. This sample should be a random subset of the FeLoBAL QSO population between $0.8<z<1.8$, with no selection on IR luminosity. The sample is listed in Table \ref{tablesample}.

We are however not certain that our sample are all FeLoBAL QSOs, since BAL features are difficult to identify (see \S\ref{discusssboutflow} and e.g. \citealt{appen05}). Two objects in particular are problematic; SDSS J033810 and SDSS J233646. SDSS J033810 is discussed in detail by \citealt{hall02}; in summary, it is not certain whether the BALs in this object are real, or a phantom of a peculiarly shaped continuum. SDSS J233646 on the other hand is clearly an FeLoBAL QSO, but is in a binary system with a separation of $<2\arcsec$ \citep{gregg02}. We have assumed that the IR emission comes solely from the FeLoBAL QSO (object B in Gregg et al), but it is probable, particularly at 24$\mu$m, that there is contamination from the companion. Two further objects are worth mentioning; SDSS J081312 \& SDSS J105748. These two objects show absorption from FeII but not FeII*; the feature originally identified as FeII*$\lambda$2600\AA\ was subsequently shown to be a blend of FeII$\lambda2580$\AA\, MnII$\lambda2577,2595,2606$\AA\, and narrow MgII$\lambda2799$\AA\ with an uncertain relation to the QSO absorption. Since the original definition of an FeLoBAL QSO requires the detection of FeII* \citep{hazard87}, these two objects would not be formally classified as FeLoBAL QSOs if extreme strictness was imposed. They do however both show FeII absorption, so we treat them as FeLoBAL QSOs.

\subsection{Observations}
A description of the observations of the six objects from F07 can be found in that paper. The 25 new objects were observed in cycle 5 with the Multiband Imaging Photometer for Spitzer (MIPS, \citealt{rieke04}) at 24$\mu$m, 70$\mu$m  and 160$\mu$m. The “small” field size was used for all three channels, using the default pixel scale at 70$\mu$m. At 70$\mu$m and 160$\mu$m we chose the same exposure parameters as used in F07; seven 10 second cycles at 70$\mu$m and four 10 second cycles at 160$\mu$m. At 24$\mu$m we used a shallower total exposure time of two 10 second cycles per source, sufficient to detect, at $\stackrel{>}{_{\sim}} 5\sigma$, a 0.5mJy source, since all the objects in F07 were strongly detected at 24$\mu$m.

\begin{figure*}
\includegraphics[width=65mm,angle=90]{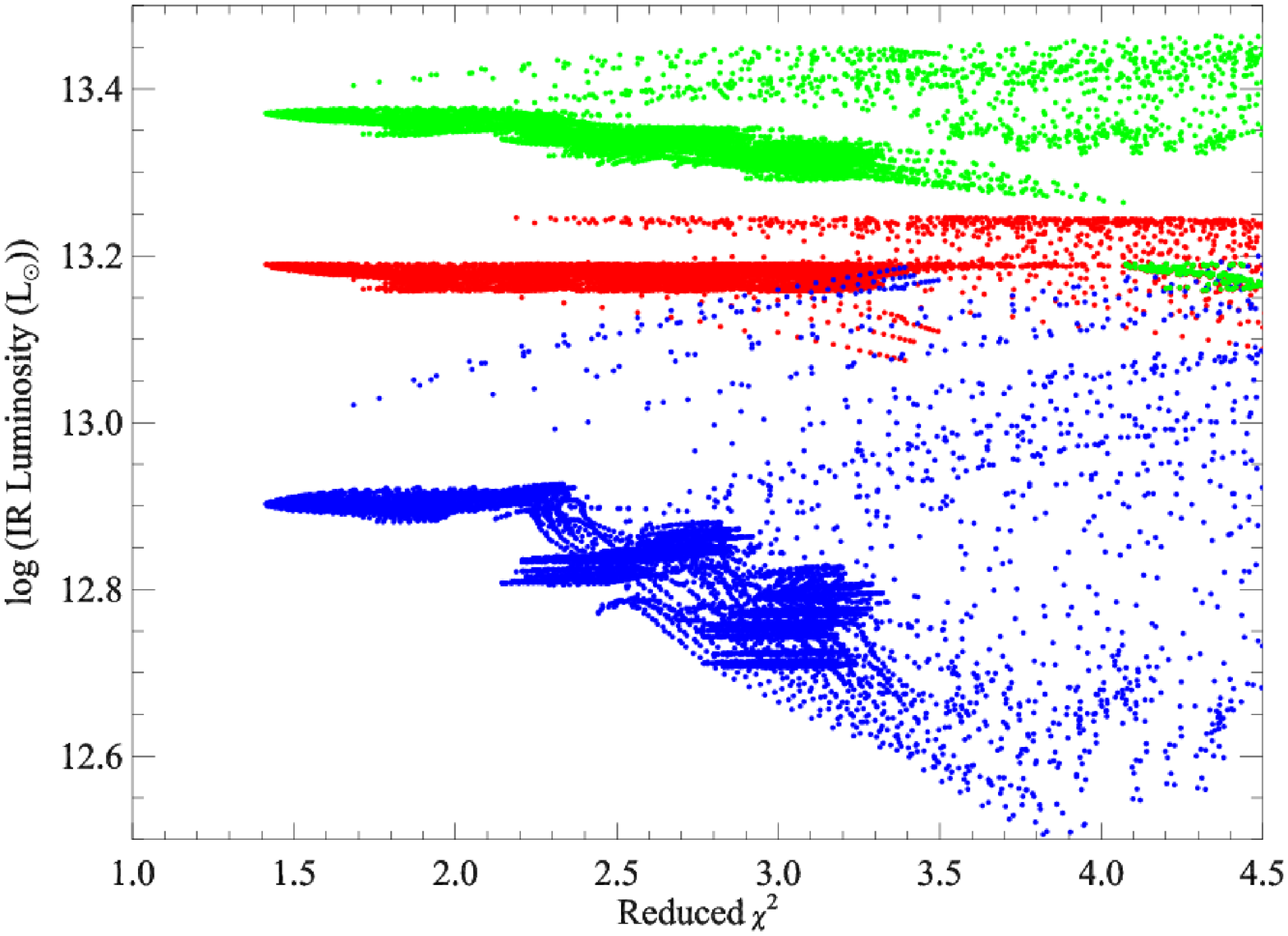}
\includegraphics[width=65mm,angle=90]{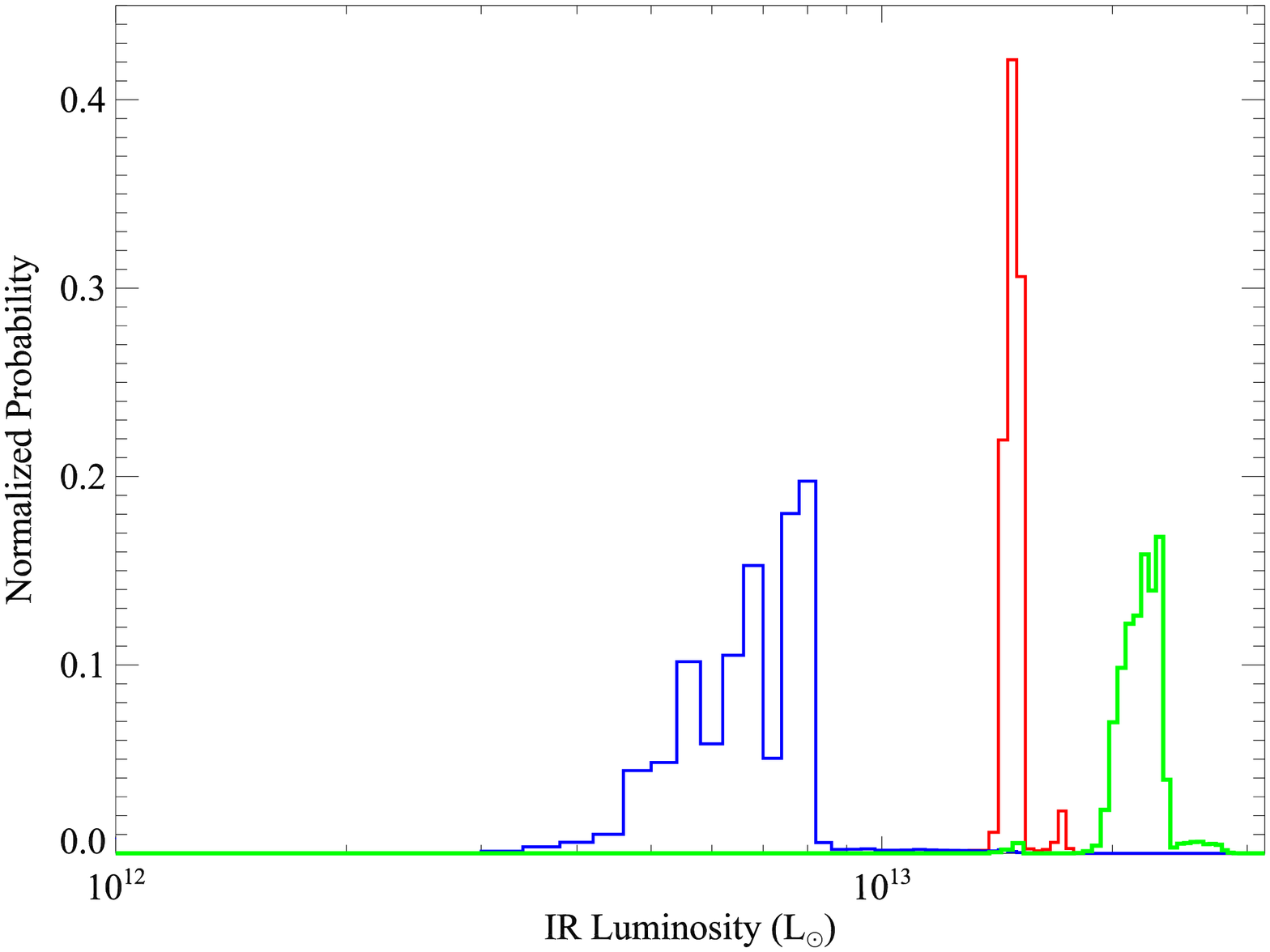}
\caption{An example of the method used to derive the IR luminosities for each object. Here we show the data for one object, SDSS J221511.93-004549.9. The left panel shows the $\chi^{2}_{red}$ values and resulting luminosities for all possible combinations of starburst and AGN template fits. The green points are the total IR luminosities, while the red and blue points are the AGN and starburst luminosities, respectively. The right panel shows the differential probability distribution functions (for total, AGN and starburst luminosity) that arise from the $\chi^{2}_{red}$ weighted combination of all the solutions in the left panel. }\label{figexamplepdf}
\end{figure*}

We reduced the data using the MOPEX software provided by the Spitzer Science Centre, which performs standard tasks such as image co-addition, sky and dark subtraction, and bias removal. We used the default MOPEX parameters for MIPS small field observations. To check the quality of our reduction we compared it to that from the automated pipeline and found them to be consistent to within a few percent. We measured fluxes in all three channels using two methods. For the observations in which the source was clearly detected - all the 24$\mu$m data and about half of the 70$\mu$m data - we used the APEX package within MOPEX with the default parameters and PRFs for point-source photometry. For the observations in which the source was weakly or not detected we found that PRF fitting photometry was not suitable, so we used the Sextractor package \citep{bertin96} in aperture photometry mode, with point-source aperture corrections provided by the Spitzer Science Center. For undetected sources we measured the flux within the aperture, and then subtracted the average background flux. We compared the results from APEX PRF fitting photometry and Sextractor aperture photometry for the brighter sources, and found that they agreed to within  a few percent, with no obvious systematic differences. We therefore use the resulting mix of aperture and PRF fluxes in our analysis.

\section{Results}
The MIPS photometry is presented in Table \ref{tableasandlirs}. We combine the MIPS data with archival photometry from SDSS, 2MASS (or UKIDSS if available), and WISE. This archival data is presented in Table \ref{tablesample}. Additional sub-mm photometry for two objects is presented in \citet{lewis03}. 

As WISE returned data only recently, we provide more detail on the WISE data. WISE completed its first full coverage of the sky in July 2010. A Preliminary Release Catalog\footnote{http://wise2.ipac.caltech.edu/docs/release/prelim/} covering 57\% of the sky, was made available to the community starting April 14, 2011. Point sources were extracted with 5-$\sigma$ sensitivities greater than 0.08, 0.11, 0.8, and 4 mJy, respectively for the four bands. Photometric calibration, source counts, colors and population statistics at the ecliptic poles are presented by \citet{jarr11}.  Of our 31 sources, ten come from the Preliminary Release Catalog, and the rest come from the proprietary WISE First Pass Internal Source Database, which performs internal verification, quality and photometric analysis. In the cases where there were multiple measurements of the same source (due to the overlap between WISE orbit-to-orbit scans) we used quality indicators to choose the best measurement. Overall, 25 of our sources have $>2\sigma$ detections in at least one of the four WISE bands.

We measure total IR luminosities, and the contributions to the total IR luminosities from star formation and AGN activity, by fitting the optical through MIPS photometry for each object with radiative transfer models for AGN and starbursts, following the methods in \citealt{farrah03} and F07. We describe the models and the fitting methods in \S\ref{subsmodels} \& \S\ref{sectfitting}.

\subsection{The Models}\label{subsmodels}
For the AGN models, we follow \citet{efs95} and use a tapered disk dust distribution, in which the disk thickness increases linearly with distance from the central source in the inner part of the disk but tapers off to a constant height in the outer part. The tapered disc models include a distribution of grain species and sizes, multiple scattering and a density distribution that follows $r^{-1}$ where $r$ is the distance from the central source. The models assumed a smooth distribution of dust, so they are a good approximation of the density distribution in the torus if the mean distance between clouds is small compared with the size of the torus. These models have been successful in fitting the SEDs of several classes of AGN, and ULIRGs \citep{alex99,ruiz01,farrah02,verma02,farrah03,efs05}. In this paper we use a grid of models with four discrete values for the equatorial 1000\AA~ optical depth (500, 750, 1000, 1250), three  values for the ratio of outer to inner disc radii (20, 60, 100) and three values for the opening angle of the disc (30$^o$, 45$^o$ and 60$^o$). The spectra are computed for 74 inclinations which are equally spaced in the range 0 to $\pi/2$. For comparison, other work on radiative transfer modeling of the dust distribution in AGN has been presented by \citet{pier92,gran94,nenkova02,dullemond05,hoenig06,schart08}.

For the starburst models, we combine the \citet{bc03} stellar population synthesis model with a radiative transfer code that includes the effect of small grains and PAHs, the updated dust model of \citet{efs09}, and a simple evolutionary scheme for the molecular clouds that constitute the starburst. The model predicts the spectral energy distributions of starburst galaxies from the ultraviolet to the millimetre as a function of the age of the starburst and the initial optical depth of the molecular clouds. These models have been successfully used to fit the SEDs of several classes of star-forming galaxy (e.g. \citealt{farrah03}), as well as the `Fork' diagram of \citealt{spoon07} \citep{mrr09}. For comparison, other starburst models have been developed by \citet{mrr89,kru94,silva98,takagi03,dopita05,sieben07}.

\begin{figure*}
\includegraphics[width=65mm,angle=90]{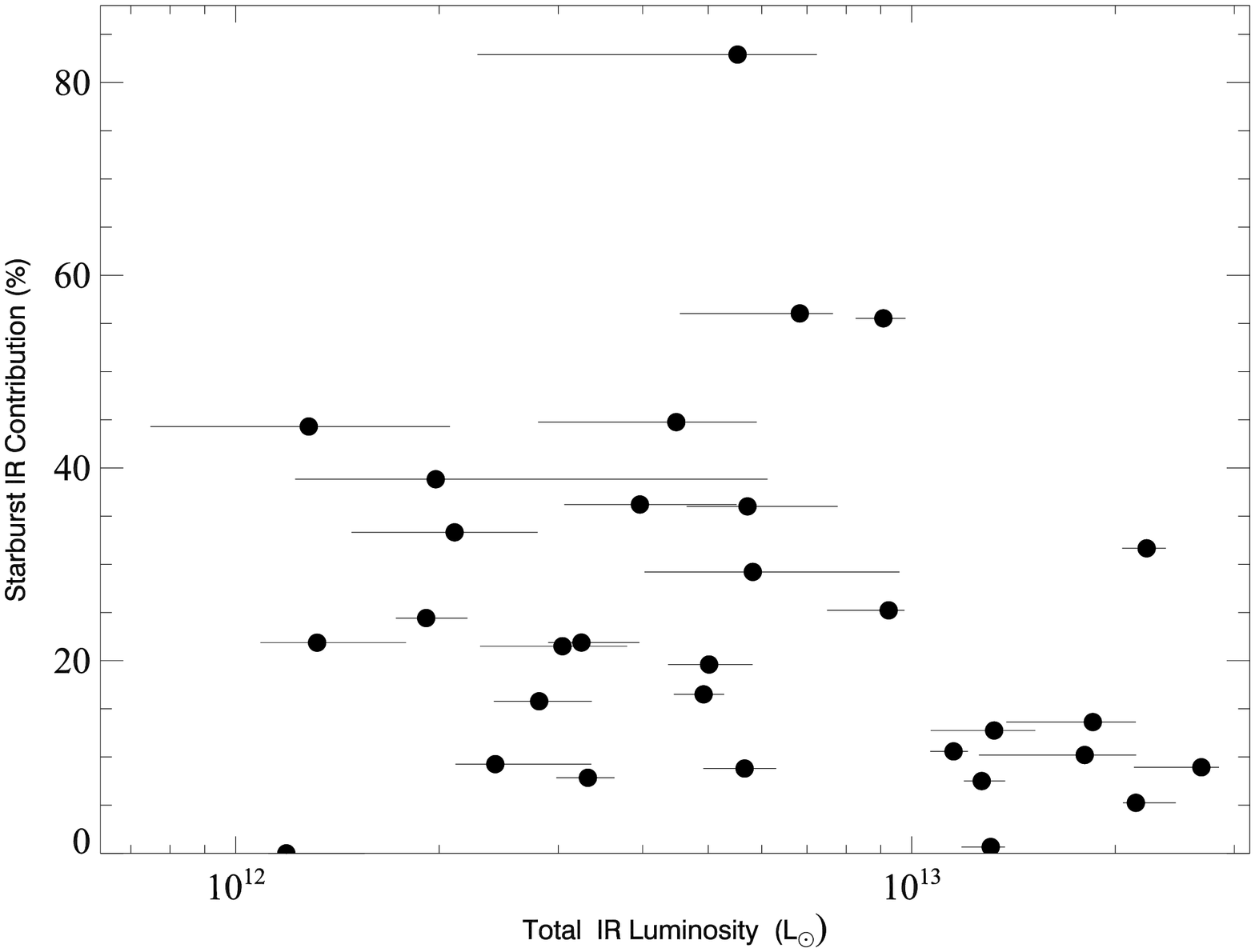}
\includegraphics[width=65mm,angle=90]{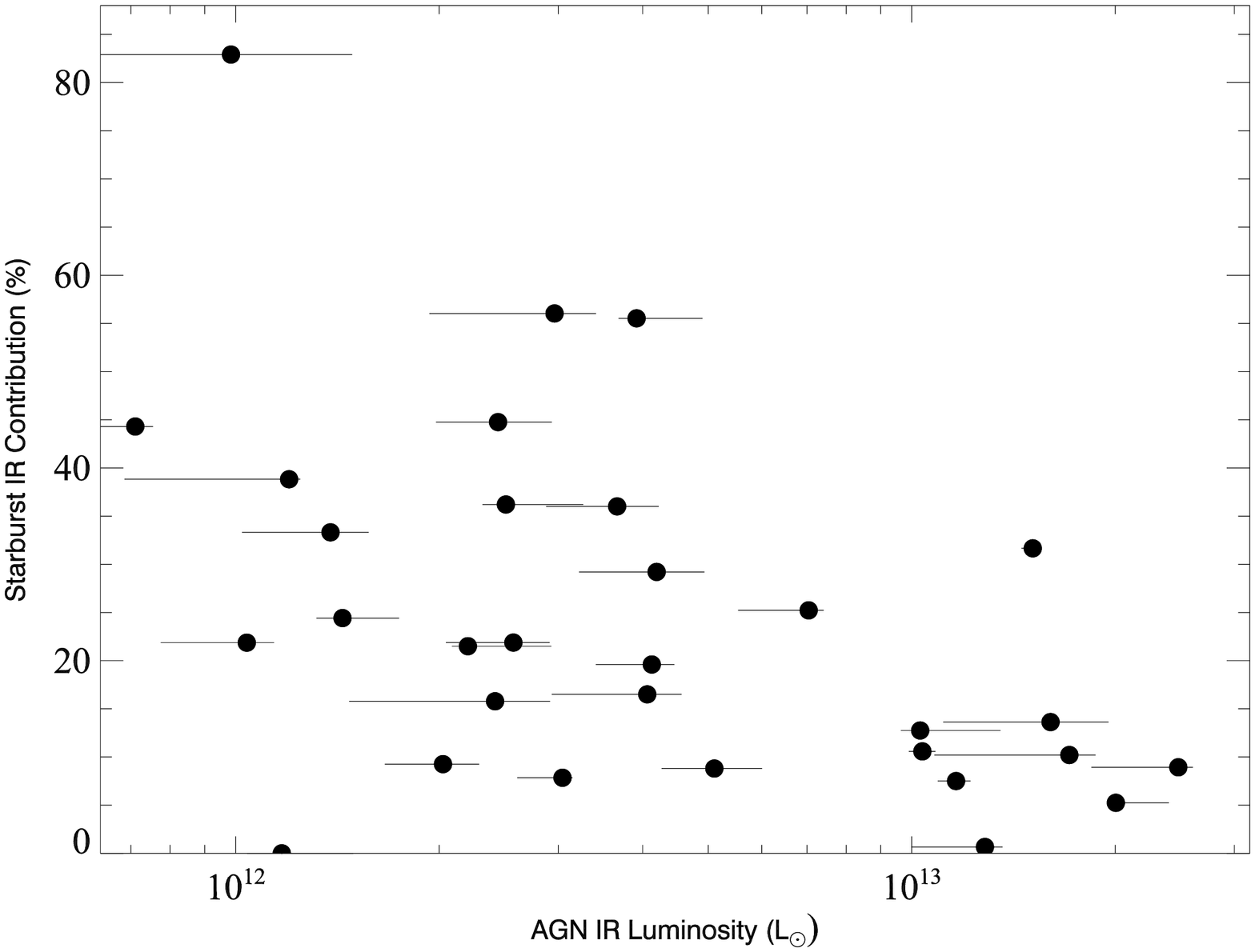}
\caption{{\itshape Left:} Total IR luminosity vs. contribution to the total IR luminosity from the starburst. {\itshape Right:} AGN luminosity vs. starburst contribution to the total IR luminosity. In neither plot do we see a clear correlation, though it is interesting that the starburst contribution drops at L$_{AGN}\gtrsim10^{13}$L$_{\odot}$.} \label{agnconttolir}
\end{figure*}

\subsection{Fitting \& Results}\label{sectfitting}
We fit all possible combinations of starburst plus AGN models to the SDSS through MIPS photometry of each object in order to extract starburst and AGN luminosities. 

The models do not include a prescription for the broad absorption features in the rest-frame UV that arise from radiatively driven outflows, so we do not include the SDSS spectra in the fits. We do not include photometry shortward of $\sim$0.35$\mu$m in the rest-frame for the same reason; this means that we do not use the $Ugri$ photometry for any object, and in some cases do not use the $z$ band data either. We do however use photometry at rest-frame wavelengths of $0.35<\lambda(\mu$m)$<1$; while we are interested in starburst and AGN luminosities at rest-frame 1-1000$\mu$m, including these data limits which models are acceptable fits in the optical, and thus limits which models can be used to fit the IR data. For the MIPS data we fit to the measurements in all cases even if they are of low significance, or negative. We allowed the contribution from each component to vary from 0$\%$ to $100\%$ to see if the IR emission was consistent with arising from a single origin. The model libraries span several free parameters (\S\ref{subsmodels}), but our data are not comprehensive enough to constrain all of them. Therefore, we use the complete model libraries to determine only the range in total, starburst and AGN luminosities that are consistent with the data. 

The best-fit SEDs are presented in Figures 1-6. In most cases the models provide an excellent fit to the data. We obtain $\chi^{2}_{red}<1$ in 12/31 objects, and $1<\chi^{2}_{red}<2$ in 15/31 objects. The remaining four objects have $2<\chi^{2}_{red}<3$. The most difficult points to fit are usually the near-IR and/or the two shorter wavelength WISE bands, which are somewhat underpredicted in many cases. This could be due to the still preliminary WISE photometric calibration, and/or the lack of a host galaxy contribution in our models, so we defer exploration of this until more comprehensive near-IR data are available, and the WISE calibration is refined. In nearly all cases the $\geq 24\mu$m photometry is well fitted by the models, though in a few objects the fit is relatively poor at these wavelengths. This effect is most noticeable in objects 3, 11, and 24, in which the 70$\mu$m flux is significantly underpredicted. This could be due to an additional `hot' dust component or an extremely strong spectral feature that is not reproduced in the models. Since the models still reproduce the slope and approximate normalization of the far-IR SEDs even for these objects, we consider the results to still be usable. 

To extract the most probable total, AGN and starburst luminosities for each object, and their confidence intervals, we first construct a discrete probability distribution function (PDF) for the total, AGN and starburst luminosities of each source. We weight each fits contribution to the PDF by its $\chi^{2}_{red}$ value. From these PDFs, we then construct cumulative distribution functions, from which the most probable luminosities, and their confidence intervals, can be derived straightforwardly. Ideally, we would show the PDFs for all objects. This would however be unwieldy, so, as an example of the method, we show the distribution of $\chi^{2}_{red}$ values and the total, starburst and AGN PDFs for one object in Figure \ref{figexamplepdf}. In most cases the PDFs have a single peak and an approximately gaussian shape. In a few cases though there are minor secondary maxima, and/or the shape is non-gaussian. We therefore quote as the positive and negative errors the 90\% confidence intervals.  The luminosities, their confidence ranges and the $\chi^{2}_{red}$ values of the best fits are presented in Table \ref{tableasandlirs}.

\section{Discussion}

\subsection{Infrared Luminosities}\label{disclums}
We find that FeLoBAL QSOs are luminous in the IR. All of our sample have (best-fit) total IR luminosities (L$_{Tot}$) in excess of $10^{12}$L$_{\odot}$, with nine objects, or 29$\%$, exceeding $10^{13}$L$_{\odot}$. There is no other sample to which we can directly compare ours, since no sample matched to ours in optical continuum luminosity and redshift has been observed in the mid/far-IR. We do however find that FeLoBAL QSOs are more IR-luminous than the Palomar-Green (PG) QSOs in \citet{haas03}. Instead, they appear to have comparable IR luminosities to the wider population of BAL QSOs \citep{gallagher07}, reddened QSOs \citep{geo09}, the most IR-luminous members of the general QSO population \citep{lutz08,orell11}, and to ULIRGs \citep{genzel98,farrah02,farrah03,desai07}.

The dominant power source behind the IR emission is, in most cases, AGN activity. A pure AGN is either the most likely power source, or consistent within the 90\% confidence interval, for $35\%$ of the sample (11/31 objects). A starburst component is required (at $\geq 90\%$ confidence) for the remaining objects, but in only twelve of these objects is the starburst more luminous than $10^{12}$L$_{\odot}$, and in only three objects is the starburst more luminous than the AGN. The mean AGN contribution to the total IR luminosity is $\sim76\%$, comparable to that seen in local ULIRGs with `warm' IR colours\footnote{That is, those objects where the IRAS 12$\mu$m to IRAS 60$\mu$m flux ratio is greater than 0.2 - this is usually interpreted as an AGN supplying at least a plurality of the IR emission.}, but lower than that seen in PG QSOs \citep{veill09}. The spread in AGN contribution to L$_{Tot}$ is however wide, spanning 0.2 to 1.0. We find a strong correlation between L$_{Tot}$ and the AGN IR luminosity (L$_{AGN}$), with a Spearman rank correlation cofficient\footnote{We also evaluated all the correlations presented here using the Kendall tau rank correlation coefficient, and obtained similar results in all cases.} of $\rho = 0.92$ and a significance of deviation from zero of $P <0.001$. We also find a correlation between L$_{Tot}$ and starburst IR luminosity (L$_{SB}$), with $\rho = 0.66$ and $P < 0.001$. Conversely, we find no trend between L$_{Tot}$ and the starburst {\itshape contribution} to L$_{Tot}$ ($f_{SB}$), with $\rho = -0.23$, $P = 0.21$ (Figure \ref{agnconttolir} left). Finally, we find a weak anticorrelation between L$_{AGN}$ and $f_{SB}$ ($\rho = -0.47$, $P = 0.007$, Figure \ref{agnconttolir} right). This is most apparent when the AGN IR luminosity exceeds $\sim10^{13}$L$_{\odot}$, at which point the starburst contribution decreases noticeably. 

\begin{figure*}
\includegraphics[width=130mm,angle=90]{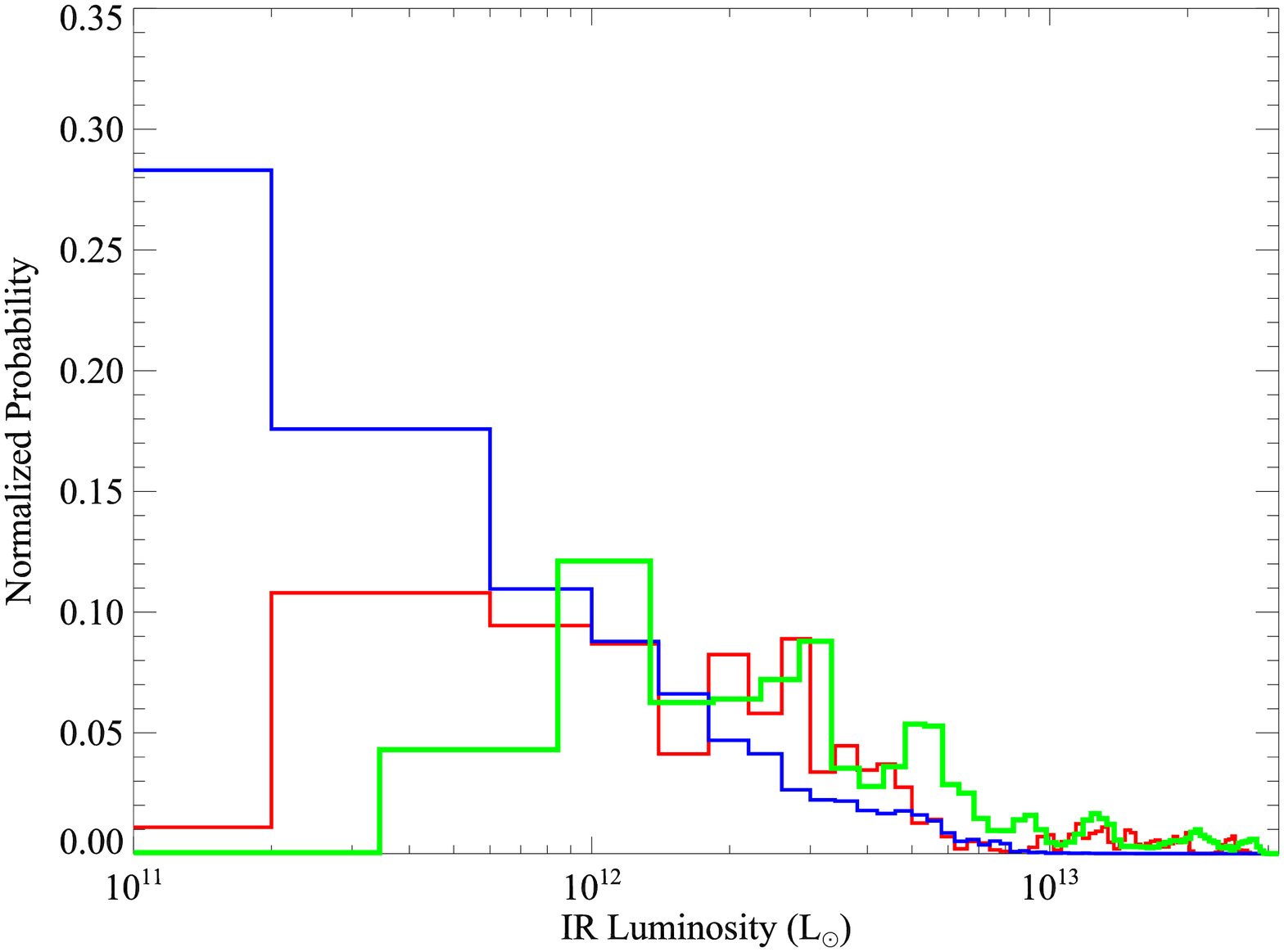}\\
\includegraphics[width=43mm,angle=90]{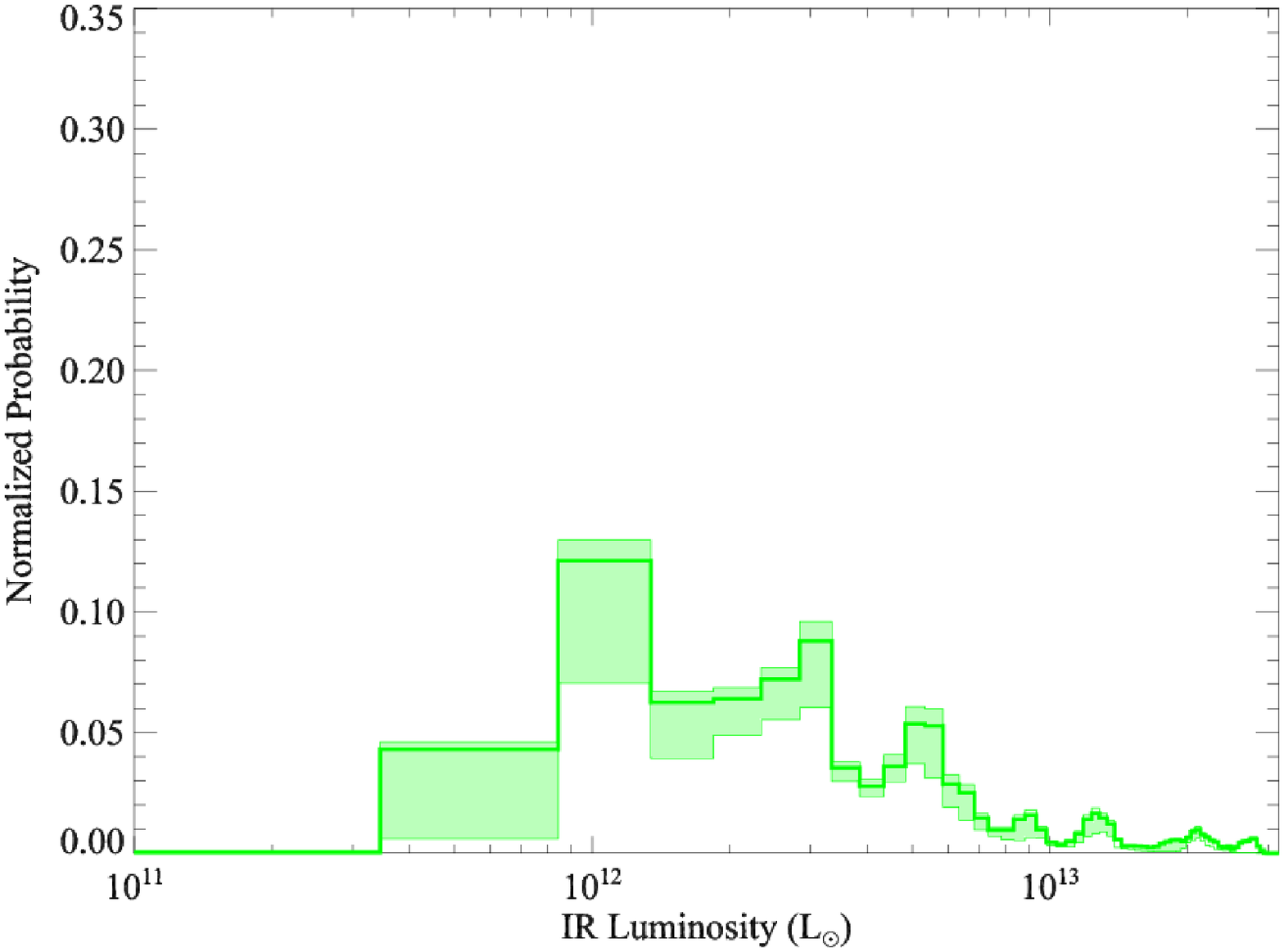} 
\includegraphics[width=43mm,angle=90]{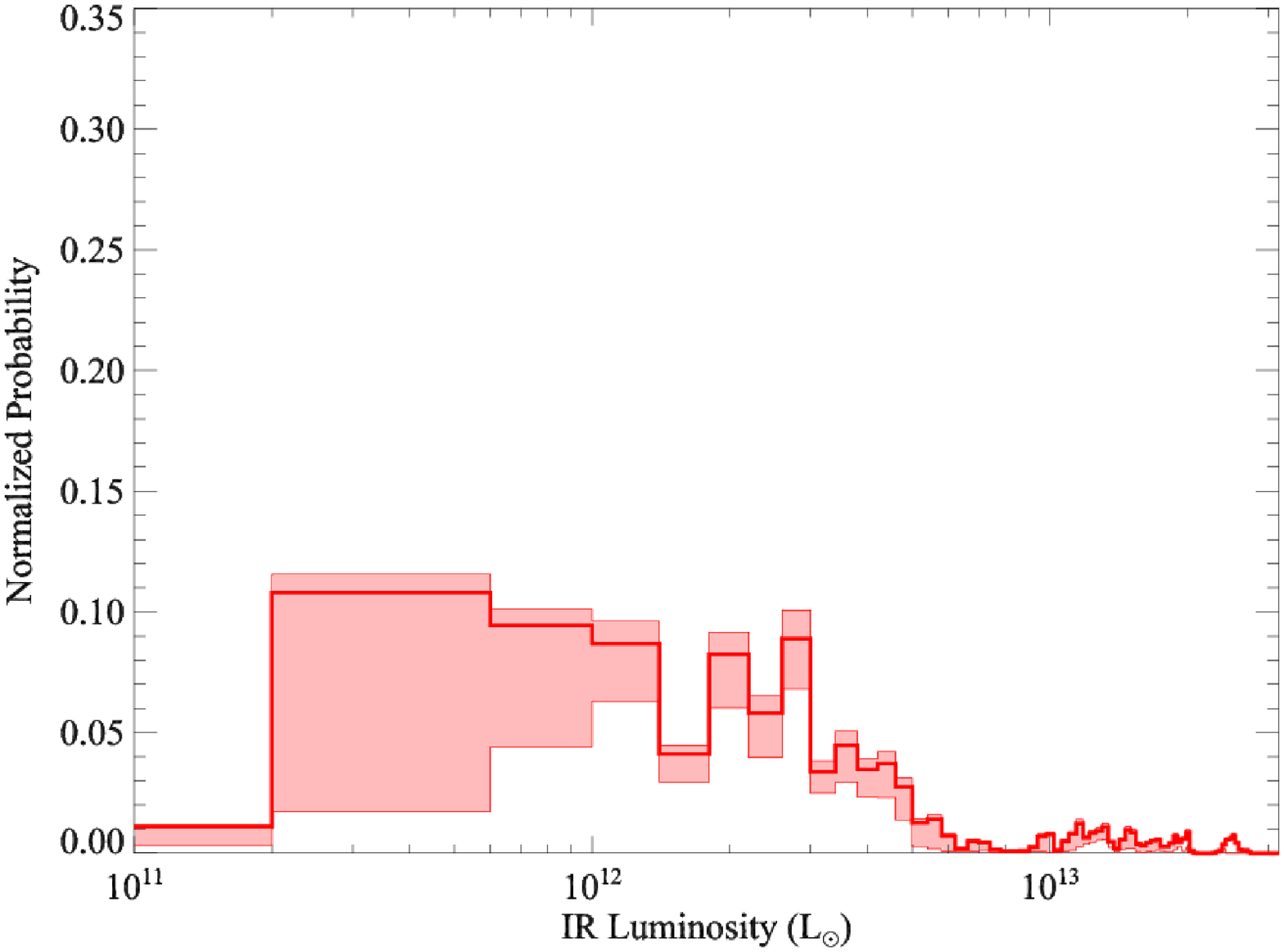}
\includegraphics[width=43mm,angle=90]{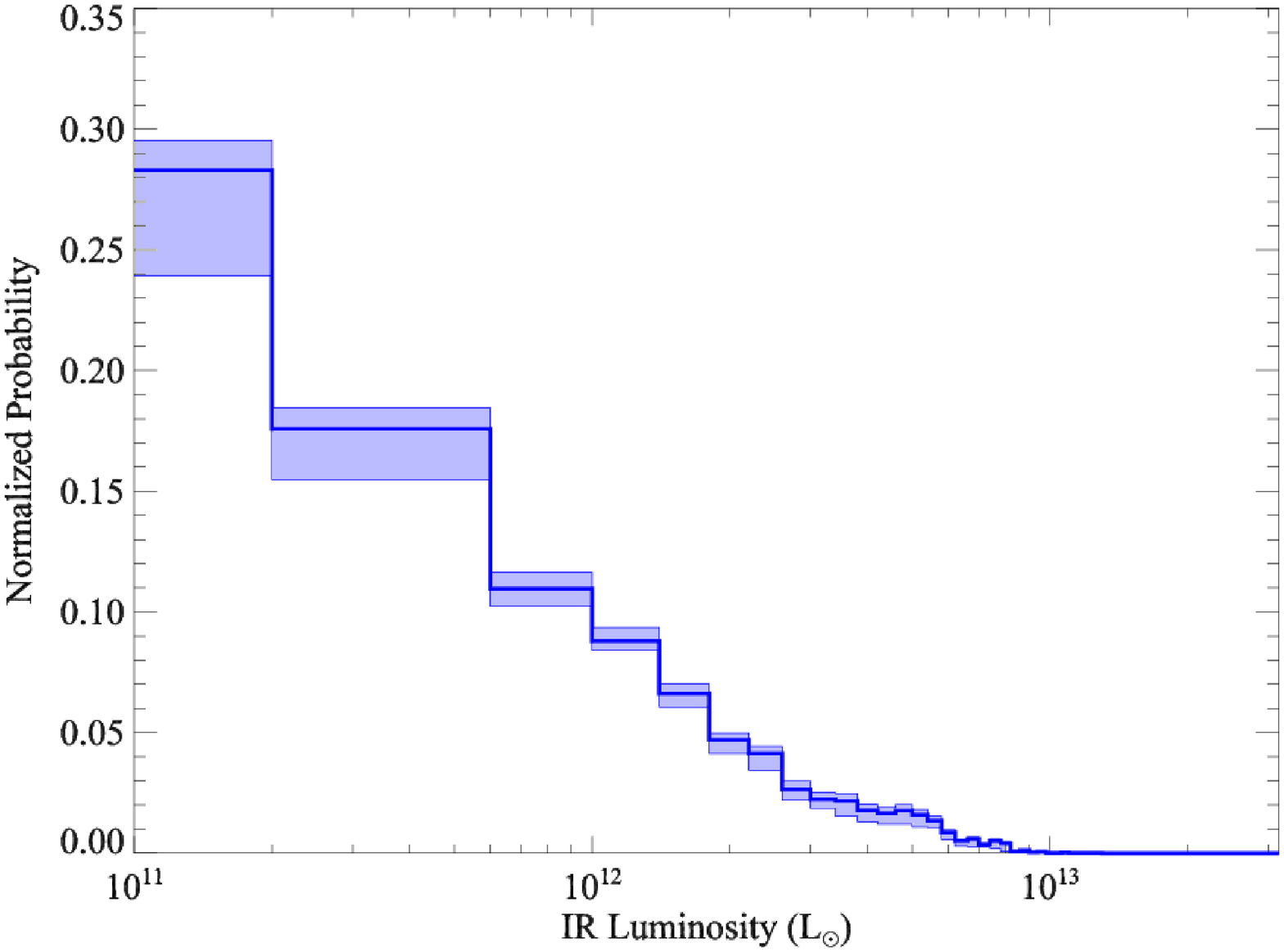}
\caption{Differential PDF for the whole sample, constructed from all fits to all objects, weighted by their individual $\chi^{2}_{Red}$ values. The PDF for the total IR luminosity is plotted in green, for the AGN luminosity in red, and for the starburst luminosity in blue. The top panel shows all three PDFs overlaid, while the bottom row shows the individual PDFs and their errors. The errors were estimated via jack-knife resampling, removing one source at a time and deriving the resulting error estimates, but were performed for the bins individually. Since the bins are correlated, these errors are an overestimate of the overall uncertainties in the PDFs.}\label{figmergedpdf}
\end{figure*}

The star formation rates for the objects with starbursts detected at $\geq 90\%$ confidence, computed by extracting the monochromatic 60$\mu$m luminosities from the SED fits and then using equation 7 of \citet{mrr97}, lie between several hundred to a few thousand solar masses per year. For the six objects in common with \citet{farrah10}, these star formation rate estimates agree with those from the IRS spectra, though the large systematics on both measurements renders this comparison of little value. 

We now examine the IR emission from AGN and starbursts in FeLoBAL QSOs using a different approach. We use all the solutions from all the individual fits to construct a single PDF for the whole sample\footnote{This is more appropriate than taking the individual PDFs and multiplying them together to make a single PDF, since the quality of the best fits from object to object varies significantly} (Figure \ref{figmergedpdf}). From this, we reach similar conclusions to those described above. An AGN supplies the bulk of the IR emission in the majority of cases. Starburst activity on the other hand is significantly less luminous; we find that the starburst is fainter than $10^{12}$L$_{\odot}$ in just over half ($56.7^{+5.2}_{-5.3}\%$) of cases, and almost never exceeds $10^{13}$L$_{\odot}$ (Table \ref{probabilities1}, first row).

Overall, we find compelling evidence from the IR properties of FeLoBAL QSOs that they are, as a class, distinct from the general QSO population. FeLoBAL QSOs have higher IR luminosities and a smaller average AGN fraction. Instead, the IR properties of FeLoBAL QSOs are consistent with those of the reddened \& LoBAL QSO populations. 

The relationship between FeLoBAL QSOs and ULIRGs is harder to discern. It has been suggested (F07) that FeLoBAL QSOs are, as a class, a ULIRG-to-QSO transition phase. There is one result from our study that supports this, namely that the mean AGN fractional IR luminosity of FeLoBAL QSOs is similar to that of `warm' ULIRGs. Conversely, if FeLoBAL QSOs were such a transition phase then we may have expected a higher fraction of them to host luminous starbursts, up to $\sim$50$\%$ if they represented the entirety of the starburst to AGN transition (see e.g. \citealt{farrah09}). Now, this fraction is an upper limit to what we might see, and such a high fraction with luminous starbursts is (just) consistent with the PDF in Figure \ref{figmergedpdf}. Other evidence for this comes from the unusual mid-IR spectral shapes of FeLoBAL QSOs in comparison to other IR-luminous QSOs \cite{farrah10}. From the results in this paper however, the scenarios that FeLoBAL QSOs are (a) randomly drawn from the reddened QSO population, and (b) a ULIRG to QSO transition phase, are both plausible. Since the former scenario is simpler, we conclude that, while some fraction of reddened/IR-luminous QSOs are almost certainly the endpoint of a ULIRG-QSO transition, we see no evidence in this paper that demands that FeLoBAL QSOs are {\itshape more} likely to be such a transition stage than the (presumably) parent reddened QSO population. 

\begin{figure*}
\includegraphics[width=66mm,angle=90]{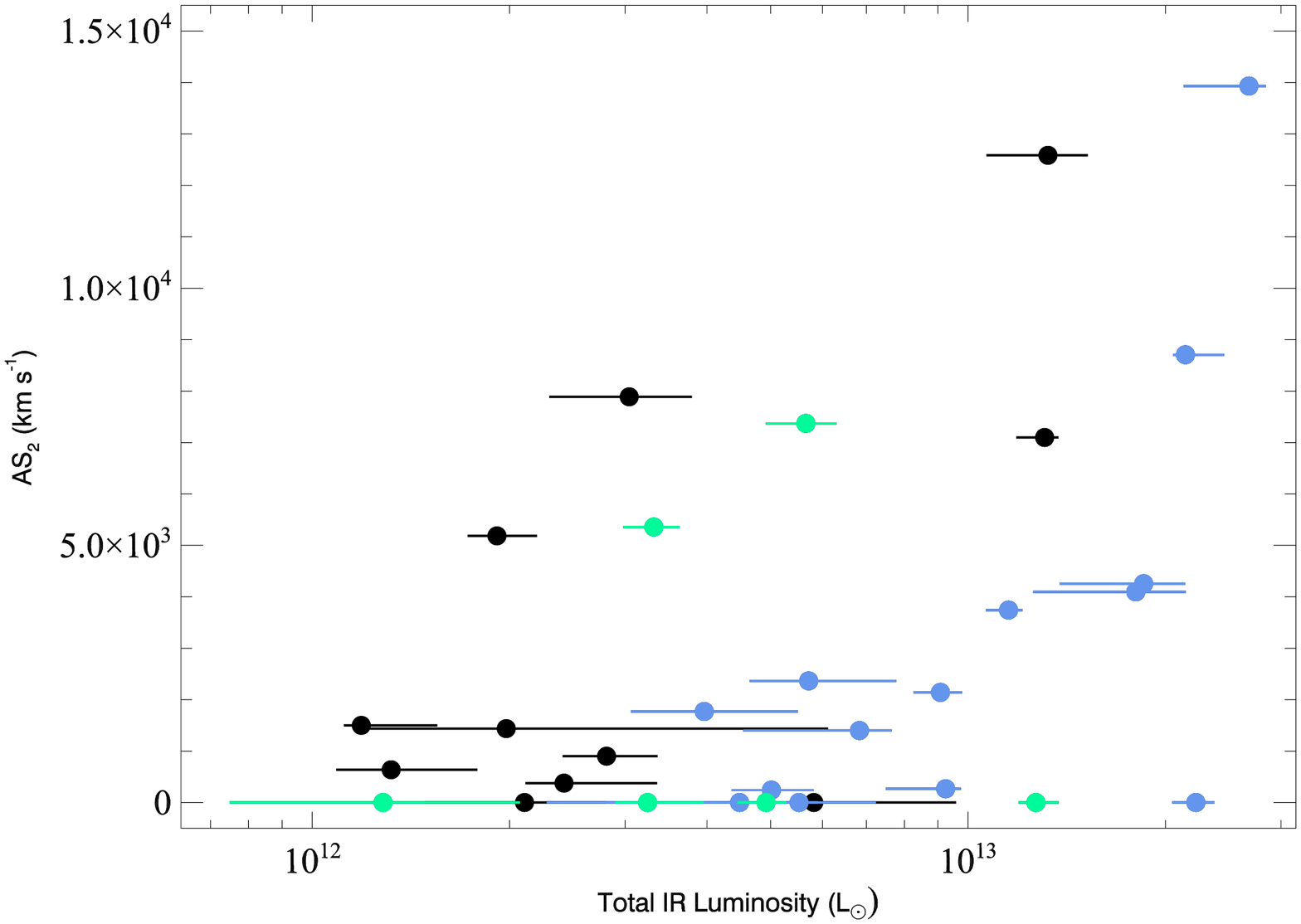}
\includegraphics[width=66mm,angle=90]{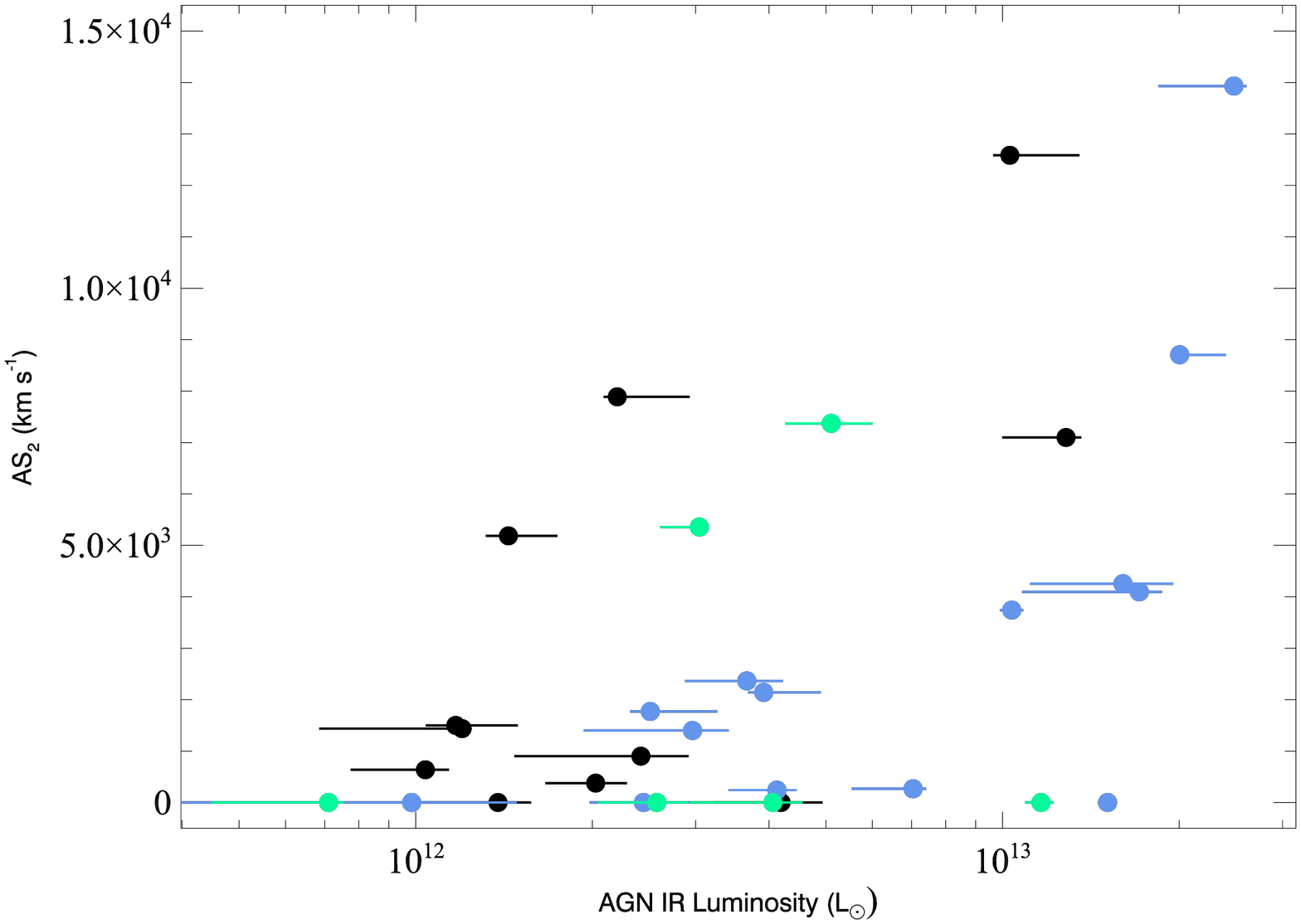} \\
\includegraphics[width=66mm,angle=90]{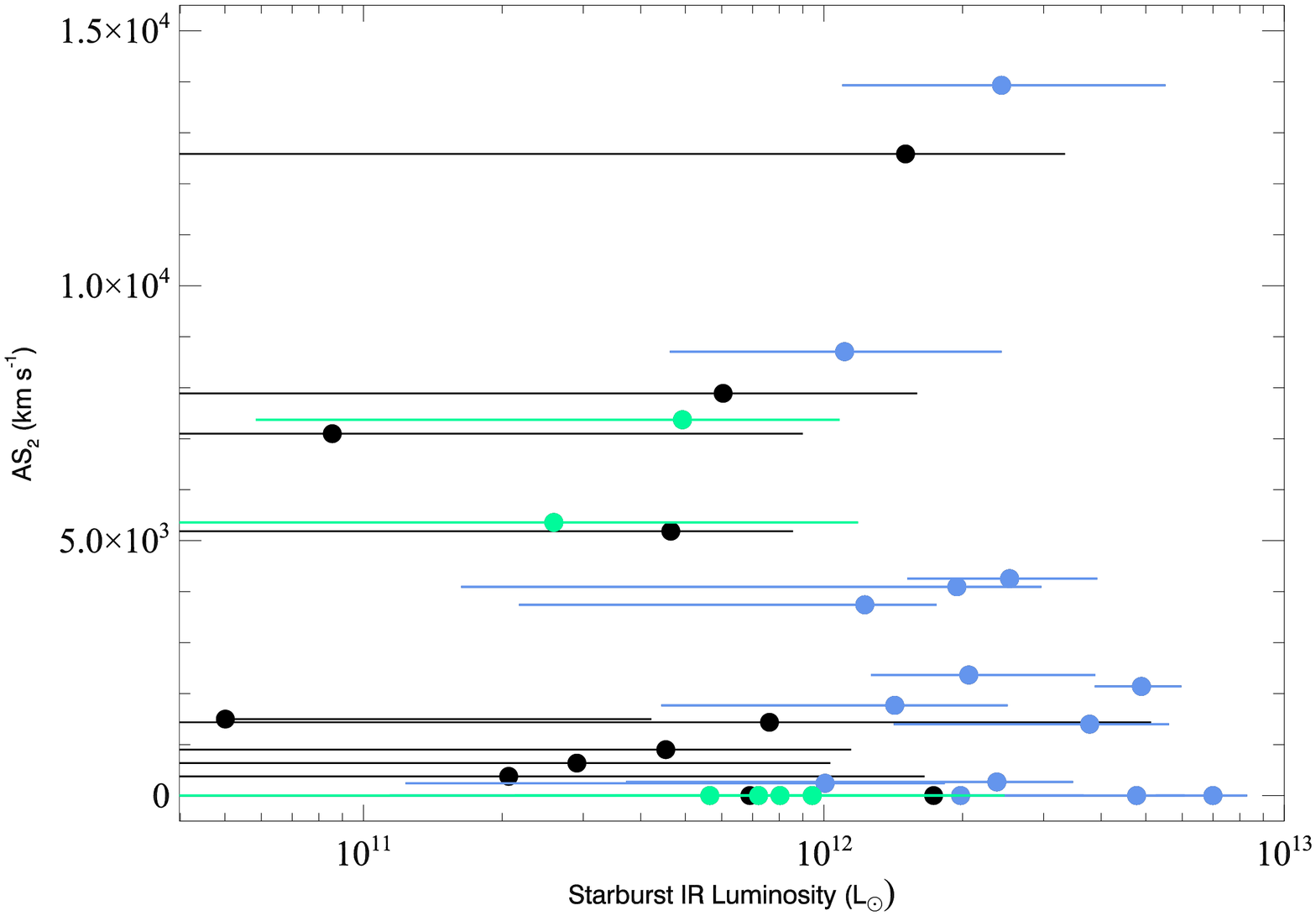}
\includegraphics[width=66mm,angle=90]{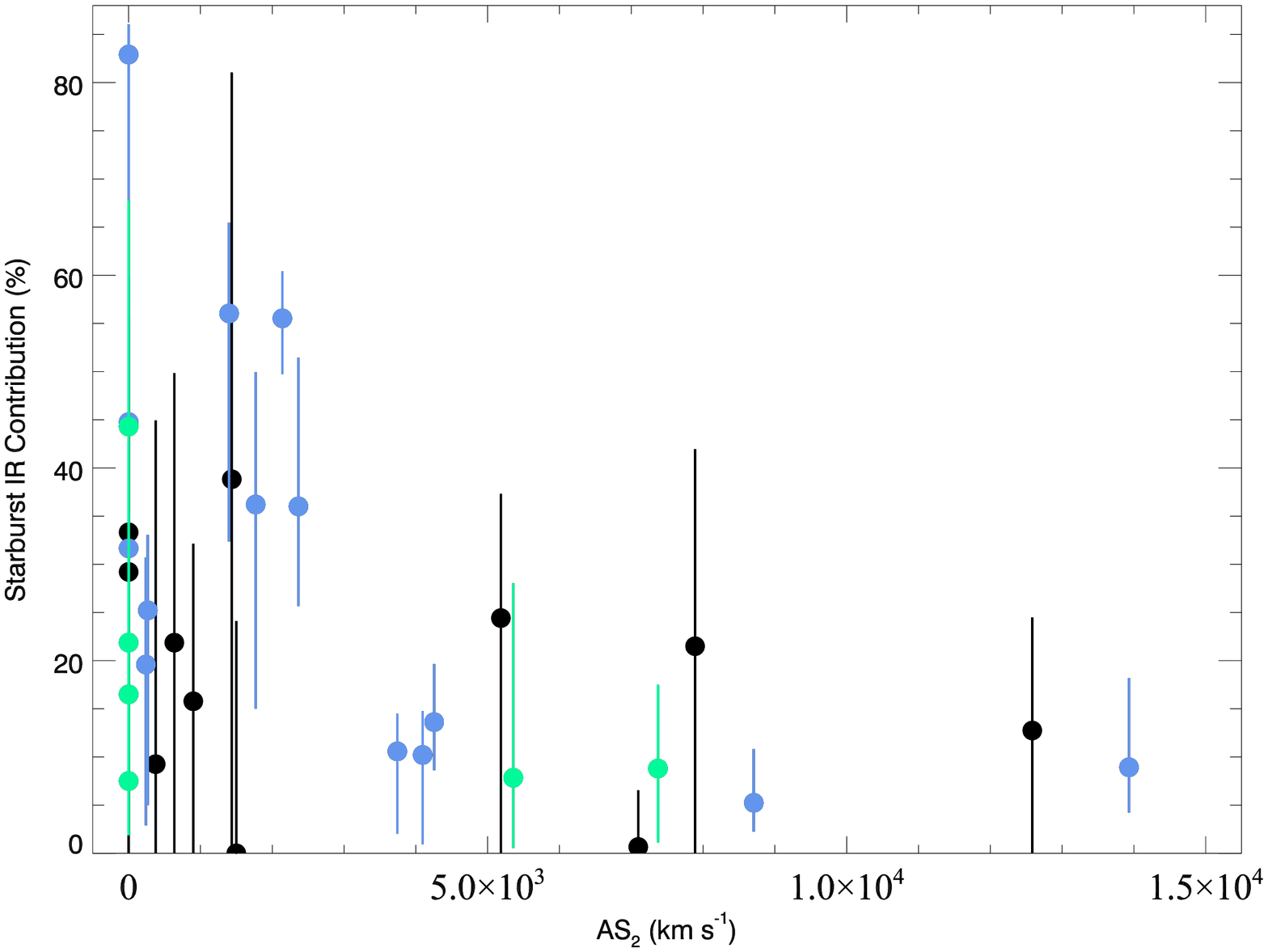}
\caption{Absorption strength vs (top left) total IR luminosity, (top right) AGN luminosity, (bottom left) starburst luminosity, and (bottom right) starburst contribution to the total IR luminosity. Objects with a starburst detected at $\geq90\%$ confidence are plotted in colour; in blue if they exceed $10^{12}$L$_{\odot}$, and in green if they are fainter than this luminosity. Objects where the 90\% confidence interval on the starburst luminosity encompasses zero are plotted in black.} \label{figabsvslir}
\end{figure*}

\subsection{Starbursts \& Outflows}\label{discusssboutflow}
We now examine whether or not there is a relationship between the AGN-driven outflows and the obscured star formation in our sample. In \S\ref{discstrengths} we discuss how to measure the strength of the BAL features. In \S\ref{discoutirs} we examine the relationship between AGN-driven outflows and obscured star formation using the best-fit luminosities and their confidence ranges to start with, and then examine it using all the information in the PDFs. We interpret our results in \S\ref{discinterp}. 

There is one point we note first. We cannot here measure how much kinetic energy the outflow is injecting into the ISM of its host galaxy. Even with well resolved BALs in multiple species, such a measurement involves in-depth radiative transfer calculations (e.g. \citealt{case08}). The SDSS optical spectra used here are however of low resolution, and usually contain BALs only for one or two species. So, even attempting such calculations for our sample is futile. It is however plausible that the depth, width, and velocity offset of BALs in a single species, as quantified in measures such as Balnicity Index and Absorption Index \citep{hazard87,hall02,trump06} do scale with increasing outflow strength. So, we here use the properties of the Mg II BAL features solely as an estimate of relative outflow strength within the sample.

\subsubsection{Measuring Outflow Strengths}\label{discstrengths}
The strength of absorption of the BALs in a given species is traditionally defined as the total velocity width over which the absorption exceeds a minimum value. This however gives rise to four problems with interpretation. First, different authors use different parametrizations for this strength, which in some cases can determine whether an object is identified as a BAL QSO (of any variety) or not. Second, the derived depths and widths of the BAL troughs are sensitive to the choice of continuum level, which can vary significantly between automated and manual measurements even for the same input data (e.g. compare the results in \citealt{trump06,gibson09,allen11}). Third, since our sample is at $z<1.8$ we see absorption only in a limited number of species. Fourth, the BALs in some QSOs show temporal variation on approximately decadal timescales \citep{arav01}.

\begin{deluxetable}{lcccc}
\tabletypesize{\scriptsize}
\tablecaption{Examples of different absorption strength measures from Equation \ref{eqn:bal}, ordered by decreasing strictness.\label{absmeasures}}
\tablewidth{0pt}
\tablehead{
\colhead{Name} & \colhead{$v_{0}$} & \colhead{$v_{1}$} & \colhead{a} & \colhead{Min. Velocity Width} 
}
\startdata
AS$_{0}$ (BI)\tablenotemark{a}       &  3000       & 25000       &  0.9   & 2000 \\
AS$_{1}$                             &  3000       & 29000       &  0.9   & 2000 \\
AS$_{2}$ (BI$_{0}$)\tablenotemark{b} &   0         & 25000       &  0.9   & 2000 \\
AS$_{3}$                             &   0         & 29000       &  0.9   & 1000 \\
AS$_{4}$ (AI)\tablenotemark{c}       &   0         & 29000       &  1.0   & 1000 
\enddata
\tablecomments{The names in parentheses are those of measures used in the literature. We do not use AS$_{1}$ or AS$_{3}$ here, but present them to illustrate other possible parametrizations.}
\tablenotetext{a}{Balnicity Index \citep{weymann91,gibson09,allen11}.}
\tablenotetext{b}{Modified Balnicity Index \citep{gibson09}.}
\tablenotetext{c}{Absorption Index \citep{hall02,trump06}.}
\end{deluxetable}

The latter two issues cannot be addressed with the data we have. To mitigate the first two as far as is possible we proceeded as follows. First, we used the following parametrization for the Absorption Strength (AS) of BALs:

\begin{equation}
AS = \int_{v_{0}}^{v_{1}}\left[1 - \frac{f(v)}{a}\right]Bdv
\label{eqn:bal}
\end{equation}

\noindent where $v$ is velocity, $f(v)$ is the (normalized) flux at that velocity, and $a$ is a scaling factor. The quantity $B$ is set to unity if the absorption is below 10\% of the continuum {\itshape and} if the width of the trough is greater than some minimum value, otherwise it is set to zero. We define the systemic redshift to be at zero velocity, and positive velocity to be blueward of this redshift. Equation \ref{eqn:bal} includes all previous definitions of BAL strength as special cases (Table \ref{absmeasures}). We chose the $AS_{2}$ parametrization as our primary measure since it is a reasonable compromise between strictness and inclusiveness. Second, we measure absorption strengths for the same, single species across the whole sample. We chose MgII$\lambda$2799\AA\ as it is the only species present in all the SDSS spectra of our sample, though we note that it can be contaminated by Fe II absorption (see Table 1 of \citealt{hall02}). We remeasured all the MgII absorption strengths by hand, using the methods described in \citet{f2ms}. Our measurements, together with one set of comparison measurements from \citet{gibson09}, are given in Table \ref{tableasandlirs}. We check our absorption measures against those in the literature, and explore the effects on our results of using different measurements of absorption strength, in the appendix. The formal error on the absorption strengths from the fits is usually of order 100 km s$^{-1}$, but there are significantly larger systematic unertainties (see appendix) that are hard to quantify. These systematic uncertainties should however not dramatically change the absorption strengths of the sample relative to each other, as long as the measurements are done in an internally consistent way.

\begin{deluxetable*}{lccccc}
\tabletypesize{\scriptsize}
\tablecaption{Probabilities of obtaining Starburst \& AGN luminosities below certain boundaries. \label{probabilities1}}
\tablewidth{0pt}
\tablehead{
{\bf Selection}                &\multicolumn{5}{c}{{\bf Probability of Obtaining}}  \\
	                           &\multicolumn{3}{c}{L$_{Sb}$(L$_{\odot}$)}                                 & \multicolumn{2}{c}{L$_{AGN}$(L$_{\odot}$)}           \\
                               &  $<10^{11}$          & $<10^{12}$           & $<10^{12.5}$               & $<10^{12}$           & $<10^{12.5}$                        
}
\startdata
All Objects                    & $28.3^{+2.8}_{-4.8}$\% & $56.7^{+5.2}_{-5.3}$\%  & $83.7^{+4.1}_{-3.2}$\% & $21.4^{+3.8}_{-10.5}$\% & $57.0^{+7.3}_{-7.9}$\%   \\
\hline
AS$_{2}<3500$ km s$^{-1}$      & $27.7^{+3.2}_{-7.0}$\% & $55.2^{+5.6}_{-6.9}$\%  & $80.0^{+5.7}_{-5.2}$\% & $32.1^{+6.2}_{-14.1}$\% & $68.7^{+8.8}_{-8.3}$\%    \\
AS$_{2}>3500$ km s$^{-1}$      & $29.6^{+6.2}_{-5.5}$\% & $59.7^{+11.0}_{-9.0}$\% & $91.0^{+4.2}_{-2.8}$\% & $0$\%                   & $33.8^{+10.8}_{-13.6}$\%  \\
\hline
L$_{AGN}<10^{12.5}$L$_{\odot}$ & $38.8^{+3.8}_{-7.1}$\% & $69.4^{+5.2}_{-4.6}$\%  & $93.7^{+2.5}_{-2.0}$\% & $37.3^{+7.4}_{-15.7}$\% & $90.9^{+5.6}_{-3.0}$\%   \\
L$_{AGN}>10^{12.5}$L$_{\odot}$ & $16.2^{+2.5}_{-4.9}$\% & $43.2^{+6.9}_{-10.4}$\% & $69.8^{+8.2}_{-7.2}$\% & $0$\%                   & $10.4^{+2.0}_{-6.2}$\%   
\enddata
\tablecomments{We give probabilities for the complete sample, the sample divided by absorption strength, and the sample divided by AGN luminosity (see also Table \ref{probabilitiesgib}). Errors were derived using jack-knife resampling, removing one source at a time and computing the $\sim1\sigma$ confidence interval from all the resulting realizations. The subsamples divided by L$_{AGN}$ were divided on their peak luminosities, hence the non-zero probabiilties of obtaining luminosities outside the boundaries.}
\end{deluxetable*}

\subsubsection{Outflows vs. Infrared Properties}\label{discoutirs}
We first use the luminosities in Table \ref{tableasandlirs} to investigate if absorption strength depends on total, AGN and starburst IR luminosity. We find no correlation between absorption strength and L$_{Tot}$ ($\rho = 0.31$, $P = 0.09$, Figure \ref{figabsvslir} top left), or between absorption strength and L$_{SB}$ ($\rho = -0.10$, $P = 0.58$, Figure \ref{figabsvslir} bottom left), though we do find a hint of a correlation between absorption strength and L$_{AGN}$ ($\rho = 0.39$, $P = 0.04$, Figure \ref{figabsvslir} top right). It is also interesting that (a) the dispersion in absorption strength is greater at higher total and AGN luminosities, and (b) all but two of the luminous ($>10^{12}$L$_{\odot}$) starbursts lie at AS$_{2}<5000$ km s$^{-1}$. We conclude that the mechanism that determines the strength of the outflows is not directly responsible for heating the dust near the AGN, and does not have a strong effect on the {\itshape absolute} luminosity of the starburst. 

If however we plot absorption strength against $f_{SB}$ then we see a weak but clear anticorrelation ($\rho = -0.49$, $P = 0.005$, Figure \ref{figabsvslir} bottom right). Moreover, there seems to be a change in the distribution of starburst contributions with absorption strength at AS$_{2}\simeq 3500$km s$^{-1}$; all the systems with AS$_{2}>3500$km s$^{-1}$ have a starburst contribution of less than $25\%$, while the systems with AS$_{2}<3500$km s$^{-1}$ have a wide dispersion in starburst contributions, from $0\%$ to $\sim80\%$. 

To estimate the probability that the distribution of starburst contributions changes at AS$_{2} = 3500$km s$^{-1}$ we employ the two-sided Kolmogorov-Smirnov test. We find the distributions of the objects above and below AS$_{2} = 3500$km s$^{-1}$ in the bottom right panel of Figure \ref{figabsvslir} are different at 99.84\% confidence\footnote{Employing the conceptually similar Mann-Whitney test gives a comparable result}. The number of objects in the two subsamples is however low enough that the Kolmogorov-Smirnov test can be unreliable. So, we employ a cruder test. We take the null hypothesis to be that there is no correlation between Absorption Strength and $f_{SB}$, and that the systems with $f_{SB}<25\%$ represent the `true' underlying distribution. The probability of finding all eight systems with $f_{SB}>25\%$ at AS$_{2}<3500$ km s$^{-1}$ is then $(8/19)^{8})\simeq1\%$, i.e. a $\simeq$99.0$\%$ probability of a difference. We regard this latter figure as a more reliable measure of the significance of a difference. Hence our finding all of the $f_{SB}>25\%$ systems at AS$_{2}<3500$ km s$^{-1}$ is only weak evidence that an anticorrelation exists.

\begin{figure*}
\includegraphics[width=130mm,angle=90]{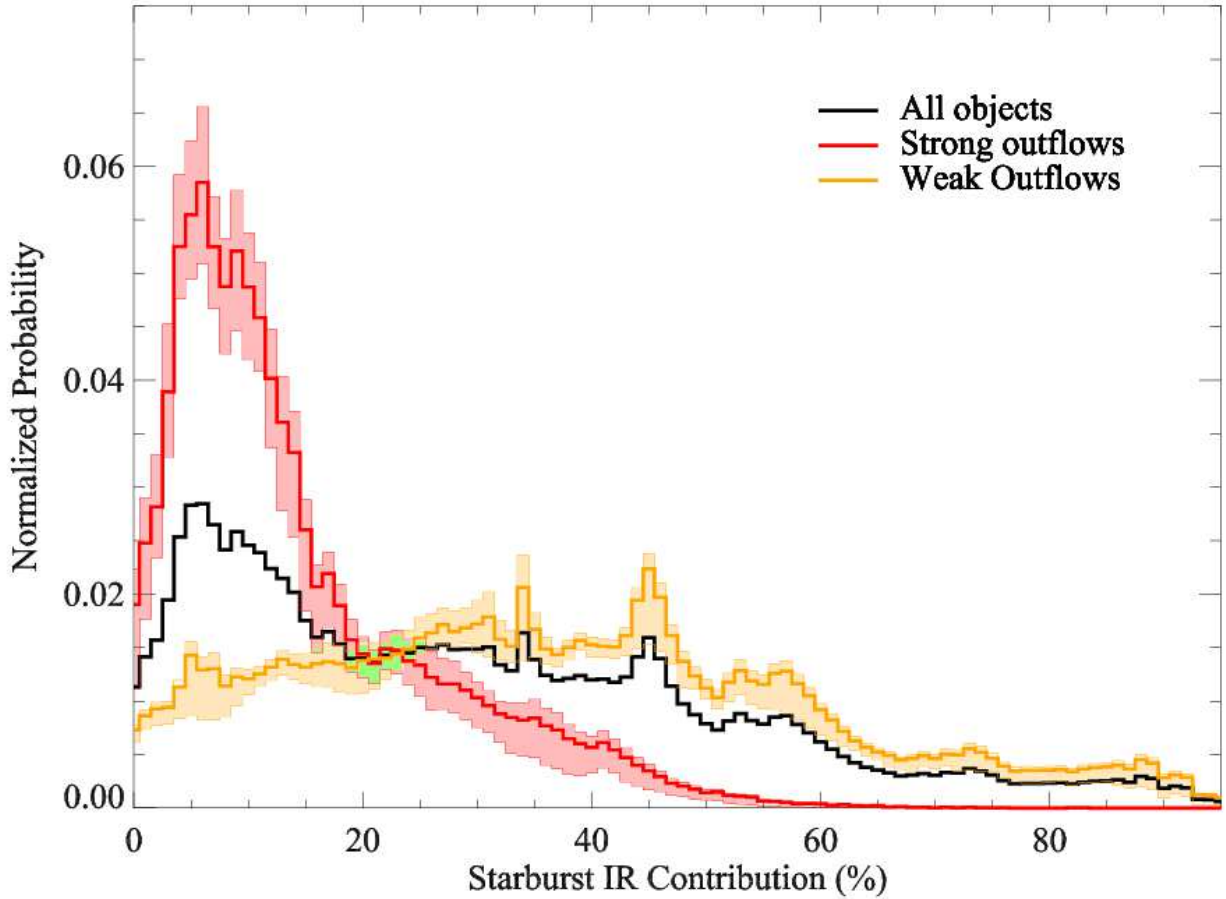}
\caption{Probability Distribution Function for starburst contribution divided according to their absorption strengths. The black line is the PDF for the whole sample, in red for the objects with AS$_{2}<3500$ km s$^{-1}$, and in orange for the objects with AS$_{2}>3500$ km s$^{-1}$. The orange line shows a much higher chance of obtaining a starburst fractional luminosity in excess of 25$\%$ than the red or black lines.}\label{figmergedpdf_agnfrac}
\end{figure*}

\begin{deluxetable}{lcc}
\tabletypesize{\scriptsize}
\tablecaption{Probabilities of obtaining Starburst contributions to the total IR luminosity, above two boundaries. \label{probabilities2}}
\tablewidth{0pt}
\tablehead{
{\bf Selection}                  &\multicolumn{2}{c}{P($f_{SB}$)} \\
                                 & $>25\%$               & $>50\%$                   
}
\startdata
All Objects                      & $50.3^{+5.3}_{-5.4}$\% & $17.9^{+2.5}_{-4.2}$\%   \\
\hline
AS$_{2}<3500$ km s$^{-1}$        & $67.3^{+4.5}_{-4.1}$\% & $26.6^{+3.2}_{-4.8}$\%   \\
AS$_{2}>3500$ km s$^{-1}$        & $17.8^{+3.7}_{-6.5}$\% & $<1.5$\%    \\
\hline
L$_{AGN}<10^{12.5}$L$_{\odot}$   & $57.8^{+6.1}_{-6.4}$\% & $18.6^{+3.1}_{-5.0}$\%   \\
L$_{AGN}>10^{12.5}$L$_{\odot}$   & $38.7^{+8.5}_{-10.0}$\% & $15.1^{+4.0}_{-7.0}$\%   
\enddata
\tablecomments{As with Table \ref{probabilities1}, we give probabilities for the complete sample, the sample divided by absorption strength, and our sample divided by AGN luminosity (see also Table \ref{probabilitiesgib}). Errors were derived using jack-knife resampling.}
\end{deluxetable}

Nevertheless, this result is consistent with the idea that the radiatively driven outflows negatively affect star formation. The systems with $f_{SB}<25\%$ would then be those in which an outflow has curtailed star formation, and those in which such an outflow has subsequently waned, making the observed dispersion in absorption strengths wide. The systems with $f_{SB}>25\%$ would be those in which an outflow has started to develop, but has not yet affected the starburst. 

To explore the relationship between absorption strength and infrared properties further, we use all of the information in the PDFs, in a manner similar to that used in \S\ref{disclums}. We adopt a boundary value of absorption strength motivated by Figure \ref{figabsvslir} of AS$_{2} = 3500$km s$^{-1}$, and divide our sample into two subsamples at this boundary. 

We first construct PDFs of L$_{AGN}$ and L$_{SB}$ for these two subsamples, and extract from them the probabilities of obtaining luminosities in excess of certain values (rows 2 \& 3 of Table \ref{probabilities1}). We see similar results to those seen in the first three panels of Figure \ref{figabsvslir}. We see no convincing differences in the starburst luminosities between the low and high absorption strength subsamples, compared to either each other or to the sample as a whole. We also see that we are only marginally more likely to see AGN with L$_{IR}<10^{12.5}$L$_{\odot}$ in the low absorption strength subsample.

If however we consider the PDFs for the contribution of the starburst to the total IR luminosity for the two subsamples (Figure \ref{figmergedpdf_agnfrac}), then we see a clear difference. The low absorption strength subsample shows a higher chance of a higher $f_{SB}$ than the sample as a whole, and a much higher chance than the high absorption strength subsample. We quantify this by extracting the probabilities of obtaining starburst contributions to the total IR luminosity in excess of 25\% and 50\% from the whole sample, and the two absorption strength subsamples (Table \ref{probabilities2}, rows 1-3). We find, at $>5\sigma$ significance, a higher chance of obtaining $f_{SB}>25\%$ and $f_{SB}>50\%$ in the low absorption strength sample compared to the high absorption strength sample. These results do not change if we exclude the objects noted as potential contaminants in \S\ref{secselect}, or if we exclude the objects with  $\chi^{2}_{Red}>2$ in Table \ref{tableasandlirs}.

\subsubsection{Interpretation}\label{discinterp}
The anticorrelation that we observe between absorption strength and contribution from star formation to the total IR emission is straightforwardly interpreted as the outflow from the AGN curtailing star formation in the host galaxy. There are however four other ways that we could see this anticorrelation. 

The first is that stronger outflows reflect an increase in the IR emission from the AGN, but have no effect on the starburst; if this is the case we would see a decline in $f_{SB}$, but without there being a direct relationship behind the two phenomena. This possibility is apposite if there is a conspiracy of fortuitous timescales, where the peak starburst luminosity precedes the peak AGN luminosity by a few Myr. The second is that starburst activity suppresses AGN outflows, so when the starburst wanes (via a cause unrelated to the AGN) an AGN driven outflow can appear. The third is an observation bias; e.g. if a high $f_{SB}$ meant that the Mg II troughs were {\it observed} to be weaker than they really are. The fourth is a selection bias, e.g. if QSOs with strong BALs {\itshape and} strong starbursts drop out of the initial SDSS QSO selection and were thus not included in \citet{trump06}.

\begin{figure*}
\includegraphics[width=130mm,angle=90]{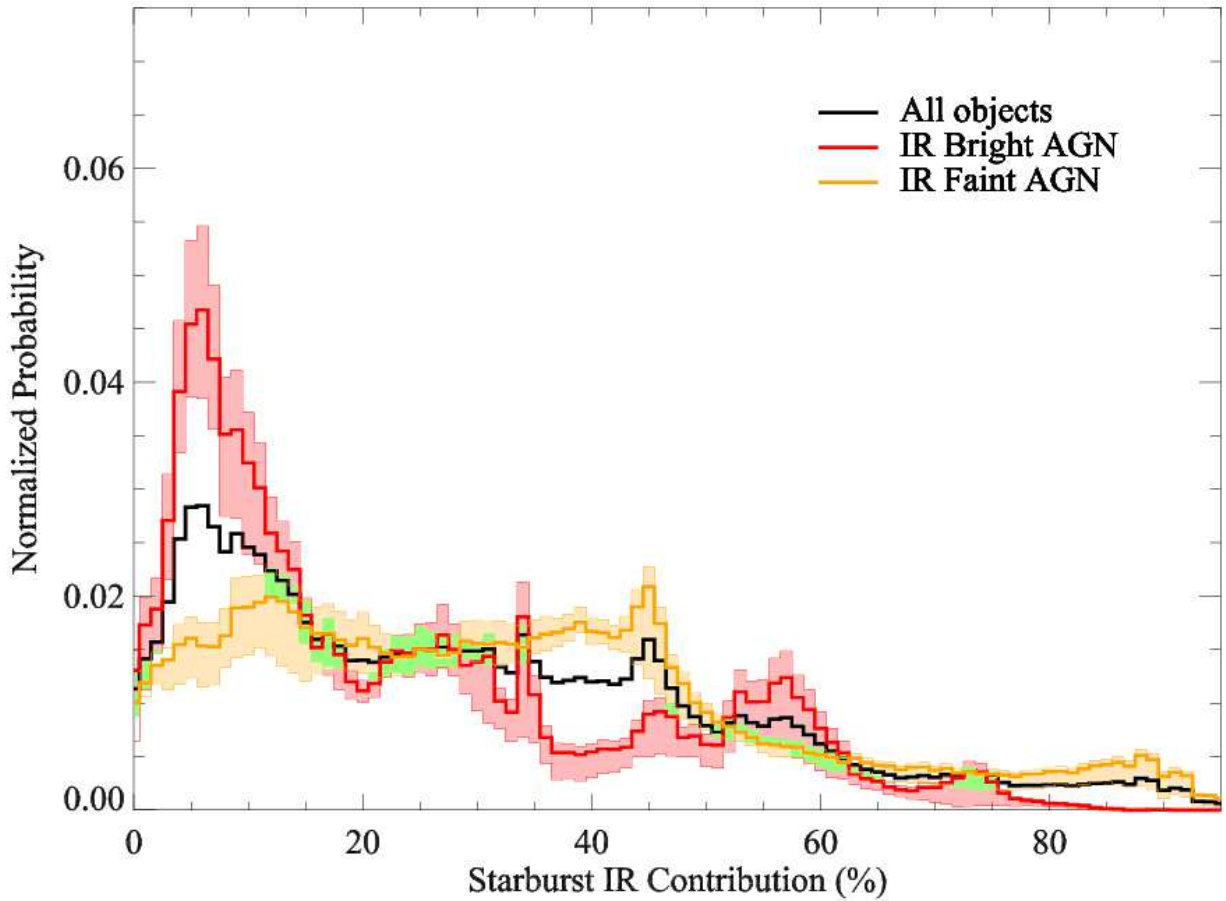}
\caption{Probability Distribution Function for starburst contribution divided according to their AGN luminosities; in black for the whole sample, in red for the objects with L$_{AGN}>10^{12.5}$L$_{\odot}$, and in orange for the objects with L$_{AGN}<10^{12.5}$L$_{\odot}$. The difference between the two PDFs is smaller than in figure \ref{figmergedpdf_agnfrac}, suggesting that the IR luminosity of the AGN is not a good proxy for outflow strength, see \S\ref{discinterp}.}\label{figmergedpdf_agnfracbylagn}
\end{figure*}

We first consider the last three of these alternatives. The second alternative has traditionally been considered unlikely, the argument being that for it to happen then the starburst would have to act near the origin of the outflow, but AGN broad-line regions are around five orders of magnitude smaller than starburst regions. Nevertheless, it is not completely implausible. One possible scenario is as follows. If the ISM was initially dense and the SMBH was initially small, then the SMBH may not be at first capable of powering outflows that extend $\sim$Kpc into the host galaxy, but as the density of the ISM was reduced by the ongoing starburst (which thus waned due to the reduction in fuel supply) and the SMBH grows, then large-scale outflows would subsequently appear. This is not a scenario we can test, but it would likely require a serendipitous conjunction of ISM and SMBH parameters, so we do not consider it further. 

The third possibility is one that we again cannot test, so we cannot formally discount the possibility of a very large population of OB stars acting to suppress the observed depth of the UV absorption troughs. Conversely, the rest-frame UV spectra of starbursts in ULIRGs reveal continua that are at least an order of magnitude too weak to provide this effect, and sometimes show absorption in the same species \citep{farrah05}. The fourth possibility is also not testable, but the SDSS is now turning up FeLoBAL QSOs in large numbers, and the initial QSO followup colour selections are fairly relaxed, so we do not consider this possibility likely either. 

The first alternative is however one that we can test, from which we propose that it is unlikely as well. The test is as follows. If it is the case that outflow strength is a proxy for AGN luminosity, then we should see a bigger difference between the starburst contribution PDFs for subsamples divided by AGN luminosity than between subsamples divided by absorption strength\footnote{Since we would be dividing on the primary driver behind the difference in starburst contribution; the timescale for an outflow is thought to be around an order of magnitude shorter than the timescale for an AGN, so the outflows would be a pseudorandom outcome of a luminous AGN.}. In Figure \ref{figmergedpdf_agnfracbylagn} we show the starburst contribution PDFs for two subsamples divided by AGN luminosity, one for objects with L$_{AGN}>10^{12.5}$L$_{\odot}$ and one for objects with L$_{AGN}<10^{12.5}$L$_{\odot}$. Qualitatively, the difference between the PDFs in Figure \ref{figmergedpdf_agnfracbylagn} is {\it weaker} than the difference between the PDFs divided by absorption strength in Figure \ref{figmergedpdf_agnfrac}. If we extract probabilities of obtaining the same starburst \& AGN luminosities, and starburst contributions as we did for the absorption strength subsets (4th \& 5th rows of Tables \ref{probabilities1} \& \ref{probabilities2}) then we see three interesting results. First, lower luminosity star formation is now more likely (at $\gtrsim3\sigma$) to be seen in the {\itshape lower} luminosity AGN subsample (e.g. for P(L$_{Sb}<10^{12}$L$_{\odot})$: the L$_{AGN}<10^{12.5}$L$_{\odot}$ subsample is $69.4^{+5.2}_{-4.6}$\% while the L$_{AGN}>10^{12.5}$L$_{\odot}$ subsample is $43.2^{+6.9}_{-10.4}$\%). Second, we are more likely (albeit only at just over 2$\sigma$) to obtain a smaller starburst contribution by selecting high absorption strength systems than we are by selecting high AGN luminosity systems (e.g. for P($f_{SB}>25\%$): the AS$_{2}>3500$km s$^{-1}$ subsample is $17.8^{+3.7}_{-6.5}$\% while the L$_{AGN}>10^{12.5}$L$_{\odot}$ subsample is $38.7^{+8.5}_{-10.0}$\%). In other words, we are more successful in finding systems with a large starburst contribution to the total IR emission by selecting on weak outflows than we are by selecting on low AGN luminosity. Third, the probabilities of a starburst contribution in excess of 25\% (or 50$\%$) are statistically indistinguishable between the high and low AGN luminosity subsamples, but are different at $>5\sigma$ between the high and low absorption strength subsamples (Table \ref{probabilities2}). 

Overall therefore, we find that radiatively driven outflows from an AGN with absorption strengths $\gtrsim3500$ km s$^{-1}$ act to curtail star formation in their host galaxies. We also find that this effect is (at least largely) relative; such outflows reduce the contribution from star formation to the total IR luminosity to less than $\sim25\%$. We also propose that the {\itshape infrared} luminosity of the AGN is not a good proxy for the degree of AGN feedback that is taking place. Finally, since the IR properties of our sample are consistent with being drawn randomly from the reddened QSO population, we conclude that this is true generally for reddened QSOs. 

These results are consistent with the idea that starburst and AGN activity crudely correlate with each other in active galaxies - a more luminous starburst means we are more likely to see a luminous AGN\footnote{this is consistent with the results in Table \ref{probabilities1} \& previous work on ULIRGs, see e.g. the luminosities in \citet{farrah03}} - but where we add that a radiatively driven outflow can curtail the relative luminosity of the starburst on much shorter timescales than the typical lifetimes of a starburst or AGN. 

We conclude on a cautionary note. Our results appear solid, but they are based on a small sample, which makes the errors difficult to estimate. We have used what we believe to be a robust error estimation method (jack-knife resampling, Tables \ref{probabilities1} \& \ref{probabilities2}), but with only 31 objects we cannot reliably measure the error function, since comprehensive resampling methods are not possible. Moreover, we cannot test the robustness of the adopted boundary for dividing the PDFs of AS$_{2}=3500$ km s$^{-1}$. This boundary was motivated by the analysis in \S\ref{discoutirs}, but we would ideally like to explore the consequences of varying this boundary by up to a few thousand km s$^{-1}$ in either direction. We did perform a basic test of this in the appendix, by adopting instead a boundary of AS$_{2}=5000$ km s$^{-1}$, but we do not consider this to be a robust assessment of how sensitive our results are to the choice of boundary. Given the distribution of AS$_{2}$ values of our sample however, we cannot perform more comprehensive tests. A similar argument applies to the choice of luminosity cut for dividing the AGN PDFs in Figure \ref{figmergedpdf_agnfracbylagn}, and to the choice of template library used to model the luminosities (since several alternatives are available, see \S\ref{subsmodels}). To explore these issues properly would require a sample at least a factor of two larger than that used here. Aside from sample size, we would also like photometry longward of 200$\mu$m so as to measure the emission from cold dust heated mainly by star formation, and higher quality optical spectra to resolve MgII kinematics and look at the absorption strengths of at least one other species.

\section{Conclusions}
We have presented a sample of 31 FeLoBAL QSOs from the SDSS at $0.8<z<1.8$. These QSOs have broad, deep absorption troughs in their rest-frame UV spectra that are unambiguous signatures of radiatively driven outflows powered by an AGN. 
Previous work has suggested that FeLoBAL QSOs are IR-luminous, and sometimes harbor star formation rates of up to a few thousand solar masses per year. Furthermore, there is evidence that the AGN-driven outflows can extend up to several kiloparsecs into the host galaxies. FeLoBAL QSOs are thus ideal laboratories for studying the effects of radiatively-driven outflows from an AGN on obscured star formation. 

We selected our sample purely on the basis of their rest-frame UV spectral properties, and assembled for them optical through far-IR photometry from the SDSS, 2MASS, UKIDSS, WISE and Spitzer. We then fit these data with radiative transfer models for the IR emission from AGN and starbursts to extract best-fit AGN and starburst IR luminosities. We then compared these luminosities to the strength of their outflows as inferred from the absorption properties of the MgII$\lambda$2799\AA\ line in their SDSS spectra. Our conclusions are:

1 - FeLoBAL QSOs are luminous in the IR. All of our sample have total IR luminosities in excess of $10^{12}$L$_{\odot}$. Nearly one-third of the sample exceed $10^{13}$L$_{\odot}$. The dominant power source behind the IR emission is, in most cases, AGN activity. A pure AGN is either the most likely power source, or consistent within the 90\% confidence interval, for eleven of the sample. A starburst component is required for the remaining objects, but in only twelve of these objects is the starburst more luminous than $10^{12}$L$_{\odot}$, and in only three objects is the starburst more luminous than the AGN. The mean AGN contribution to the total IR luminosity is $\sim76\%$. The spread in AGN contribution to the total IR luminosity is however wide, spanning 0.2 to 1.0. Overall, the IR properties of FeLoBAL QSOs appear consistent with those of the general red/dusty QSO population. We do not however find convincing evidence that FeLoBAL QSOs are more likely to be a ULIRG to QSO transition phase than the general red QSO population. 

2 - We find no convincing relationship between the strength of the outflows and the total IR luminosity, or between the strength of the outflows and either the starburst or AGN luminosities. Conversely, we find a clear relationship between the strength of the outflows and the contribution from star formation to the total IR luminosity. If we divide our sample in two at an outflow strength boundary of AS$_{2}=3500$km s$^{-1}$ and construct probability distribution functions for the starburst contribution for the two subsamples, then the low absorption strength subsample shows a higher chance of a higher starburst contribution than the sample as a whole, and a much higher chance of a higher starburst contribution than the high absorption strength subsample (Figure \ref{figmergedpdf_agnfrac}). We quantify this by extracting the probabilities of obtaining starburst contributions to the total IR luminosity in excess of 25\% (Table \ref{probabilities2}, rows 1-3). We find, at $>5\sigma$ significance, a higher chance of obtaining a starburst contribution in excess of $25\%$ in the low absorption strength sample compared to the high absorption strength sample. 

3 - This anticorrelation between outflow strength and the contribution from star formation to the total IR luminosity is straightforwardly interpreted as the outflow from the AGN curtailing star formation in the host galaxy. There are however several other ways that it could arise. The most obvious alternative is that stronger outflows reflect an increase in the IR emission from the AGN, but have no effect on the starburst; if this is the case then the starburst contribution would decline with stronger outflows, but without there being a direct relationship behind the decline. To test this possibility we divided our sample into two at an AGN IR luminosity boundary of L$_{AGN}<10^{12.5}$L$_{\odot}$ and constructed probability distribution functions for the starburst contribution for these two subsamples (Figure \ref{figmergedpdf_agnfracbylagn}). If it is the case that outflow strength is a proxy for AGN luminosity, then we should see a bigger difference between these two PDFs than that seen between the PDFs in Figure \ref{figmergedpdf_agnfrac}. Instead, the PDFs show a smaller difference. Furthermore, we find that (a) we are (marginally) more successful in finding systems with a large starburst contribution to the total IR emission by selecting on weak outflows than we are by selecting on low AGN luminosity, and (b) the probabilities of a starburst contribution in excess of 25\% are statistically indistinguishable between the high and low AGN luminosity subsamples, but are different at $>5\sigma$ between the high and low absorption strength subsamples (Table \ref{probabilities2}). We considered several further alternative possibilities (\S\ref{discinterp}) but did not find any of them convincing. 

4 - We therefore conclude that strong, radiatively driven outflows in FeLoBAL QSOs can have a dramatic, negative effect on obscured star formation in their host galaxies. This is the most direct evidence yet obtained for 'quasar mode' feedback in the QSO population at high redshifts. We find that outflows with an absorption strength in MgII$\lambda$2799\AA\ of greater than $AS_{2}=3500$km s$^{-1}$ (Table \ref{absmeasures}) act to curtail luminous star formation in their host galaxies. We also find that this effect is at least largely relative; the starburst luminosity is reduced to less than about 25\% of the total luminosity. Finally, we propose that the magnitude of this effect is not deducible from the IR luminosity of the AGN.

\section*{Acknowledgments}
We thank the referee for a very helpful report, Sandra Blevins for help with reducing the Spitzer data, and Karen Leighly, Amanda Truitt and Adrian Lucy for helpful comments. This work is based on observations made with the Spitzer Space Telescope, the Wide-field Infrared Survey Explorer (WISE), the Two Micron All-Sky Survey (2MASS) and the Sloan Digital Sky Survey (SDSS). Spitzer is operated by the Jet Propulsion Laboratory, California Institute of Technology under a contract with NASA. WISE is a joint project of the University of California, Los Angeles, and the Jet Propulsion Laboratory/California Institute of Technology, funded by the National Aeronautics and Space Administration. 2MASS is a joint collaboration between the University of Massachusetts and the Infrared Processing and Analysis Center (JPL/Caltech), with funding provided primarily by NASA and the NSF. The SDSS and SDSS-II are funded by the Alfred P. Sloan Foundation, the Participating Institutions, the National Science Foundation, the U.S. Department of Energy, the National Aeronautics and Space Administration, the Japanese Monbukagakusho, the Max Planck Society, and the Higher Education Funding Council for England. The SDSS Web Site is http://www.sdss.org/. This research has made extensive use of the NASA/IPAC Extragalactic Database (NED) which is operated by the Jet Propulsion Laboratory, California Institute of Technology, under contract with NASA, and of NASA's Astrophysics Data System. This research has also made use of Ned Wrights online cosmology calculator \citep{wright06}. DF acknowledges support from the Science \& Technology Facilities Council via an Advanced Fellowship (PP/E005306/1). JA acknowledges support from the Science and Technology Foundation (FCT, Portugal) through the research grant PTDC/CTE-AST/105287/2008.

{\it Facilities:} \facility{Spitzer}, \facility{WISE}, \facility{2MASS}, \facility{SDSS}, \facility{UKIDSS}

\appendix

\section{The Effects of Different Absorption Strength Measures}\label{appendixbunny}

As described in \S\ref{discstrengths}, measuring the properties of BALs is not straightforward. So, we here summarize the checks we performed to see if our results are robust against the choice and method of BAL measurement. 

\begin{figure*}
\includegraphics[width=63mm,angle=90]{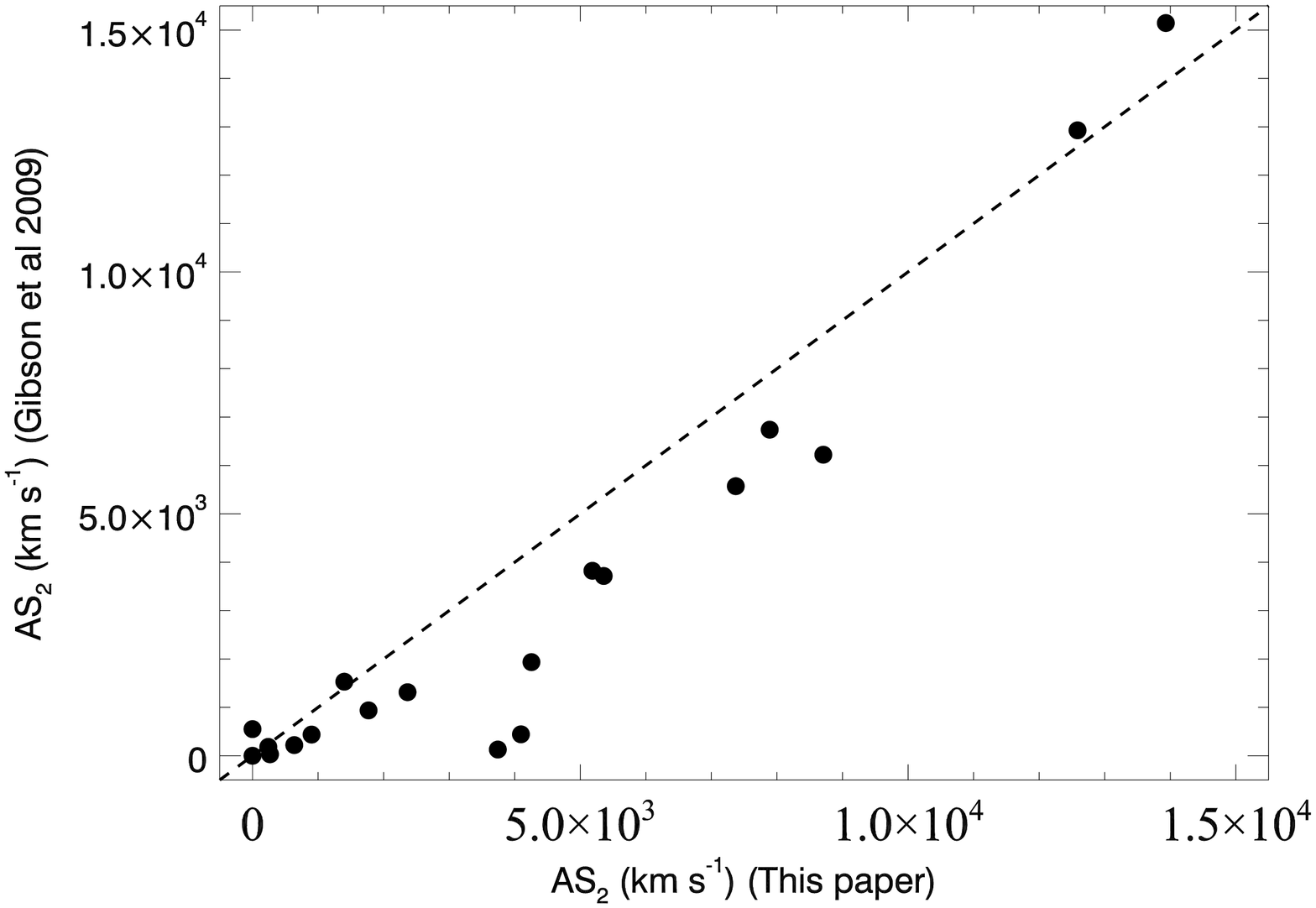}
\includegraphics[width=63mm,angle=90]{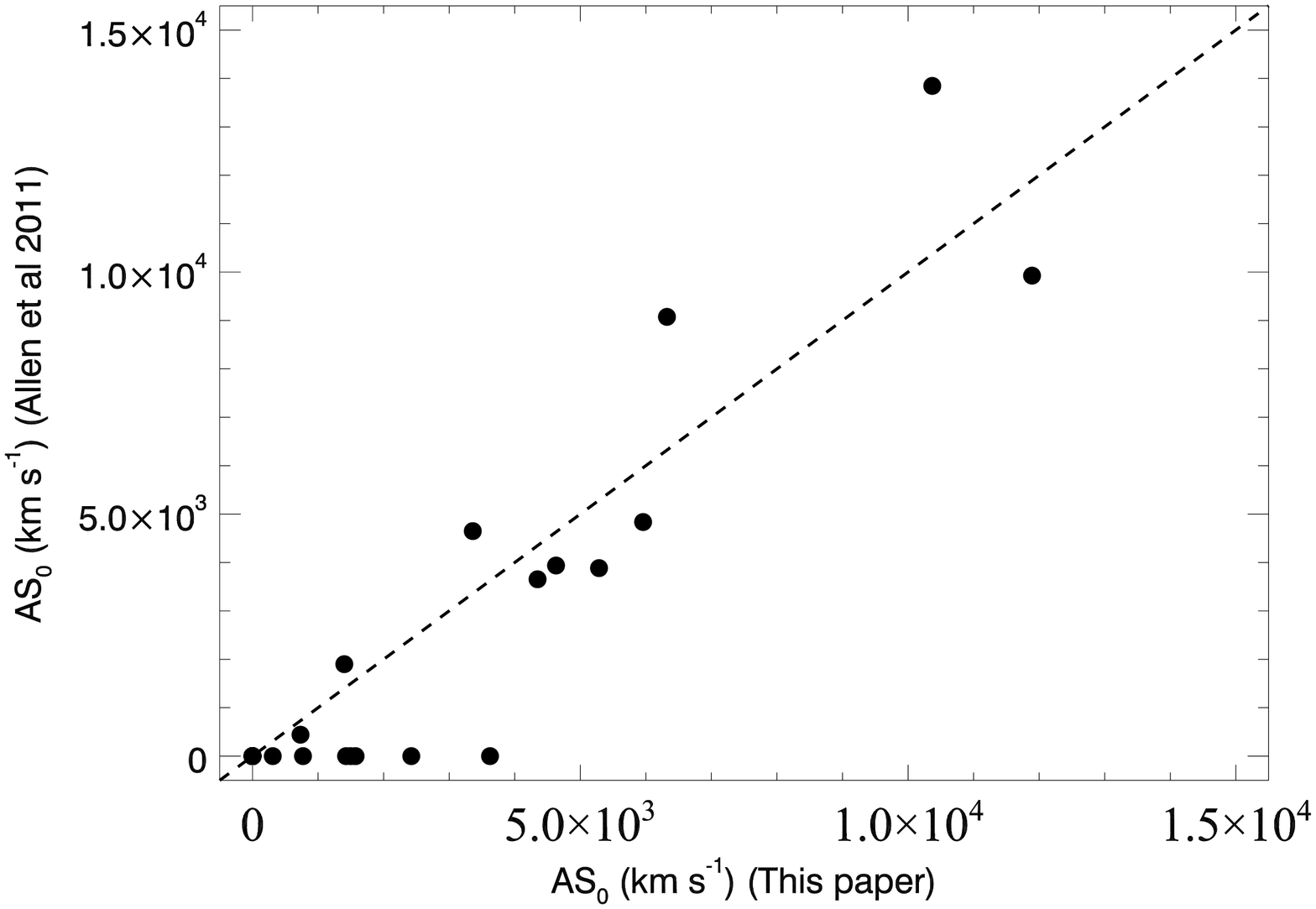}
\caption{Two examples of comparisons between our absorption strength measures, and those for the same objects but by different authors. {\itshape Left:} AS$_{2}$ measured by us and by \citet{gibson09}. {\itshape Right:} AS$_{0}$ measured by us and by \citet{allen11}. In general the measures are consistent, albeit with some scatter, see \S\ref{discstrengths}.}\label{figascomparisons}
\end{figure*}

First, we check our measures by comparing them to the three independent measurements already in the literature (\citealt{trump06,gibson09,allen11}, though we note that the different absorption strength measures used by these authors together with the evolving reduction of the SDSS spectra means the samples in them are not identical). We obtain generally reasonable agreement (Figure \ref{figascomparisons}). We conclude that our measurements of absorption strength are acceptable. We note though that the scatter in this figure is larger than the formal errors on the fits, by a factor of two or more in some cases. This likely reflects systematic uncertainties arising from continuum placement. We do not attempt to account for these systematics, or quote them, as they are difficult to estimate robustly. Instead, we use our absorption strength measurements only to make comparisons to each other, where they should be reliable. As a consequence, we do not quote individual errors on the fits (which are usually of order $100-200$ km s$^{-1}$) as we feel these numbers are misleadingly small. 

\begin{figure*}
\includegraphics[width=66mm,angle=90]{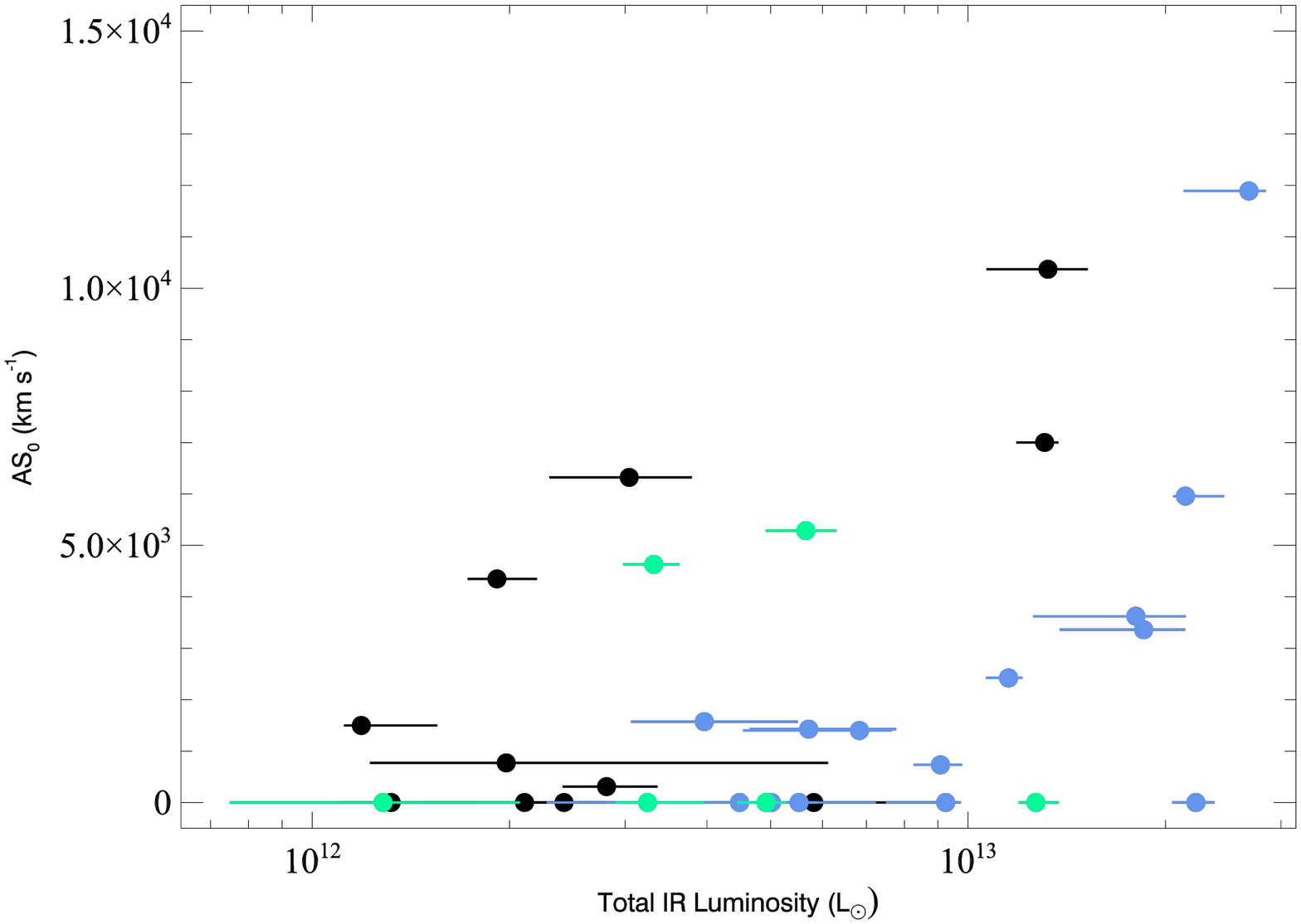}
\includegraphics[width=66mm,angle=90]{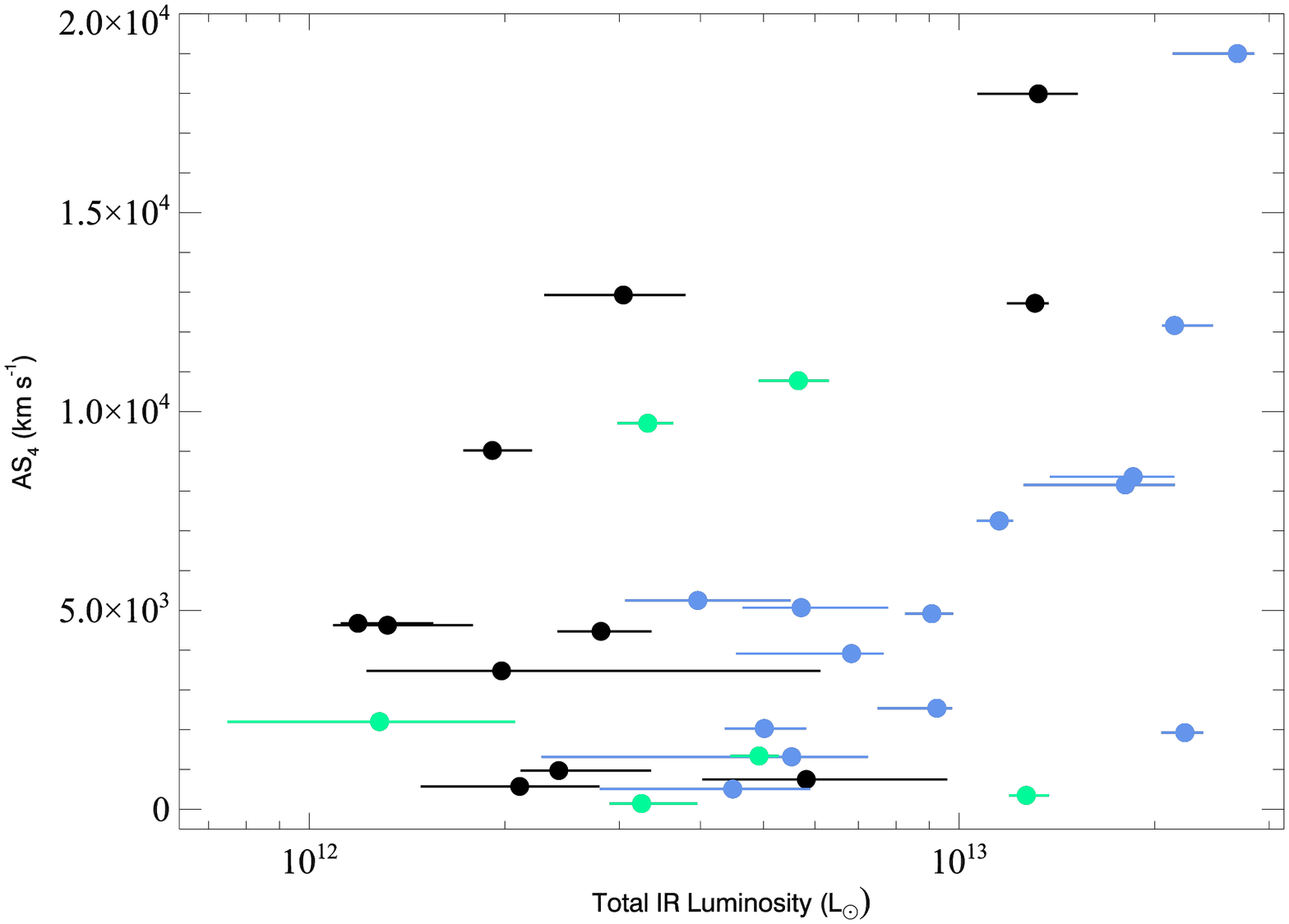} \\
\includegraphics[width=66mm,angle=90]{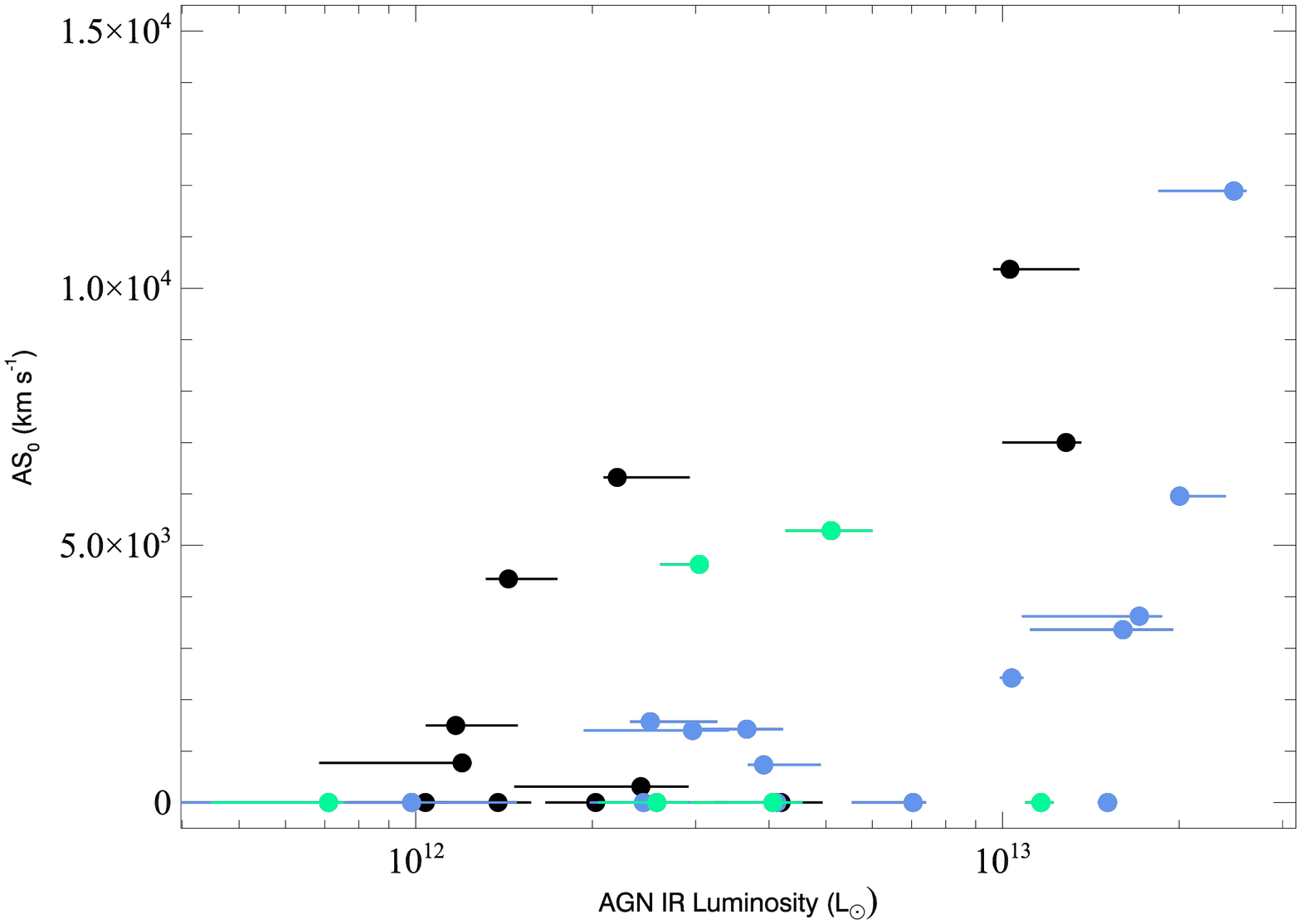}
\includegraphics[width=66mm,angle=90]{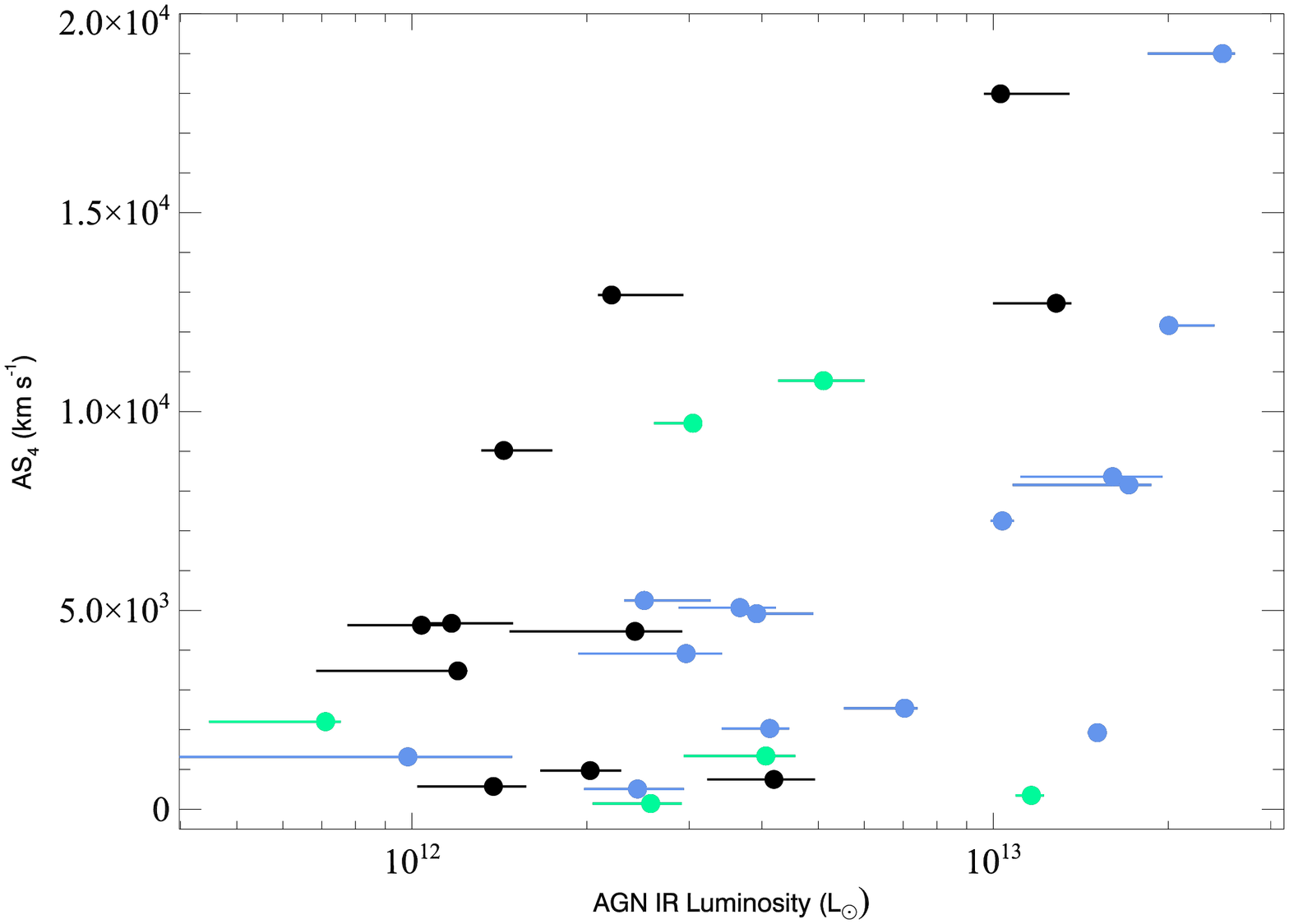} \\
\caption{Consistency checks to see if the results in Figure \ref{figabsvslir} depend on the parametrization of absorption strength used. The left column uses AS$_{0}$ (a strict measure) while the right column uses AS$_{4}$ (a relaxed measure). Top row: total IR luminosity. Bottom row: AGN luminosity. We obtain consistent results in all cases.}\label{figconschecka}
\end{figure*}

\begin{figure*}
\includegraphics[width=66mm,angle=90]{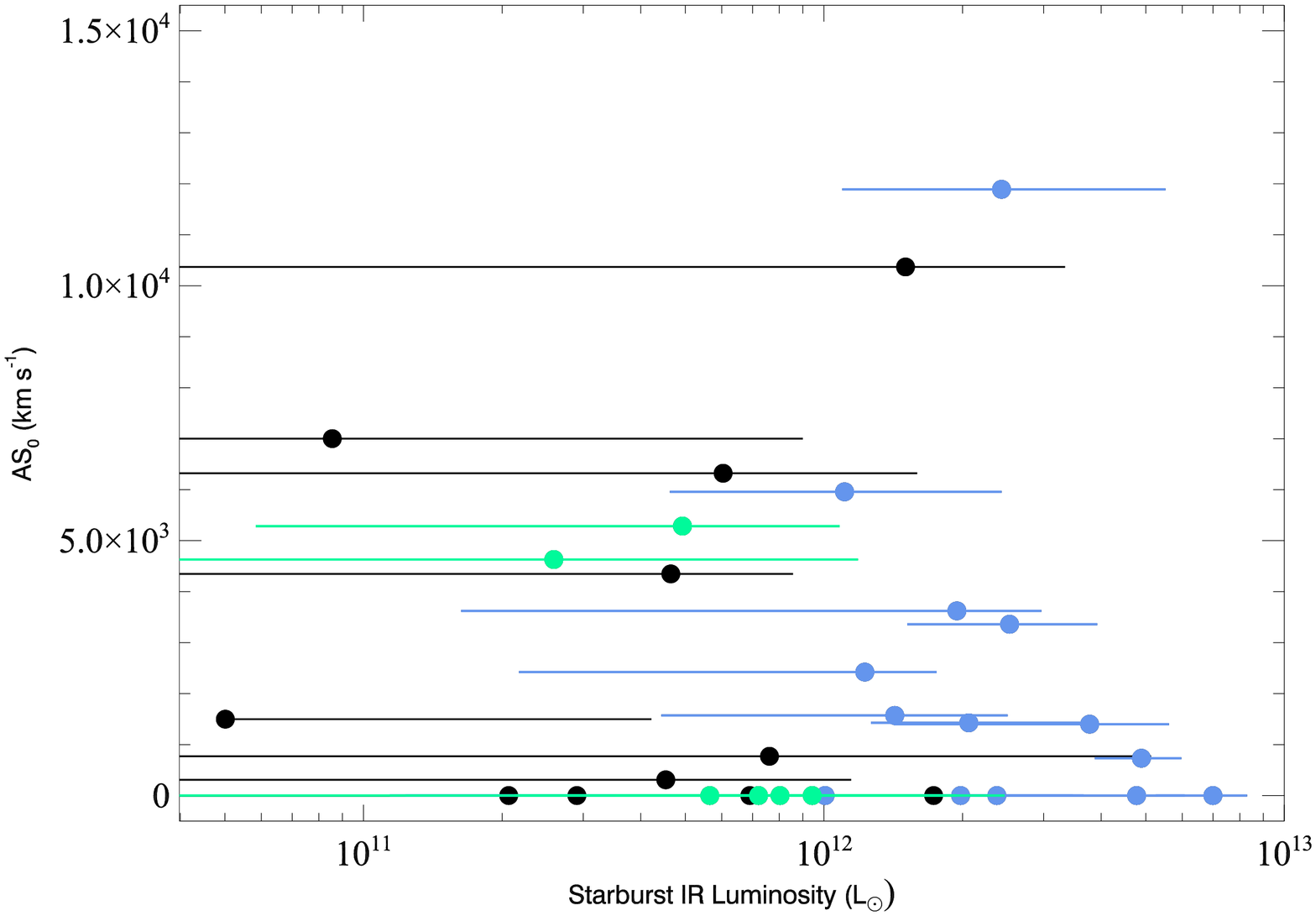}
\includegraphics[width=66mm,angle=90]{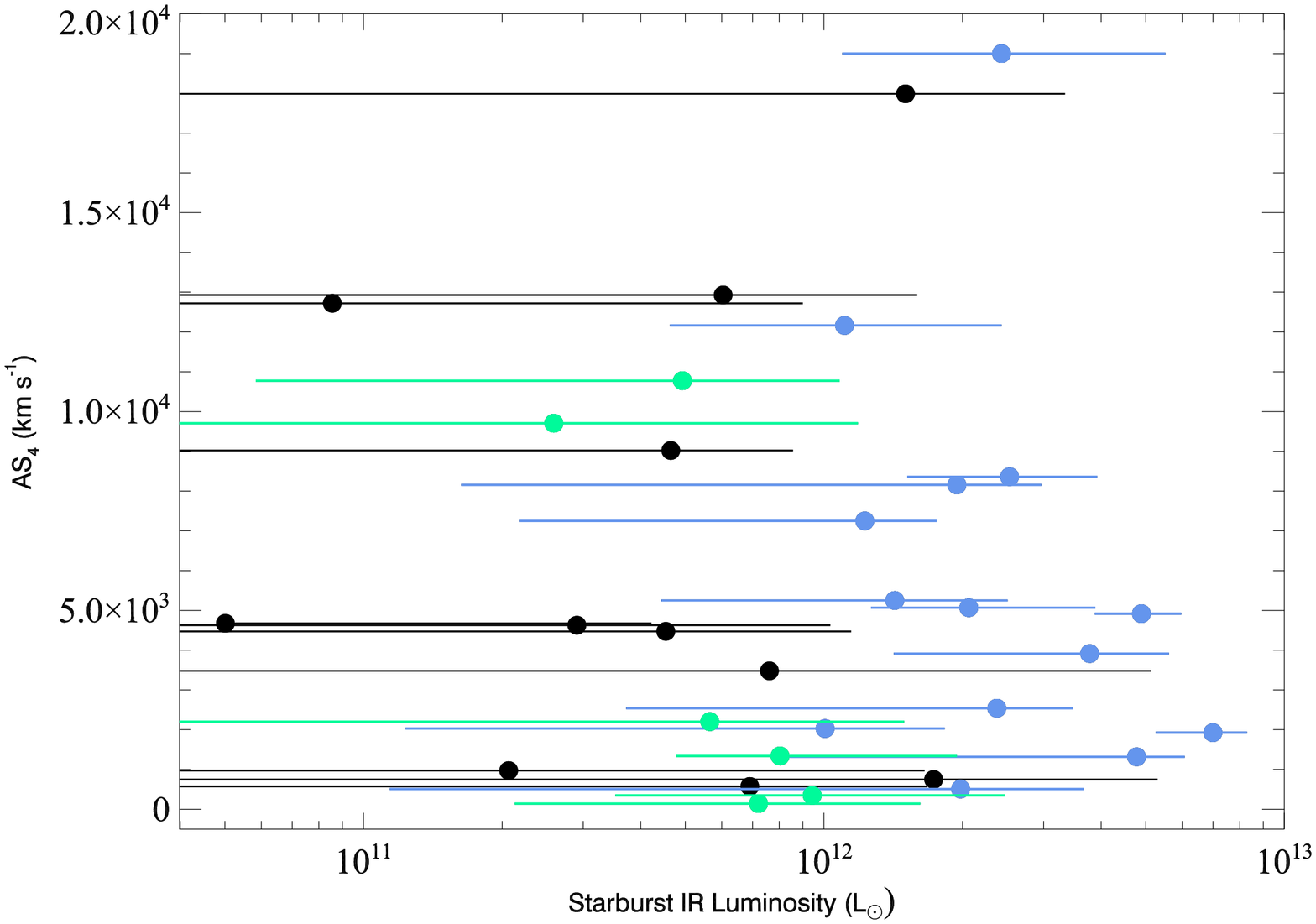} \\
\includegraphics[width=66mm,angle=90]{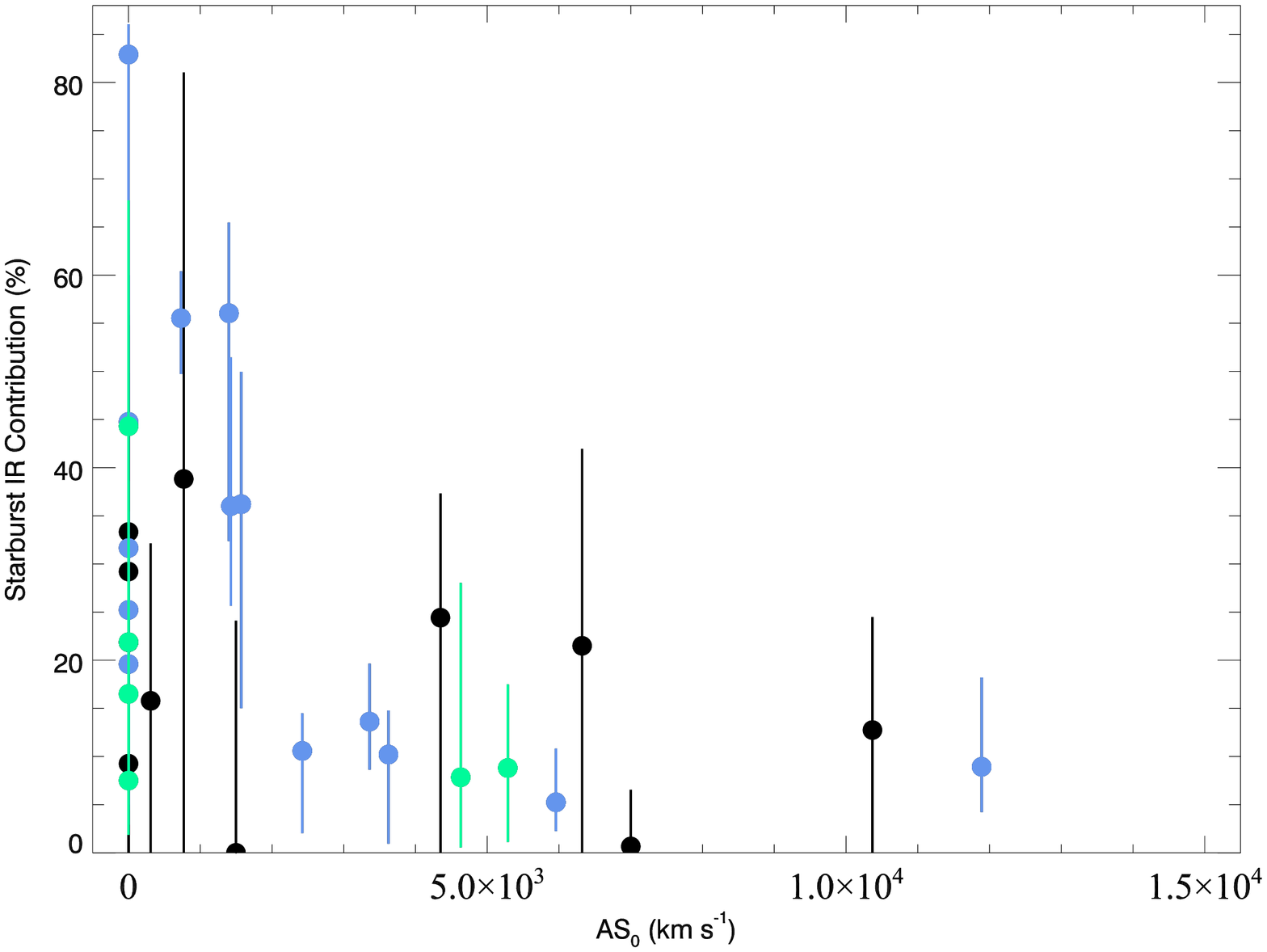}
\includegraphics[width=66mm,angle=90]{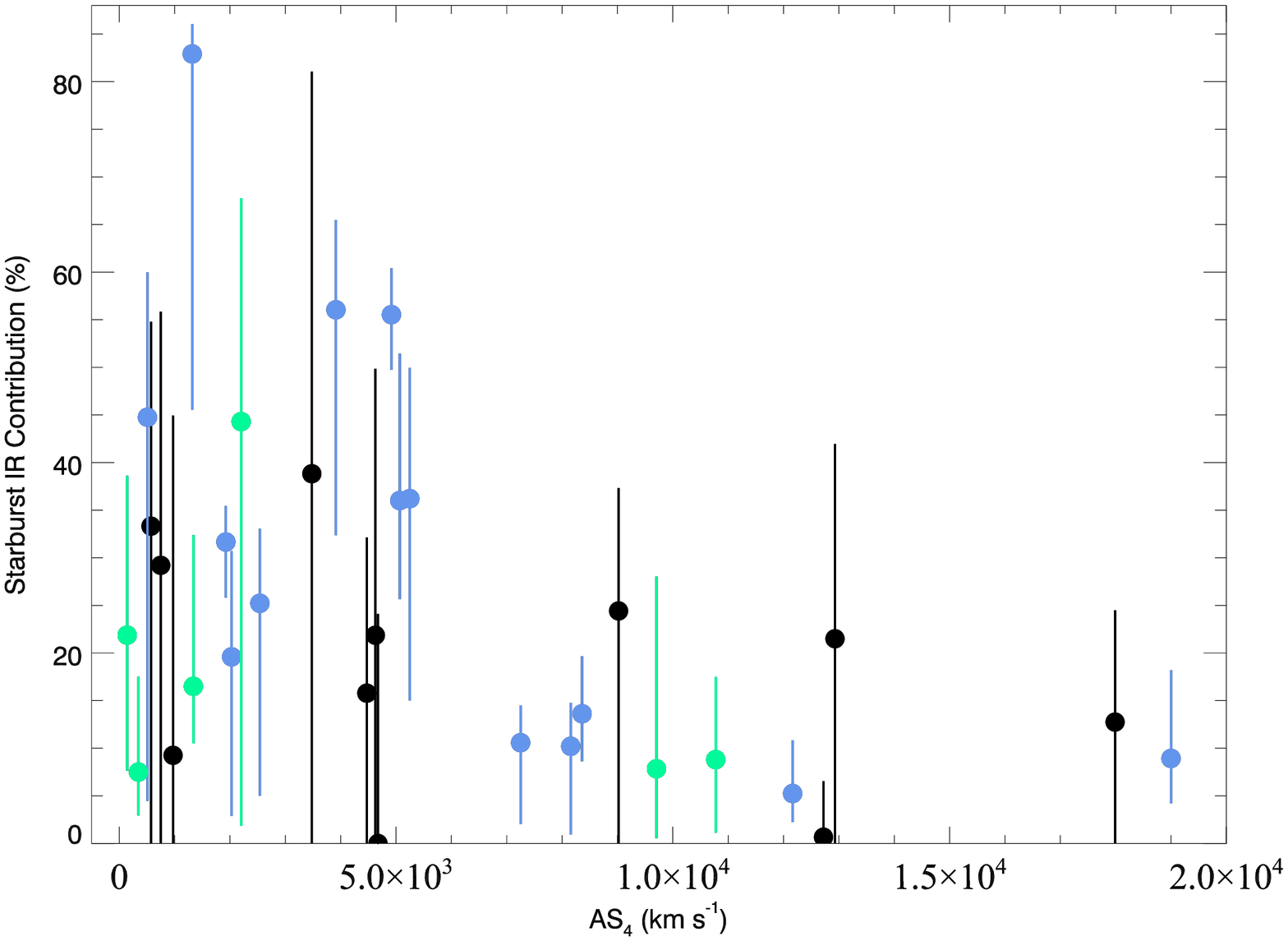}
\caption{Further consistency checks to see if the results in Figure \ref{figabsvslir} depend on the parametrization of absorption strength used. The left column uses AS$_{0}$ (a strict measure) while the right column uses AS$_{4}$ (a relaxed measure). Top row: starburst luminosity. Bottom row: AGN fraction. We obtain consistent results in all cases.}\label{figconscheckb}
\end{figure*}

Next, we check whether the results in Figure \ref{figabsvslir} depend on the adopted parametrization of absorption strength. In Figures \ref{figconschecka} \& \ref{figconscheckb} we replot these figures with both a stricter and more relaxed definition of absorption strength (AS$_{0}$ \& AS$_{4}$, see Table \ref{absmeasures}). In all cases we recover comparable results. It is interesting that the relation is slightly offset from zero in the bottom right panel of Figure \ref{figconscheckb}, perhaps suggesting that the anticorrelation is not driven by troughs much narrower than 2000km s$^{-1}$. We conclude that our results are reasonably robust to the choice of parametrization of absorption strength. 

\begin{deluxetable*}{lccccccc}
\tabletypesize{\scriptsize}
\tablecaption{As tables \ref{probabilities1} \& \ref{probabilities2}, but using (a) a boundary of 5000 km s-1, and (b) only those objects that appear in \citet{gibson09}, using their AS$_{0}$ measures. \label{probabilitiesgib}}
\tablewidth{0pt}
\tablehead{
{\bf Selection}                  &\multicolumn{7}{c}{{\bf Probability of Obtaining}}  \\
	                             &\multicolumn{3}{c}{L$_{Sb}$(L$_{\odot}$)}                                  & \multicolumn{2}{c}{L$_{AGN}$(L$_{\odot}$)}               &\multicolumn{2}{c}{P(f$_{Sb}$)} \\
                                 &  $<10^{11}$            & $<10^{12}$              & $<10^{12.5}$           & $<10^{12}$           & $<10^{12.5}$                      & $>25\%$               & $>50\%$                          
}
\startdata
AS$_{2}<5000$ km s$^{-1}$        & $26.0^{+3.2}_{-6.7}$\% & $51.9^{+6.7}_{-7.3}$\%  & $79.6^{+5.3}_{-4.9}$\% & $30.4^{+5.7}_{-13.4}$\% & $64.2^{+8.7}_{-8.8}$\%   & $62.9^{+5.7}_{-4.4}$\%  & $24.7^{+3.2}_{-4.8}$\%   \\
AS$_{2}>5000$ km s$^{-1}$        & $34.1^{+6.2}_{-5.4}$\% & $68.6^{+10.7}_{-5.5}$\% & $93.7^{+4.4}_{-2.3}$\% & $0$\%                   & $39.2^{+12.2}_{-14.6}$\% & $19.5^{+4.7}_{-7.4}$\%  & $<2$\%    \\
\hline
All \citep{gibson09}             & $23.5^{+4.6}_{-4.2}$\% & $48.6^{+6.9}_{-7.4}$\% & $81.8^{+4.8}_{-3.5}$\% & $<2$\%                   & $34.8^{+7.2}_{-8.5}$\%   & $39.8^{+6.7}_{-6.8}$\%  & $8.8^{+1.9}_{-4.4}$\%    \\
\hline
AS$_{0}^{Gib}<4000$ km s$^{-1}$  & $12.4^{+2.7}_{-3.6}$\% & $27.5^{+6.0}_{-7.7}$\% & $69.2^{+7.7}_{-4.4}$\% & $3.2^{+1.2}_{-3.0}$\%    & $30.1^{+9.3}_{-10.4}$\% & $60.1^{+10.7}_{-7.5}$\%  & $16.4^{+4.1}_{-7.7}$\%  \\
AS$_{0}^{Gib}>4000$ km s$^{-1}$  & $27.9^{+6.6}_{-5.5}$\% & $61.5^{+12.2}_{-11.2}$\% & $92.0^{+5.1}_{-5.5}$\% & $0$\%                  & $23.4^{+9.0}_{-18.4}$\% & $17.8^{+5.6}_{-10.1}$\% & $1.3^{+0.5}_{-1.0}$\%    
\enddata
\end{deluxetable*}

\begin{figure*}
\includegraphics[width=64mm,angle=90]{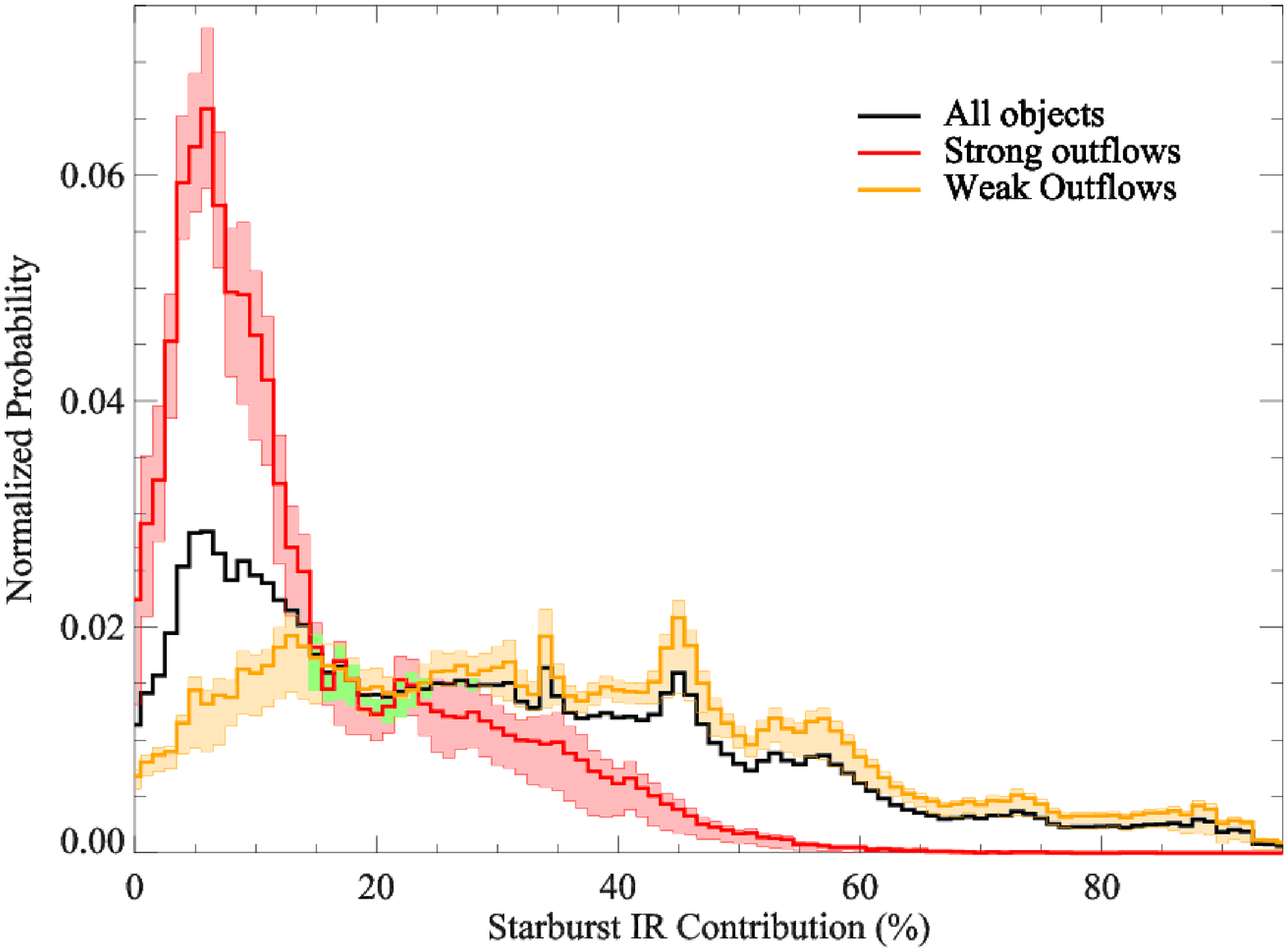}
\includegraphics[width=64mm,angle=90]{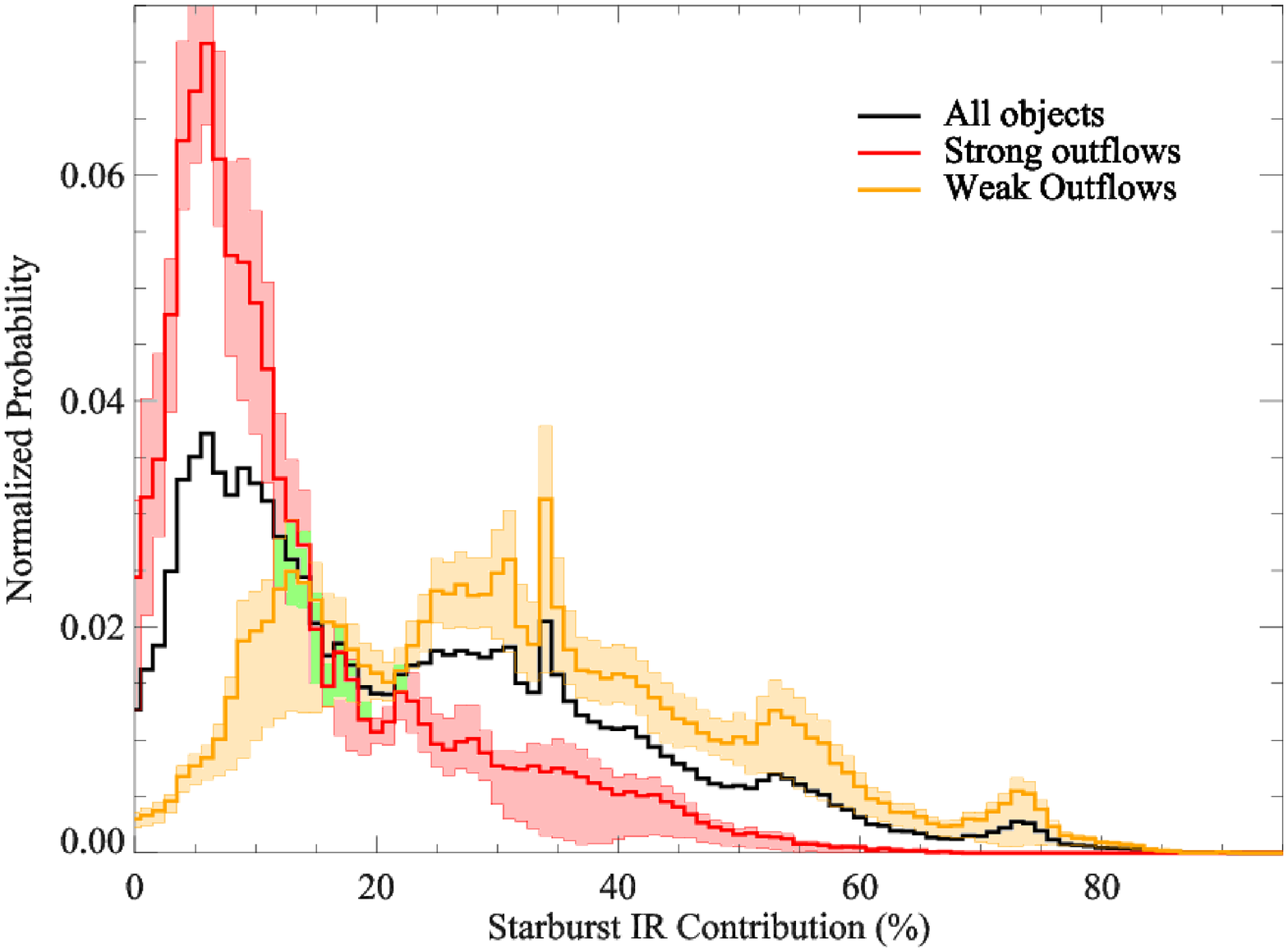}
\caption{Consistency checks to see if the results in Figure \ref{figmergedpdf_agnfrac} depend on the parametrization of absorption strength used. The left panel uses the same parametrization, but a different boundary of AS$_{2}=5000$ km s$^{-1}$. The right panel substitutes the AS$_{2}$ measurements performed by us with the AS$_{0}$ measurements performed by \citet{gibson09}. We obtain the same results in both cases (see also Table \ref{probabilitiesgib}).}\label{figpdfconscheck}
\end{figure*}

Finally, we check whether the results in \S\ref{discoutirs} depend on the method used to measure the absorption strengths. As our sample is small, we cannot robustly test the effects of different techniques to measure absorption strength. We can however do two basic checks. First, we keep our AS$_{2}$ measurements, but change the boundary for the low vs high absorption strength samples from 3500 km s$^{-1}$ to 5000 km s$^{-1}$. Second, we substitute our AS$_{2}$ measurements with the AS$_{0}$ measurements for the same objects in \citet{gibson09}, adjusting the boundary to AS$_{0}=4000$ km s$^{-1}$. Since \citealt{gibson09} are more stringent about what constitutes a BAL QSO in comparison to \citealt{trump06} (from where we chose our sample), this is also a test on a smaller, `golden' sample. We plot the resulting PDFs for AGN fractions in Figure \ref{figpdfconscheck} and present the confidences on luminosity ranges and AGN fractions in Table \ref{probabilitiesgib} for both tests. We obtain the same results - an anticorrelation between absorption strength and fractional AGN luminosity - at a similar degree of significance. We conclude that our result is robust against the method used to measure absorption strengths.

\begin{thebibliography}{}

\bibitem[Alexander et al.(1999)]{alex99} Alexander, D.~M., 
Efstathiou, A., Hough, J.~H., Aitken, D.~K., Lutz, D., Roche, P.~F., 
\& Sturm, E.\ 1999, \mnras, 310, 78 

\bibitem[Alexander et al.(2008)]{alex08} 
Alexander, D.~M., et al.\ 2008, \aj, 135, 1968 

\bibitem[Alexander et al.(2010)]{alex10} 
Alexander, D.~M., Swinbank, A.~M., Smail, I., McDermid, R., 
\& Nesvadba, N.~P.~H.\ 2010, \mnras, 402, 2211 

\bibitem[Allen et al.(2011)]{allen11} 
Allen, J.~T., Hewett, P.~C., Maddox, N., Richards, G.~T., 
\& Belokurov, V.\ 2011, \mnras, 410, 860 

\bibitem[Ammons et al.(2011)]{ammons11} 
Ammons, S.~M., Rosario, D.~J.~V., Koo, D.~C., et al.\ 2011, \apj, 740, 3 

\bibitem[Antonuccio-Delogu \& Silk(2008)]{ant08} 
Antonuccio-Delogu, V., \& Silk, J.\ 2008, \mnras, 389, 1750 

\bibitem[Appenzeller et al.(2005)]{appen05} Appenzeller, I., Stahl, O., Tapken, C., Mehlert, D., \& Noll, S.\ 
2005, \aap, 435, 465 

\bibitem[Arav et al.(1994)]{arav94} 
Arav, N., Li, Z.-Y., \& Begelman, M.~C.\ 1994, \apj, 432, 62 

\bibitem[Arav et al.(2001)]{arav01} 
Arav, N., et al.\ 2001, \apj, 561, 118
 
\bibitem[Arav et al.(2008)]{arav08} 
Arav, N., Moe, M., Costantini, E., et al.\ 2008, \apj, 681, 954 
 
\bibitem[Austermann et al.(2009)]{aus09} 
Austermann, J.~E., et al.\ 2009, \mnras, 1513 

\bibitem[Babbedge et al.(2006)]{babbedge06} Babbedge, T.~S.~R., 
Rowan-Robinson, M., Vaccari, M., et al.\ 2006, \mnras, 370, 1159 

\bibitem[Baugh et al.(2005)]{baugh05} 
Baugh, C.~M., Lacey, C.~G., Frenk, C.~S., Granato, G.~L., Silva, L., Bressan, A., Benson, A.~J., 
\& Cole, S.\ 2005, \mnras, 356, 1191 

\bibitem[Bautista et al.(2010)]{bautista10} Bautista, M.~A., Dunn, 
J.~P., Arav, N., et al.\ 2010, \apj, 713, 25 

\bibitem[Becker et al.(1997)]{becker97} 
Becker, R.~H., Gregg, M.~D., Hook, I.~M., et al.\ 1997, \apjl, 479, L93 

\bibitem[Benson et al.(2003)]{benson03} 
Benson, A.~J., Bower, R.~G., Frenk, C.~S., Lacey, C.~G., Baugh, C.~M., 
\& Cole, S.\ 2003, \apj, 599, 38 

\bibitem[Bertin \& Arnouts(1996)]{bertin96} 
Bertin, E., \& Arnouts, S.\ 1996, \aap, 117, 393 

\bibitem[Best et al.(2006)]{best06} Best, P.~N., Kaiser, 
C.~R., Heckman, T.~M., \& Kauffmann, G.\ 2006, \mnras, 368, L67 

\bibitem[Blain et al.(2002)]{blain02} Blain, A.~W., Smail, I., 
Ivison, R.~J., Kneib, J.-P., \& Frayer, D.~T.\ 2002, Physics Reports, 369, 111 

\bibitem[Booth \& Schaye(2009)]{booth09} 
Booth, C.~M., \& Schaye, J.\ 2009, \mnras, 398, 53 

\bibitem[Bower et al.(2006)]{bower06} Bower, R.~G., Benson, 
A.~J., Malbon, R., Helly, J.~C., Frenk, C.~S., Baugh, C.~M., Cole, S., 
\& Lacey, C.~G.\ 2006, \mnras, 370, 645 

\bibitem[Bower et al.(2008)]{bower08} Bower, R.~G., McCarthy, 
I.~G., \& Benson, A.~J.\ 2008, \mnras, 390, 1399 

\bibitem[Braito et al.(2007)]{braito07} Braito, V., Reeves, 
J.~N., Dewangan, G.~C., et al.\ 2007, \apj, 670, 978 

\bibitem[Bruzual \& Charlot(2003)]{bc03} 
Bruzual, G., \& Charlot, S.\ 2003, \mnras, 344, 1000 

\bibitem[Canalizo \& Stockton(2001)]{canalizo01} Canalizo, G. \& Stockton, A. 
2001, \apj, 555, 719

\bibitem[Casebeer et al.(2008)]{case08} 
Casebeer, D., Baron, E., Leighly, K., Jevremovic, D., \& Branch, D.\ 2008, \apj, 676, 857 

\bibitem[Chapman et al.(2005)]{chapman05} 
Chapman, S.~C., Blain, A.~W., Smail, I., \& Ivison, R.~J.\ 2005, \apj, 622, 772 

\bibitem[Chartas et al.(2003)]{chartas03} 
Chartas, G., Brandt, W.~N., \& Gallagher, S.~C.\ 2003, \apj, 595, 85 

\bibitem[Chatterjee et al.(2011)]{chat11} Chatterjee, S., 
Degraf, C., Richardson, J., et al.\ 2011, \mnras, 1869 

\bibitem[Chung et al.(2011)]{chung11} 
Chung, A., Yun, M.~S., Naraynan, G., Heyer, M., \& Erickson, N.~R.\ 2011, \apjl, 732, L15 

\bibitem[Cid Fernandes et al.(2001)]{cidfernandes01} Cid Fernandes, R., 
Heckman, T., Schmitt, H., González Delgado, R. M., Storchi-Bergmann, T. 2001, 
\apj, 558, 81

\bibitem[Ciotti et al.(1991)]{ciotti91} 
Ciotti, L., D'Ercole, A., Pellegrini, S., \& Renzini, A.\ 1991, \apj, 376, 380 

\bibitem[Ciotti \& Ostriker(2007)]{ciotti07} 
Ciotti, L., \& Ostriker, J.~P.\ 2007, \apj, 665, 1038 

\bibitem[Cole et al.(2000)]{cole00} 
Cole, S., Lacey, C.~G., Baugh, C.~M., \& Frenk, C.~S.\ 2000, \mnras, 319, 168 

\bibitem[Coppin et al.(2006)]{coppin06} 
Coppin, K., et al.\ 2006, \mnras, 372, 1621 

\bibitem[Crenshaw et al.(2003)]{crenshaw03} 
Crenshaw, D.~M., Kraemer, S.~B., \& George, I.~M.\ 2003, AR\aap, 41, 117 

\bibitem[Croton et al.(2006)]{croton06} 
Croton, D.~J., et al.\ 2006, \mnras, 365, 11 

\bibitem[Dai et al.(2008)]{dai08} 
Dai, X., Shankar, F., Sivakoff, G. R. 2008, \apj, 672, 108

\bibitem[De Lucia et al.(2006)]{deluc06} 
De Lucia, G., Springel, V., White, S.~D.~M., Croton, D., 
\& Kauffmann, G.\ 2006, \mnras, 366, 499 

\bibitem[Debuhr et al.(2011)]{debuhr11} Debuhr, J., Quataert, 
E., \& Ma, C.-P.\ 2011, \mnras, 412, 1341 

\bibitem[Desai et al.(2007)]{desai07} 
Desai, V., Armus, L., Spoon, H.~W.~W., et al.\ 2007, \apj, 669, 810 

\bibitem[Di Matteo et al.(2005)]{dimatteo05} Di Matteo, T., 
Springel, V., \& Hernquist, L.\ 2005, 
\nat, 433, 604 

\bibitem[Dopita et al.(2005)]{dopita05} Dopita, M.~A., et al.\ 
2005, \apj, 619, 755 

\bibitem[Dullemond \& van Bemmel(2005)]{dullemond05} 
Dullemond, C.~P., \& van Bemmel, I.~M.\ 2005, \aap, 436, 47 

\bibitem[Dunlop et al.(1996)]{dun96} 
Dunlop, J., Peacock, J., Spinrad, H., Dey, A., Jimenez, R., Stern, D., 
\& Windhorst, R.\ 1996, 
\nat, 381, 581 

\bibitem[Dunn et al.(2010)]{dunn10} 
Dunn, J.~P., et al.\ 2010, \apj, 709, 611 

\bibitem[Eales et al.(2010)]{eales10} 
Eales, S.~A., et al.\ 2010, \aap, 518, L23 

\bibitem[Efstathiou \& Rowan-Robinson(1995)]{efs95} 
Efstathiou, A., \& Rowan-Robinson, M.\ 1995, \mnras, 273, 649 

\bibitem[Efstathiou et al.(2000)]{efs00} Efstathiou, A., 
Rowan-Robinson, M., \& Siebenmorgen, R.\ 2000, \mnras, 313, 734 

\bibitem[Efstathiou \& Siebenmorgen(2005)]{efs05} 
Efstathiou, A., \& Siebenmorgen, R.\ 2005, \aap, 439, 85 

\bibitem[Efstathiou \& Siebenmorgen(2009)]{efs09} Efstathiou, A., \& Siebenmorgen, R.\ 2009, \aap, 502, 541 

\bibitem[Ehlert et al.(2011)]{ehlert11} Ehlert, S., et al.\ 
2011, \mnras, 411, 1641 

\bibitem[Ellis et al.(1997)]{ell97} 
Ellis, R.~S., Smail, I., Dressler, A., Couch, W.~J., Oemler, A.~J., Butcher, H., 
\& Sharples, R.~M.\ 1997, \apj, 483, 582 

\bibitem[Fabian(1999)]{fabian99} Fabian, A.~C.\ 1999, \mnras, 
308, L39 

\bibitem[Fabian et al.(2006)]{fabian06} 
Fabian, A.~C., Celotti, A., \& Erlund, M.~C.\ 2006, \mnras, 373, L16 

\bibitem[Farrah et al.(2002)]{farrah02} Farrah, D., Serjeant, 
S., Efstathiou, A., Rowan-Robinson, M., 
\& Verma, A.\ 2002, \mnras, 335, 1163

\bibitem[Farrah et al.(2003)]{farrah03} Farrah, D., Afonso, J., 
Efstathiou, A., Rowan-Robinson, M., Fox, M., 
\& Clements, D.\ 2003, \mnras, 343, 585 

\bibitem[Farrah et al.(2005)]{farrah05} Farrah, D., Surace, 
J.~A., Veilleux, S., Sanders, D.~B., \& Vacca, W.~D.\ 2005, \apj, 626, 70 

\bibitem[Farrah et al.(2007)]{farrah07} Farrah, D.; Lacy, M.; Priddey, R.; 
Borys, C.; Afonso, J. 2007, \apj, 662, 59

\bibitem[Farrah et al.(2009)]{farrah09} Farrah, D., et al.\ 
2009, \apj, 700, 395 

\bibitem[Farrah et al.(2010)]{farrah10} Farrah, D., et al.\ 
2010, \apj, 717, 868 

\bibitem[Ferrarese \& Merritt(2000)]{ferrarese00} Ferrarese, L. \& Merritt, D. 
2000, \apj, 539, 9

\bibitem[Ferrarese(2002)]{ferrarese02} 
Ferrarese, L.\ 2002, \apj, 578, 90 

\bibitem[Feruglio et al.(2010)]{ferug10} 
Feruglio, C., Maiolino, R., Piconcelli, E., Menci, N., Aussel, H., Lamastra, A., \& Fiore, F.\ 2010, \aap, 518, L155 

\bibitem[Fontana et al.(2006)]{fon06} 
Fontana, A., Salimbeni, S., Grazian, A., et al.\ 2006, \aap, 459, 745 

\bibitem[Gallagher et al.(2007)]{gallagher07} Gallagher, S. C., Hines, D. C., 
Blaylock, M., Priddey, R. S., Brandt, W. N., Egami, E. E. 2007, \apj, 665, 157

\bibitem[Gaspari et al.(2011)]{gasp11} 
Gaspari, M., Brighenti, F., D'Ercole, A., \& Melioli, C.\ 2011, \mnras, 415, 1549 

\bibitem[Gebhardt et al.(2000)]{gebhardt} Gebhardt K. et al. 2000, \apj, 539, 13

\bibitem[Genzel et al.(1998)]{genzel98} 
Genzel, R., Lutz, D., Sturm, E., et al.\ 1998, \apj, 498, 579 

\bibitem[Georgakakis et al.(2009)]{geo09} Georgakakis, A., 
Clements, D.~L., Bendo, G., Rowan-Robinson, M., Nandra, K., 
\& Brotherton, M.~S.\ 2009, \mnras, 394, 533 

\bibitem[Gibson et al.(2009)]{gibson09} Gibson, R.~R., et al.\ 
2009, \apj, 692, 758 

\bibitem[Glazebrook et al.(2004)]{glaz04} 
Glazebrook, K., Abraham, R.~G., McCarthy, P.~J., et al.\ 2004, \nat, 430, 181 

\bibitem[Glikman et al.(2007)]{f2m} Glikman, E., Helfand, D. J., White, R. 
L., Becker, R. H., Gregg, M. D.; Lacy, M. 2007, \apj, 667, 673

 \bibitem[Gonzalez-Perez et al.(2009)]{gonzalez09} Gonzalez-Perez, 
V., Baugh, C.~M., Lacey, C.~G., \& Almeida, C.\ 2009, \mnras, 398, 497 

\bibitem[Granato \& Danese(1994)]{gran94} Granato, G.~L., \& Danese, L.\ 1994, \mnras, 268, 235 

\bibitem[Granato et al.(2004)]{granato04} Granato, G.~L., De 
Zotti, G., Silva, L., Bressan, A., \& Danese, L.\ 2004, \apj, 600, 580 

\bibitem[Green \& Mathur(1996)]{green96} 
Green, P.~J., \& Mathur, S.\ 1996, \apj, 462, 637 

\bibitem[Gregg et al.(2002)]{gregg02} Gregg, M.~D., Becker, 
R.~H., White, R.~L., Richards, G.~T., Chaffee, F.~H., 
\& Fan, X.\ 2002, \apjl, 573, L85 

\bibitem[Haas et al.(2003)]{haas03} Haas, M., et al.\ 2003, \aap, 402, 87 

\bibitem[Haiman et al.(1996)]{haiman96} Haiman, Z., Rees, M.~J., 
\& Loeb, A.\ 1996, \apj, 467, 522 

\bibitem[Hall et al.(2002)]{hall02} Hall, P.~B., et al.\ 2002, 
\apjs, 141, 267 

\bibitem[Hambrick et al.(2011)]{hambrick11} 
Hambrick, D. C., et al.\ 2011, \apj, accepted, astroph 1106.1405 

\bibitem[H{\"a}ring \& Rix(2004)]{har04} 
H{\"a}ring, N., \& Rix, H.-W.\ 2004, \apjl, 604, L89 

\bibitem[Hazard et al.(1987)]{hazard87} Hazard, C., McMahon, R. G., Webb, J. 
K., Morton, D. C. 1987, \apj, 323, 263

\bibitem[Heavens et al.(2004)]{heavens04} 
Heavens, A., Panter, B., Jimenez, R., \& Dunlop, J.\ 2004, 
at, 428, 625 

\bibitem[Hern{\'a}n-Caballero et al.(2009)]{hernan09} 
Hern{\'a}n-Caballero, A., P{\'e}rez-Fournon, I., Hatziminaoglou, E., et 
al.\ 2009, \mnras, 395, 1695 

\bibitem[Hewett \& Foltz(2003)]{hewett03} Hewett, P.~C., \& Foltz, C.~B.\ 2003, \aj, 125, 1784 

\bibitem[H{\"o}nig et al.(2006)]{hoenig06} H{\"o}nig, S.~F., Beckert, T., Ohnaka, K., 
\& Weigelt, G.\ 2006, \aap, 452, 459 

\bibitem[Hopkins et al.(2008)]{hopkins08} Hopkins, P. F., Hernquist, L., Cox, 
T. J., Keres, D. 2008, \apjs, 175, 356

\bibitem[Hopkins \& Elvis(2010)]{hopkins10} Hopkins, P. F. \& Elvis, M. 2010, 
\mnras, 401, 7

\bibitem[Ilbert et al.(2010)]{ilb10} 
Ilbert, O., et al.\ 2010, \apj, 709, 644 

\bibitem[Jahnke \& Macci{\`o}(2011)]{jahnke11} 
Jahnke, K., \& Macci{\`o}, A.~V.\ 2011, \apj, 734, 92 

\bibitem[Jarrett et al.(2011)]{jarr11} 
Jarrett, T.~H., et al.\ 2011, \apj, 735, 112 

\bibitem[Kauffmann et al.(1999)]{kauff99} 
Kauffmann, G., Colberg, J.~M., Diaferio, A., \& White, S.~D.~M.\ 1999, \mnras, 307, 529

\bibitem[Kaviraj et al.(2011)]{kav11} 
Kaviraj, S., Schawinski, K., Silk, J., \& Shabala, S.~S.\ 2011, \mnras, 415, 3798 

\bibitem[Kitzbichler \& White(2007)]{kitz07} 
Kitzbichler, M.~G., \& White, S.~D.~M.\ 2007, \mnras, 376, 2 

\bibitem[Kruegel \& Siebenmorgen(1994)]{kru94} 
Kruegel, E., \& Siebenmorgen, R.\ 1994, \aap, 288, 929 

\bibitem[Lacy et al.(2002)]{lacy02} 
Lacy, M., Gregg, M., Becker, R.~H., White, R.~L., Glikman, E., Helfand, D., 
\& Winn, J.~N.\ 2002, \aj, 123, 2925 

\bibitem[Lagache et al.(2005)]{lag05} 
Lagache, G., Puget, J.-L., \& Dole, H.\ 2005, AR\aap, 43, 727 

\bibitem[Lawrence et al.(2007)]{law07} 
Lawrence, A., Warren, S.~J., Almaini, O., et al.\ 2007, \mnras, 379, 1599 

\bibitem[Le Floc'h et al.(2005)]{leflo05} Le Floc'h, E., et 
al.\ 2005, \apj, 632, 169 

\bibitem[Leighly et al.(2011)]{leigh11} 
Leighly, K.~M., Dietrich, M., \& Barber, S.\ 2011, \apj, 728, 94 

\bibitem[Lewis et al.(2003)]{lewis03} 
Lewis, G.~F., Chapman, S.~C., \& Kuncic, Z.\ 2003, \apjl, 596, L35 

\bibitem[Lonsdale et al.(2006)]{lonsdale06} Lonsdale, C. J., Farrah, D., 
Smith, H. E. 2006, Astrophysics Update 2, 285

\bibitem[Lutz et al.(2008)]{lutz08} Lutz, D., et al.\ 2008, \apj, 684, 853 

\bibitem[Lutz et al.(2010)]{lutz10} Lutz, D., et al.\ 2010, \apj, 712, 1287 

\bibitem[Lynds(1967)]{lyn67} 
Lynds, C.~R.\ 1967, \apj, 147, 396 

\bibitem[Marchesini et al.(2009)]{march09} Marchesini, D., van 
Dokkum, P.~G., F{\"o}rster Schreiber, N.~M., et al.\ 2009, \apj, 701, 1765 

\bibitem[McCarthy et al.(2011)]{mccar11} McCarthy, I.~G., 
Schaye, J., Bower, R.~G., Ponman, T.~J., Booth, C.~M., Dalla Vecchia, C., 
\& Springel, V.\ 2011, \mnras, 412, 1965 

\bibitem[McNamara \& Nulsen(2007)]{macnam07} McNamara, B.~R., \& Nulsen, P.~E.~J.\ 2007, AR\aap, 45, 117 

\bibitem[Ma et al.(2011)]{ma11} Ma, C.-J., McNamara, B.~R., 
Nulsen, P.~E.~J., Schaffer, R., \& Vikhlinin, A.\ 2011, \apj, 740, 51 

\bibitem[Magorrian et al.(1998)]{mag98} Magorrian, J., Tremaine, S., 
Richstone, D., Bender, R., Bower, G., Dressler, A., Faber, S., Gebhardt, K., 
Green, R., Grillmair, C., Kormendy, J., Lauer, T. 1998, \aj, 115, 2285

\bibitem[Marconi \& Hunt(2003)]{marc03} 
Marconi, A., \& Hunt, L.~K.\ 2003, \apjl, 589, L21 

\bibitem[Mart{\'{\i}}nez-Sansigre et al.(2005)]{mart05} 
Mart{\'{\i}}nez-Sansigre, A., et al.\ 2005, 
at, 436, 666 

\bibitem[Menci et al.(2008)]{menci08} 
Menci, N., Fiore, F., Puccetti, S., \& Cavaliere, A.\ 2008, \apj, 686, 219 

\bibitem[Mittal et al.(2009)]{mit09} 
Mittal, R., Hudson, D.~S., Reiprich, T.~H., \& Clarke, T.\ 2009, \aap, 501, 835 

\bibitem[Moe et al.(2009)]{moe09} 
Moe, M., Arav, N., Bautista, M.~A., \& Korista, K.~T.\ 2009, \apj, 706, 525 

\bibitem[Monaco et al.(2007)]{monaco07} 
Monaco, P., Fontanot, F., \& Taffoni, G.\ 2007, \mnras, 375, 1189 

\bibitem[M{\"u}ller-S{\'a}nchez et al.(2011)]{muller11} 
M{\"u}ller-S{\'a}nchez, F., Prieto, M.~A., Hicks, E.~K.~S., et al.\ 2011, 
\apj, 739, 69 

\bibitem[Nenkova et al.(2002)]{nenkova02} Nenkova, M., 
Ivezi{\'c}, {\v Z}., \& Elitzur, M.\ 2002, \apjl, 570, L9 

\bibitem[Nesvadba et al.(2010)]{nesv10} 
Nesvadba, N.~P.~H., et al.\ 2010, \aap, 521, A65
 
\bibitem[Omma et al.(2004)]{omma04} 
Omma, H., Binney, J., Bryan, G., \& Slyz, A.\ 2004, \mnras, 348, 1105 

\bibitem[Orellana et al.(2011)]{orell11} 
Orellana, G., Nagar, N.~M., Isaak, K.~G., Priddey, R., Maiolino, R., McMahon, R., 
Marconi, A., \& Oliva, E.\ 2011, \aap, 531, A128 

\bibitem[Ostriker et al.(2010)]{ost10} 
Ostriker, J.~P., Choi, E., Ciotti, L., Novak, G.~S., \& Proga, D.\ 2010, \apj, 722, 642 

\bibitem[P{\'e}rez-Gonz{\'a}lez et al.(2008)]{per08} 
P{\'e}rez-Gonz{\'a}lez, P.~G., Rieke, G.~H., Villar, V., et al.\ 2008, 
\apj, 675, 234 

\bibitem[Peterson et al.(2003)]{peterson03} 
Peterson, J.~R., Kahn, S.~M., Paerels, F.~B.~S., et al.\ 2003, \apj, 590, 207 

\bibitem[Peterson \& Fabian(2006)]{peterson06} 
Peterson, J.~R., \& Fabian, A.~C.\ 2006, \physrep, 427, 1 

\bibitem[Pier \& Krolik(1992)]{pier92} 
Pier, E.~A., \& Krolik, J.~H.\ 1992, \apj, 401, 99

\bibitem[Pope(2009)]{pope09} 
Pope, E.~C.~D.\ 2009, \mnras, 395, 2317 

\bibitem[Power et al.(2011)]{power11} 
Power, C., Nayakshin, S., \& King, A.\ 2011, \mnras, 412, 269 

\bibitem[Proga et al.(2000)]{proga00} 
Proga, D., Stone, J.~M., \& Kallman, T.~R.\ 2000, \apj, 543, 686 

\bibitem[Puchwein et al.(2008)]{puch08} 
Puchwein, E., Sijacki, D., \& Springel, V.\ 2008, \apjl, 687, L53 

\bibitem[Rakos et al.(2007)]{rak07} 
Rakos, K., Schombert, J., \& Odell, A.\ 2007, \apj, 658, 929 

\bibitem[Rieke et al.(2004)]{rieke04} 
Rieke, G.~H., et al.\ 2004, \apjs, 154, 25

\bibitem[Roseboom et al.(2009)]{rose09} 
Roseboom, I.~G., Oliver, S., \& Farrah, D.\ 2009, \apjl, 699, L1 

\bibitem[Rowan-Robinson \& Crawford(1989)]{mrr89} Rowan-Robinson, M., \& Crawford, J.\ 1989, \mnras, 238, 523 

\bibitem[Rowan-Robinson et al.(1997)]{mrr97} 
Rowan-Robinson, M., et al.\ 1997, \mnras, 289, 490 

\bibitem[Rowan-Robinson \& Efstathiou(2009)]{mrr09} 
Rowan-Robinson, M., \& Efstathiou, A.\ 2009, \mnras, 399, 615 

\bibitem[Ruiz et al.(2001)]{ruiz01} Ruiz, M., Efstathiou, A., 
Alexander, D.~M., \& Hough, J.\ 2001, \mnras, 325, 995 

\bibitem[Rupke \& Veilleux(2011)]{rupke11} 
Rupke, D.~S.~N., \& Veilleux, S.\ 2011, \apjl, 729, L27 

\bibitem[Sales et al.(2010)]{sales10} 
Sales, L.~V., Navarro, J.~F., Schaye, J., et al.\ 2010, \mnras, 409, 1541 

\bibitem[Sanders et al.(1988)]{sanders88} Sanders, D. B., Soifer, B. T., 
Elias, J. H., Madore, B. F., Matthews, K., Negebauer, G., Scoville, N. Z. 
1988, \apj, 325, 74

\bibitem[Sanders \& Mirabel(1996)]{sanders96} Sanders, D. B. \& Mirabel, I. 
F. 1996, AR\aap, 34, 749

\bibitem[Sazonov et al.(2005)]{saz05} 
Sazonov, S.~Y., Ostriker, J.~P., Ciotti, L., \& Sunyaev, R.~A.\ 2005, \mnras, 358, 168 

\bibitem[Scannapieco \& Oh(2004)]{scannapieco04} Scannapieco, E. \& Oh, S. P. 
2004, \apj, 608, 62

\bibitem[Schartmann et al.(2008)]{schart08} Schartmann, M., Meisenheimer, K., 
Camenzind, M., Wolf, S., Tristram, K.~R.~W., \& Henning, T.\ 2008, \aap, 482, 67 

\bibitem[Schawinski et al.(2007)]{schaw07} Schawinski, K., 
Thomas, D., Sarzi, M., Maraston, C., Kaviraj, S., Joo, S.-J., Yi, S.~K., 
\& Silk, J.\ 2007, \mnras, 382, 1415 

\bibitem[Schmidt \& Hines(1999)]{sch99} 
Schmidt, G.~D., \& Hines, D.~C.\ 1999, \apj, 512, 125 

\bibitem[Schweitzer et al.(2006)]{schweitzer06} 
Schweitzer, M. et al. 2006, \apj, 649, 79

\bibitem[Shabala et al.(2011)]{shab11} 
Shabala, S.~S., Kaviraj, S., \& Silk, J.\ 2011, \mnras, 413, 2815 

\bibitem[Shields et al.(2003)]{shields03} 
Shields, G.~A., Gebhardt, K., Salviander, S., Wills, B.~J., Xie, B., Brotherton, M.~S., 
Yuan, J., \& Dietrich, M.\ 2003, \apj, 583, 124 

\bibitem[Shin et al.(2011)]{shin11} Shin, M.-S., Strauss, 
M.~A., \& Tojeiro, R.\ 2011, \mnras, 410, 1583 

\bibitem[Short \& Thomas(2009)]{short09} 
Short, C.~J., \& Thomas, P.~A.\ 2009, \apj, 704, 915 

\bibitem[Shupe et al.(2008)]{shupe08} 
Shupe, D.~L., et al.\ 2008, \aj, 135, 1050 

\bibitem[Siebenmorgen \& Kr{\"u}gel(2007)]{sieben07} Siebenmorgen, R., \& Kr{\"u}gel, E.\ 2007, \aap, 461, 445 

\bibitem[Sijacki et al.(2007)]{sija07} Sijacki, D., Springel, 
V., Di Matteo, T., \& Hernquist, L.\ 2007, \mnras, 380, 877 

\bibitem[Silk \& Rees(1998)]{silk98} Silk, J., \& Rees, M.~J.\ 1998, \aap, 331, L1 

\bibitem[Silk \& Nusser(2010)]{silk10} 
Silk, J., \& Nusser, A.\ 2010, \apj, 725, 556 

\bibitem[Silva et al.(1998)]{silva98} Silva, L., Granato, 
G.~L., Bressan, A., \& Danese, L.\ 1998, \apj, 509, 103 

\bibitem[Skrutskie et al.(2006)]{skrutskie06} Skrutskie, M.~F., et al.\ 2006, \aj, 131, 1163 

\bibitem[Snyder et al.(2011)]{snyd11} Snyder, G.~F., Cox, 
T.~J., Hayward, C.~C., Hernquist, L., \& Jonsson, P.\ 2011, \apj, 741, 77 

\bibitem[Somerville et al.(2001)]{somer01} 
Somerville, R.~S., Primack, J.~R., \& Faber, S.~M.\ 2001, \mnras, 320, 504 

\bibitem[Somerville et al.(2008)]{somer08} 
Somerville, R.~S., Hopkins, P.~F., Cox, T.~J., Robertson, B.~E., 
\& Hernquist, L.\ 2008, \mnras, 391, 481 

\bibitem[Spoon et al.(2007)]{spoon07} Spoon, H.~W.~W., 
Marshall, J.~A., Houck, J.~R., Elitzur, M., Hao, L., Armus, L., Brandl, 
B.~R., \& Charmandaris, V.\ 2007, \apjl, 654, L49 

\bibitem[Springel et al.(2005)]{sprdimher05} Springel, V., Di 
Matteo, T., \& Hernquist, L.\ 2005, \mnras, 361, 776 

\bibitem[Sturm et al.(2011)]{sturm11} Sturm, E., et al.\ 2011, 
\apjl, 733, L16 

\bibitem[Takagi et al.(2003)]{takagi03} Takagi, T., Arimoto, N., 
\& Hanami, H.\ 2003, \mnras, 340, 813 

\bibitem[Teyssier et al.(2011)]{teyss11} Teyssier, R., Moore, 
B., Martizzi, D., Dubois, Y., \& Mayer, L.\ 2011, \mnras, 414, 195 

\bibitem[Tombesi et al.(2010)]{tombesi10} Tombesi, F., Sambruna, 
R.~M., Reeves, J.~N., et al.\ 2010, \apj, 719, 700 

\bibitem[Tortora et al.(2009)]{tort09} 
Tortora, C., Antonuccio-Delogu, V., Kaviraj, S., et al.\ 2009, \mnras, 396, 61 

\bibitem[Tremaine et al.(2002)]{tre02} 
Tremaine, S., et al.\ 2002, \apj, 574, 740 

\bibitem[Trump et al.(2006)]{trump06} Trump, J.~R., et al.\ 
2006, \apjs, 165, 1 

\bibitem[Urrutia et al.(2008)]{redqsohst} Urrutia, T., Lacy, M., Becker, R. 
2008, \apj, 674, 80

\bibitem[Urrutia et al.(2009)]{f2ms} Urrutia, T.; Becker, R. H.; White, R. 
L.; Glikman, E.; Lacy, M.; Hodge, J.; Gregg, Michael D. 2009, \apj, 698, 1095

\bibitem[Veilleux et al.(2009)]{veill09} Veilleux, S., et al.\ 
2009, \apjs, 182, 628 

\bibitem[Verma et al.(2002)]{verma02} 
Verma, A., Rowan-Robinson, M., McMahon, R., \& Efstathiou, A.\ 2002, \mnras, 335, 574 

\bibitem[Voit et al.(1993)]{voit93} Voit, G. M., Weymann, R. J., Korista, K. 
T. 1993, \apj, 413, 95

\bibitem[Wagner \& Bicknell(2011)]{wagner11} 
Wagner, A.~Y., \& Bicknell, G.~V.\ 2011, \apj, 728, 29 

\bibitem[Werner et al.(2004)]{werner04} 
Werner, M.~W., et al.\ 2004, \apjs, 154, 1 

\bibitem[Werner et al.(2010)]{werner10} 
Werner, N., et al.\ 2010, \mnras, 407, 2063 

\bibitem[Werner et al.(2011)]{wern11} 
Werner, N., Sun, M., Bagchi, J., et al.\ 2011, \mnras, 415, 3369 

\bibitem[Weymann et al.(1991)]{weymann91} Weymann, R. J., Morris, S. L, 
Foltz, C. B., Hewett, P. C. 1991, \apj, 373, 23

\bibitem[Wright(2006)]{wright06} Wright, E.~L.\ 2006, PASP, 118, 1711 

\bibitem[Wright et al.(2010)]{wright10} Wright, E.~L., et al. 2010, \aj, 140, 1868 

\bibitem[Wyithe \& Loeb(2003)]{wyithe03} Wyithe, J.~S.~B., \& Loeb, A.\ 2003, \apj, 595, 614 

\bibitem[Xu et al.(2002)]{xu02} 
Xu, H., et al.\ 2002, \apj, 579, 600 

\bibitem[York et al.(2000)]{york00} 
York, D.~G., et al.\ 2000, \aj, 120, 1579 

\end{thebibliography}
\end{document}